\documentclass[12pt]{JHEP3}

\usepackage{graphicx}
\usepackage{cite}
 \usepackage{ae}
 \usepackage{aecompl}
\usepackage{amsmath,epsfig}
\usepackage{amssymb,amsfonts}
\usepackage{latexsym}
\usepackage{epsfig}
\usepackage{subfigure}

\relax
\renewcommand{\theequation}{\arabic{section}.\arabic{equation}}

\def\be{\begin{equation}}
\def\ee{\end{equation}}

\newcommand{\<}{\langle}
\renewcommand{\>}{\rangle}
\newcommand{\de}{\partial}
\newcommand{\bear}{\begin{eqnarray}}
\newcommand{\bea}{\begin{eqnarray}}
\newcommand{\eear}{\end{eqnarray}}
\newcommand{\eea}{\end{eqnarray}}
\newbox\pippobox

\def\II{\relax{\rm I\kern-.18em I}}

\def\cO{{\cal O}}
\def\e{\epsilon}

\def\sp{\;\;\;,\;\;\;}

\def\tr{\ensuremath{\mathrm{Tr}}}

\def\A{{\cal A}}

\title{Holographic RG flow and the Quantum Effective Action}

\author{Elias Kiritsis$^{1,2,3}$, Wenliang Li$^2$ and Francesco Nitti$^2$\\
 ~\\
 $^1$
 \href{http://hep.physics.uoc.gr/}
 {Crete Center for Theoretical Physics}, Department of Physics, University of Crete
 71003 Heraklion, Greece\\
  ~\\
$^2$
\href{http://www.apc.univ-paris7.fr}
{APC, Universit\'e Paris 7}, CNRS/IN2P3, CEA/IRFU, Obs. de Paris, Sorbonne Paris Cit\'e, B\^atiment Condorcet, F-75205, Paris Cedex 13, France (UMR du CNRS 7164).\\
 ~\\
 $^3$ \href{http://wwwth.cern.ch/}{Theory Group, Physics Department, CERN}, CH-1211, Geneva 23, Switzerland }

\abstract{The calculation of the full (renormalized) holographic action is undertaken in general Einstein-scalar theories. The appropriate formalism is developed and the renormalized effective action is calculated up to two derivatives in the metric and scalar operators. The holographic RG equations involve a generalized Ricci flow for the space-time metric as well as the standard $\beta$-function flow for scalar operators.
Several examples are analyzed and the effective action is calculated. A set of conserved quantities of the holographic flow is found, whose interpretation is not yet understood.    }


\preprint{CCTP-2013-23\\ CCQCN-2013-13\\CERN-PH-TH/2013-321}

\begin{document}

\maketitle 

\section{Introduction, Summary and Outlook} \label{intro}

The holographic correspondence between Quantum Field Theory (QFT) and String Theory (ST) has provided new insights for both of them. It was realized early on, that the UV divergences of QFT, corresponded to the IR divergences of ST near the AdS boundary\footnote{This matches previous expectations in string theory where the one-loop $\beta$-functions arise from the IR running of the amplitudes, \cite{beta}.}  , \cite{witten}. This correspondence, and the associated holographic renormalization has been made precise in a series of works, \cite{hs,BK,bianchi,ps,ps2,papa}, where the foundations  of holographic renormalization were laid in the case of Lorentz-invariant holographic QFTs.

The UV divergences in QFT are intimately related to the Renormalization group (RG) concept.
These particular items of the QFT tool box, namely the renormalization group flows, have an elegant description in the string theory/gravity language. They correspond to bulk solutions of the equations of motion with appropriate boundary conditions.

The relation of the second order equations of string theory and the first order equations of the QFT RG has been debated for quite a while, (see \cite{tseytlin} and references therein). In the context of holography
 there is a way of writing the bulk equations using the Hamilton-Jacobi formalism, so that they look formally similar to RG equations, \cite{deboer,ps,ps2,papa,bourdier}, a fact that has been exploited in numerous situations. There have been various takes on the form of the RG/coarse graining procedure in the holographic case, \cite{holorg}, including the Wilsonian approach to IR physics \cite{cgkkm}.

However, the difference between first and second order equations is of crucial importance and the gap between the two descriptions seems still open.
Holography suggests that in an appropriate large-N limit the string theory equations must be related to the QFT RG equations. This particular relation
can be found by considering the generalized source functional for QFT, defined properly so that the global symmetries of the QFT are realized non-linearly as local symmetries.
It has been argued that string theory is the dynamics of sources of QFT, \cite{LV,string,sslee}. In particular,  in the QFT source functional
the sources become dynamical variables if multitrace operators are integrated out. This leads to a theory that is reminiscent of string theory in the appropriate limits, \cite{sslee}.
This defines a quantum version of the QFT RG group that is second order in derivatives and is expected to match with the string theory equations in holographic contexts.

In \cite{papa1} the holographic effective action for scalar-tensor theories was calculated to second order in derivatives as a functional of the UV sources for Lorentz invariant states using the Hamilton Jacobi formalism.

In \cite{KN} a step was taken towards calculating the effective action  beyond the Lorentz-invariant case. In particular, the problem that was addressed is the (holographic) calculation of the quantum renormalized effective potential for scalar operators. When this is calculated in states that are Lorentz invariant (like the vacuum state), then the Hamilton-Jacobi (HJ) formalism developed for holographic renormalization, is sufficient in order to calculate the effective potential.

For more general applications though, especially for physics at finite temperature and density, it is not yet known how to apply the  HJ formalism in order to calculate the effective potential. In \cite{KN} a different method was used that works also in non-trivial states that break Lorentz invariance, and the calculation of the effective potential was reduced to the (generically numerical) solution of several non-linear first order ordinary differential equations. In the scaling regions, this equation can be solved analytically and the effective potential calculated. This provides, among others,  tools to calculate the presence and parameters of non-trivial phase transitions at strong coupling.\\

The purpose of the present paper is to go beyond the effective potential, and provide a formalism and practical algorithm in order to calculate the renormalized effective action as explicitly as possible, in a simple Einstein-scalar theory.

Our algorithm will  make use of  a derivative expansion, involving the derivatives with respect to transverse (i.e. non-holographic) coordinates. Here, we will stop at second order in derivatives, which includes the Einstein-Hilbert term and scalar kinetic term.  The result will be expressed in terms of  {\em covariant, universal terms} whose functional form is  scheme-independent.  Each of these terms depends on a constant which completely  encodes the scheme dependence of the renormalization procedure.
We will explicitly write the renormalized  generating functional (and the corresponding effective action) as expressed in terms of the coupling (and respectively, of the classical operator) at a fixed finite scale, rather than in terms of bare UV source and operator vev. This allows to apply our results  to theories which do not admit an $AdS$ fixed point solution in the UV.

In this paper, we will assume that the state in which the calculation is done is Lorentz invariant, however our procedure is easily generalized to less symmetric homogeneous configurations, and in a subsequent paper, we will also develop the calculations to non-Lorentz invariant setups, \cite{kln}.

\subsection{Summary of results}

For simplicity\footnote{The case with several scalars can be treated with the same techniques and involves no further issues. The addition of vector fields, of direct interest to condensed matter/finite density applications, can be also addressed and will be undertaken in \cite{kln}.}  we focus on a $d+1$-dimensional\footnote{Since we are going to write expressions valid in gerneral for any $d$, we assume $d>2$. For $d=2$, one finds a logarithmic divergence (and correspondingly a Weyl anomaly) already at second derivative order. Therefore, although our methods can still be used,  this case must be considered separately.} gravity theory,  with a metric $g_{\mu\nu}$, a  single scalar field $\phi$, and an arbitrary potential $V(\phi)$ whose bulk action is, schematically:
\be
S_{bulk} = \int d^{d+1} x\, \sqrt{-g}\left( R^{(d+1)} - {1\over 2}(\de \phi)^2 + V(\phi) \right).
\ee
In the dual field theory,  the scalar field $\phi(x^\mu,u)$ and the $d$-dimensional induced metric $\gamma_{\mu\nu}(x^\mu,u)$  represent spacetime-dependent (and RG-scale dependent) sources for the field theory stress tensor and a scalar operator $\cO$. The goal will be  to  compute the finite, renormalized generating functional of connected correlators,  $S^{(ren)}[\phi,\gamma]$,  and its Legendre transform, i.e. the quantum effective action,  $\Gamma[\cO, \gamma],$ to all orders in $\phi$ and $\gamma$ and up to second order in their space-time derivatives.  We will take into account the full backreaction of the scalar field on the metric, and we will express the final result in a fully $d$-dimensionally covariant form, but we will neglect higher-derivative and higher-curvature terms. Explicitly, the result will  take the form of two-derivative covariant actions:
\bea
S^{(ren)}[\phi,\gamma] &&= \int d^d x\sqrt{-\gamma} \left[Z_0(\phi) + Z_1(\phi) R + Z_2(\phi) \gamma^{\mu\nu} \de_\mu \phi \de_\nu \phi\right] , \label{genfun}\\
\Gamma[\cO,\gamma] &&= \int d^d x\sqrt{-\gamma} \left[\tilde{Z}_0(\cO) + \tilde{Z}_1(\cO) R + \tilde{Z}_2(\cO) \gamma^{\mu\nu} \de_\mu \cO \de_\nu \cO\right]. \label{qea}
\eea
where $R$ is the intrinsic curvature of the $d$-dimensional induced metric $\gamma_{\mu\nu}$.
The functionals $S^{(ren)}[\phi,\gamma]$ and $\Gamma[\cO,\gamma]$ are related by Legendre transform:
\be
\cO = {\delta S^{(ren)} \over \delta \phi}, \qquad \Gamma[\cO,\gamma] = \cO \phi(\cO) - S^{(ren)}[\gamma, \phi(\cO)],
\ee
and we  use the shorthand $\cO$ for the classical field $\<\cO\>$. We will give the explicit form of the functions $Z_i(\phi)$ and $\tilde{Z}_i(\cO)$, in terms of the leading order homogeneous solution of the bulk gravity theory, specified by the form of the scale factor $\A(\phi)$, or equivalently by the lowest-order superpotential function.

Throughout the process, we will first need to compute the divergent, bare on-shell action, and identify the counterterms. This has already been done, for general dilaton-gravity theories,  in \cite{papa} and our result reproduces the same counterterms that were found there.

Here however,  we will go one step further and write explicitly the finite terms, in a way which can be directly used to compute correlation functions or expectation values and that makes manifest both the scheme dependence (coming from the choice of the counterterms) and the dependence on the renormalization conditions. In  other words, the  holographic quantum generating functionals (\ref{genfun}-\ref{qea})  will be   expressed as   functionals  of  the  couplings at a finite, arbitrary holographic RG-scale $\mu$ (not necessarily in the UV) in a way which matches what is done in field theory.

This allows, among other things, to derive the renormalized trace identities directly,  to identify unambiguously the {\em exact} holographic $\beta$-function for the  source $\phi$, and to determine unambiguously the relation between a change in the  holographic coordinate on the gravity side, and a RG transformation on the the field theory side. In the special case of a homogeneous background, the  RG-scale $\mu$ has to be  identified,  {\em anywhere in the bulk } (i.e. not only close to an AdS fixed point, but for a generic RG flow solution away from any fixed point),  with the metric warp factor  of the leading order homogeneous solution,
\be\label{intrometric}
ds^2 = du^2 + e^{2A(u)} \gamma_{\mu\nu} dx^\mu dx^\nu, \qquad \mu \equiv \mu_0 e^{A(u)}
\ee
up to an arbitrary normalization constant $\mu_0$.

It is important to stress that our assumptions on the bulk theory and on the types of solutions are extremely  general: in particular we will {\em not} assume that the theory represents a deformation of an $AdS$ fixed point, nor any particular type of asymptotics for the potential. The minimal requirement is that the  bulk theory admits solutions which contain a UV region, in which to lowest order,  the scale factor in (\ref{intrometric}) goes to infinity. This of course applies to  deformations of  $AdS$ by a relevant operator, but also includes theories with asymptotically free  $AdS$ solutions like Improved Holographic QCD \cite{gkn}, or with scaling solutions associated with exponential potentials (see e.g. \cite{cgkkm}) displaying hyperscaling violations\footnote{For such theories, renormalization has been discussed   in \cite{kst}.}.

In all these cases,  the success of the holographic  renormalization procedure depends on the fact that the {\em UV solution is an attractor.} This will be discussed more extensively in Sections 2 and 5, but essentially it means that there must be a continous family of inequivalent bulk solutions (i.e. such that they cannot be simply obtained from each other by a change in the initial condition of the RG-flow)  which have the same UV asymptotics.  If this is not the case, then one cannot identify counterterms which universally subtract the divergences and holographic renormalization fails.

  Although this is a very interesting problem, we will not explore here the most general conditions that allow the theory to be renormalized, but we will analyze case by case whether this happens in examples.

In order to compute the quantum generating functionals (\ref{genfun}-\ref{qea}), we first rewrite the bulk Einstein's equations as covariant flow equations, order by order in a derivative expansion. We will  then use these equations to write explicitly the on-shell action, which in holographic theories is the bare generating functional. The latter has the same form as in (\ref{genfun}), with different functions $Z_i^{bare}(\phi)$ but the sources are evaluated in the UV limit, and is typically divergent.

Next, we identify the counterterms to subtract the divergences, which agree with  those that were found for general Einstein-dilaton theories in \cite{papa} using a procedure similar to ours but using  the Hamilton-Jacobi method.

 We then proceed one step further and write explicitly the finite part which remains after the divergence is subtracted, i.e. the renormalized generating functional (\ref{genfun}).

With our ansatz for the flow equations we  are only  able to tackle bulk geometries  which respect the symmetry of the leading homogeneous term: these symmetries restrict  the terms we include in the flow equation ansatz. For example, only metrics which to lowest order have a space-time-isotropic  radial evolution can be treated  with our ansatz. This does not include finite temperature or density solutions (on the other hand, the counterterms are universal and will renormalize these solutions as well). However,  our method can be easily applied to these solutions as well, once a less symmetric ansatz for the flow equations is assumed, and this will be pursued in  future work \cite{kln}.

The first order holographic flow equations as well as  the coefficient of the bare and renormalized effective action, up to two derivatives, depend only on two functions ({\em generalized superpotentials}) of the scalar field, $W(\phi)$ and $U(\phi)$, which satisfy a simple system of first order ordinary differential equations:
\begin{eqnarray}
V&=&\frac d {4(d-1)}W^2-\frac 1 2{W^\prime}^2, \label{flow-eq-3-intro}\\
1&=&W'U'-\frac {d-2} {2(d-1)}WU. \label{flow-eq-4-intro}
\end{eqnarray}
where $V(\phi)$ is the scalar potential appearing in the bulk action. These equations had already  appeared in \cite{papa,papa1}, where they were derived using the  the Hamilton-Jacobi method, and were shown to govern the zero and second order on-shell action and counterterms.

The first superpotential function $W(\phi)$ is the familiar one that it is often used to find  the lowest order, or background (i.e. homogeneous in the space-time coordinate), solution   of Einstein's equation: to lowest order in the derivative expansion and for a flat space-time metric $\gamma_{\mu\nu} = \eta_{\mu\nu}$, the scale factor in (\ref{intrometric}) and the scalar field solve the first order system:
\be\label{bg-intro}
\de_u A(u) = -{1\over2(d-1)}W(\phi), \qquad \de_u \phi(u) = W'(\phi)
\ee

The flow equations are the covariant generalization of  (\ref{bg-intro}), that include the effect of the space-time dependence of $\phi$ and $\gamma_{\mu\nu}$, and are given in  (\ref{flow-eq-1}-\ref{flow-eq}). These equations  can also be put in a form of geometric RG-flow equations:
\be \label{rgflow}
\Delta \gamma_{\rho\sigma}(x) = 2\gamma_{\rho\sigma} + \beta^{(2)}_{\rho\sigma}  \,, \qquad \Delta \phi(x) = \beta^{(0)}_\phi + \beta^{(2)}_\phi
\ee
Equations (\ref{rgflow}) can be  interpreted as RG-flow equations for the  space-time dependent coupling and the four-dimensional metric  in the dual field theory,  where $\Delta$ is the generator of an RG transformation, $\beta^{(0)}_\phi$ is the $\beta$-function for the space-time independent coupling in flat spacetime, and  $\beta^{(2)}_\phi$ and $\beta^{(2)}_{\rho\sigma}$ are constructed  from covariant,  two-derivative terms, with coefficients which are functions of $\phi$ and are determined by  $W(\phi)$ and $U(\phi)$.
 The $\beta$-functions  are explicitly given in equations (\ref{beta0}-\ref{beta2-2}) and (\ref{X}-\ref{Z}) . The flow of the metric is a generalization of the Ricci flow, \cite{friedan}, that emerged first from 2d $\sigma$-models.

Using the flow equations, we compute the  bare on-shell action, which takes the simple form,
\be\label{introS}
S_{on-shell}=\int d^d x\sqrt\gamma\left(W(\phi)-  U(\phi) R-\left(\frac W{W'}  {U}'(\phi)\right)\frac 1 2\gamma^{\mu\nu}\partial_\mu\phi\partial_\nu\phi+...\right)_{u_{UV}}
\ee
and matches the result found in \cite{papa1} using the Hamilton-Jacobi method.
A similar contribution from the IR  region is in general possible, but it turns out that the derivative expansion breaks down unless $W$ and $U$ satisfy suitable regularity conditions in the IR. Under this condition, the IR contribution vanishes and the on-shell action can be completely written  in
terms of the UV data.

One interesting phenomenon we find in the course of our analysis is the existence of  $(d+1)$-dimensional currents which are conserved as a consequence of the flow equations. Up to two derivative order, these conserved quantities are in one-to-one correspondence with the independent covariant terms in the $d$-dimensional effective action, i.e. we find three of them. The first one, associated to the potential term, already appears at the homogeneous level, and receives higher derivative corrections at second order.
Up to this point, it is unclear to us what is the physical  meaning of these conserved quantities, but it would be extremely interesting to investigate whether this phenomenon extend to higher orders in the derivative expansions,
with the new conserved quantities appearing at each order.

 The existence of the conserved currents allows to write explicitly the renormalized generating functional, which in fact is expressed as a linear combination of the radial ``charges'' associated to these currents. Their  indepedence of the holographic coordinate allows us to write the renormalized generating functional in terms of the coupling and metric at {\em any point} along the holographic RG-flow. From the dual field theory point of view, this means having an explicit expression for the generating funcional as a function of the coupling at an arbitrary finite scale.  Renormalization group invariance of $S^{(ren)}$  is manifested by the fact that the functional is constant along a holographic RG-flow trajectory, and we show that it is  expressed by the local RG-invariance equation:
\be\label{intro-RG}
\left(2\gamma^{\mu\nu}{\delta \over \delta \gamma^{\mu\nu}} - \beta^{(2)}_{\mu\nu}{\delta \over \delta \gamma_{\mu\nu}} -  \beta_\phi {\delta \over \delta \phi} \right) S^{(ren)} = 0,
\ee
which holds  up to four-derivative terms.

As anticipated, the renormalized generating functional and quantum effective action take the form (\ref{genfun}-\ref{qea}), where the functions $Z_i(\phi)$ are expressed in terms of the  same superpotential functions $W$ and $U$.   Each of the three independent terms is multiplied by an  arbitrary constant,  which reflects the scheme dependence of the holographic renormalization procedure for the three independent operators of order up to two derivatives, i.e. the potential, Ricci and scalar kinetic term. The explicit result is given in equation (\ref{running}) for the generating functional of connected correlators, and in equation (\ref{1PI}) for the quantum effective action.

Equation (\ref{flow-eq-4-intro})  for the second superpotential, which governs the
2-derivative terms, is linear, and $U(\phi)$ can be easily written in terms of $W$ up to an integration constant. This integration constant can be fixed by assuming a suitable regularity condition in the infrared (essentially, that the derivative expansion of the flow equations and of the on-shell action  holds in the limit where $e^A\to 0$).

Therefore, effectively, the superpotential $W(\phi)$ is the crucial ingredient that determines both the lowest order background solution, {\em and} the higher derivative terms. It is also what determines the $\beta$-function of the theory: as we  have anticipated, the renormalized trace identities will lead us to  identify the metric scale factor with the RG energy scale, thus the $\beta$-function for a homogeneous coupling is, from equation (\ref{bg-intro}), \cite{gkn,bourdier}:
\be
 \beta^{(0)}_\phi = {d\phi \over dA} =  -2(d-1) {W' \over W}.
  \ee

Different $W$ corresponds therefore to different classes of RG-flows (i.e. different $\beta$-functions), and  it seems that to a given bulk theory, specified by the potential $V(\phi)$,  there correspond an infinity
of boundary field theories. This is a reformulation of the usual puzzle about holographic RG-flows.

The question of how one particular superpotential function is selected is closely related to the IR regularity condition: if one imposes some (mild) conditions in the IR  then one finds one or at most a finite number of ``good'' superpotential solutions. For example,  Gubser's criterion, that the IR solution can be  uplifted to an arbitrarily small black hole,  is not expected to  hold for a general solution at large $\phi$ of equation (\ref{flow-eq-3-intro}).  This was observed for example in \cite{thermo2} (in particular see  Appendix F of that work), in the case of an exponential superpotential at large $\phi$. Similarly, for a generic solution the fluctuation equations need extra boundary conditions in the IR to completely specify the spectral problem, which means that the dynamics of  these solutions is  driven by some extra unknown features localized in the IR (IR branes, or higher curvature terms) that are not included in the 2-derivative Einstein-dilaton action \cite{5dgraviton}. It remains an open question whether one can find a {\em dynamical} selection criterion for the superpotential in the full quantum  gravity theory which allows to discard the generic solutions of (\ref{flow-eq-3-intro}) and to pick only the ``regular'' ones.

In the last part of this work we discuss the calculation of the effective action in a few explicit examples:
\begin{enumerate}
\item{ $AdS$ {\em deformation}}\\
 The first one is the standard holographic setup of a deformation of a UV  $AdS$ fixed point by a relevant operator, realized around an extremum of the scalar potential $V$,
\be
V\simeq {1\over \ell^2} \left[d(d-1) + {\Delta (d-\Delta)\over 2}\phi^2 + O(\phi^4)\right]
\ee
where we have chosen $\phi_{UV}=0$ without loss of generality.

This example was discussed  in detail in  \cite{papa1}. Compared to more general situations, due to asymptotic conformal invariance in this case it is possible to define finite  {\em bare} couplings, i.e. the   source term which governs the leading UV asymptotics of the  scalar field, and represents the UV CFT deformation parameter.  The scalar field can be related to the UV limit of the induced metric $\tilde{\gamma}_{\mu\nu} = \lim_{u\to -\infty} e^{2u} \gamma_{\mu\nu}$, and of the UV  coupling of the CFT, $\alpha = \lim_{u\to -\infty} e^{-(d-\Delta)u} \phi(u)$, where $\Delta$ is the dimension of the operator dual to $\phi$. As a consequence, the renormalized generating functional assumes a simple form in terms of the dimensionful coupling $\alpha$ which defines the deformation,
\be\label{introads}
S^{(ren)}[\alpha, \tilde{\gamma}] = \int d^d x \sqrt{\tilde{\gamma}}\left[c_0 \alpha^{d\over d-\Delta} + \alpha^{d-2\over d-\Delta}\left( c_1  R^{(\tilde{\gamma})} + c_2  \tilde{\gamma}^{\mu\nu}(\de_\mu \log \alpha)(\de_\nu \log \alpha ) \right)\right]
\ee
which up to the scheme-dependent constants $c_i$ is completely dictated by the UV conformal invariance. This expression has the same form as the one that was  obtained in \cite{papa1}.

When expressed in terms of the running coupling and metric, the  generating functional has the same form as (\ref{introads}), but with $\alpha$ and $\tilde{\gamma}_{\mu\nu}$ replaced by $\phi(u)$ and $\gamma_{\mu\nu}(u)$. This form will change as we go deeper in the bulk towards the IR, where it will depend on the details of the bulk solution.

\item{{\em Eponential potentials}}\\
Another class of examples we analyze involves potentials which asymptote a simple exponential in the UV (chosen to be at $\phi\to -\infty$),
\be
V(\phi) \simeq V_0 e^{b\phi} \qquad \phi \to -\infty.
\ee
 These asymptotics generate scaling solutions, which in some cases can be obtained by a ``generalized dimensional reduction''  on torii or  on spheres of pure Einstein gravity in higher dimensions \cite{kst,gout} and therefore have  a hidden conformal symmetry. These solutions are  holographically-renormalizable for $b<\sqrt{2d/(d-1)}$, in the sense that the UV attractor condition is realized. This in fact coincides with Gubser's bound and with the case when the exponential potentials can be realized by generalized  reduction from a higher-dimensional pure gravity theories.

 However, these theories are not  ``UV complete,'' and there is no natural definition of a UV bare coupling as in asymptotically AdS solutions.  Therefore,  they intrinsically need to  be defined at some finite scale. We can still write the generating functional and effective action  as a function of the metric and the coupling at a given finite scale, and the result in the UV is again very simple and dictated by scaling (see equation (\ref{Sexp}):
\be
S^{(ren)}[\gamma,\phi] = \int d^d x\sqrt{-\gamma}\left(D_0 e^{\frac d {b(d-1)}\phi}+e^{\frac {d-2} {b(d-1)}\phi}\left(D_1 R +\tilde D_2\frac 1 2 \gamma^{\mu\nu}\partial_\mu \phi\partial_\nu \phi\right)\right)
\ee
In this case it is exponentials of $\phi$ which have definite scaling dimension, i.e. $e^{\phi}\sim e^{-b(d-1)A}$, thus the coupling $e^{\phi}$ has dimension  $b(d-1)$.  With this counting, we see that all terms are again  fixed by covariance and dimensional analysis in  the absence of other dimensionful parameters.

Interestingly, the constant $\tilde{D}_2$ is fixed in terms of $D_1$, so in these theories scheme dependence is encoded in only two arbitrary constants. This is probably due to the relation of these models to dimensionally reduced pure gravity theories, where only two independent covariant counterterms  are possible up to  two-derivative order, i.e. the cosmological constant and the Einstein-Hibert term.

\item{{\em Asymptotically free fixed points (i.e. IHQCD)}}\\
Next, we analyze the case of asymptotically free fixed points that includes Improved Holographic QCD (IHQCD) \cite{gkn}.  In the UV, i.e.  as $\phi\to -\infty$ the potential has an  expansion in exponentials, $V\sim V_0 + V_1e^{\phi}+ \ldots$ and the attractor  solutions are $AdS$ with logarithmic corrections, which mimic asymptotic freedom of the coupling $\lambda \propto e^\phi$. In this case our most interesting result is the computation of the two-derivative quantum effective action in flat space for the {\em canonically normalized} operator $\cO = \tr F^2$ . At one-loop order (i.e. stopping at the first non-trivial  term in the exponential  expansion of $V(\phi)$, we find:
\be
\Gamma(\cO) \, \sim_{UV}\,   - \int d^d x \left(\cO^4 \log\cO + {1\over 2}(\de \cO)^2 \right)
\ee
The effective potential $\sim \cO^4\log \cO$ is exactly the one which is needed to reproduce the one-loop Yang-Mills trace anomaly, as shown for example in \cite{preskill}.

In the IR, i.e. the  region of large positive $\phi$ the potential of IHQCD has a (corrected) exponential  asymptotics,
\be
V\simeq V_{\infty} (\log \phi)^{P}e^{2\phi/\sqrt{6}}
\ee
In this case, the renormalized effective potential for the canonically normalized operator $\cO$ has a similar expansion in the IR,
\be
\Gamma(\cO) \, \sim_{IR} \,  - \int d^d x \left(\cO^4 (\log\cO)^{6P} + {1\over 2}(\de \cO)^2 \right)
\ee
we see that the logarithmic term is fixed by the subleading behavior in the IR parametrized by $P$. In IHQCD the value $P=1/2$ is assumed, leading to  $V_{eff}\sim \cO^4(\log \cO)^3$.

\item{{\em $AdS$-to-$AdS$ flow}}\\
Finally,  we analyze numerically a full flow from a UV to an IR AdS fixed points, assuming a potential with a simple polynomial form. We explicitly show the space of solutions of the superpotential equation for $W(\phi)$ (which was already studied in detail in \cite{martelli}) and the one for $U(\phi)$, and we reproduce the known fact that there is a single solution among a continuous infinity which interpolates between the UV and IR fixed points. The other solutions either overshoot or undershoot the IR fixed point. We compute the full effective potential numerically as a function of the running coupling and the energy scale, showing how the simple power-law potential in equation (\ref{introads}) gets modified as the theory flows to the IR.

 \end{enumerate}

\subsection{Relation to previous work}

There exists a substantial literature on holographic renormalization (see \cite{hs,BK,bianchi,ps,ps2,papa} and references therein). In particular,  boundary counterterms  for asymptotically $AdS$ spacetimes with  non-trivial bulk scalar field profiles were  discussed  in detail in  \cite{bianchi}\footnote{An earlier analysis of the Einstein-scalar field system can be found in   \cite{odintsov}, which correctly captured  the leading divergent part of the on-shell action.}. The most general divergence structure and  counterterms in an Einstein-dilaton setup with Lorentz invariance were given in \cite{papa}, using the HJ method. Our methods agree and reproduce the divergences and counterterms that appeared in that work.

In \cite{papa1} the effective action for Lorentz invariant states was calculated to second order in derivatives as a function of the UV sources for deformations of an $AdS$ fixed point using the HJ formalism, and several non-linear examples were worked out.
Here, we rather put the emphasis on the calculation of the generating functional  as a function of the renormalized fields, defined at a finite energy scale, giving  particular importance to the renormalization scheme dependence and the RG flow.  This allows us to treat general theories which do not have an $AdS$ fixed point in the UV and in which there exists no  good definition of finite UV sources. In the special case of a relevant deformation of  an $AdS$ fixed point,  we arrive at an explicit expression for the action that agrees  with \cite{papa1}.

To compute the generating functional, we  use two different techniques. The first one is close to the HJ method used in \cite{papa,papa1}, and highlights the role played by IR regularity of the bulk solution.
Our second method is the generalization of the one used in \cite{KN} for the calculation of the effective potential, and it consists in writing the on-shell action as a total derivative. This method can be  easily generalized to more complicated setups which display finite temperature and density, which seem not to be easily approached via the HJ method.

Both of our methods are based on writing the bulk field equations in the form of first order flow equations in a derivative expansion.  In this sense, the present work is similar in spirit to what was done in the  recent work \cite{stan}, in which a similar expansion was used to derive the flow from UV to IR effective hydrodynamic in the holographic description. The difference lies in the fact that here we use an expansion in  small  deviations from a Lorentz-invariant, but non-conformal background, whereas in \cite{stan} the expansion is a long-wavelength expansion in the fluid velocity in a pure gravity bulk theory.

As we already mentioned, there has been recently a renewed interest in understanding the  holographic analog of the Wilsonian picture of effective field theory, and on the definition of the Wilsonian effective action \cite{holorg}. In this work we do not address this problem directly, as our effective actions are calculated integrating over the whole bulk down to the IR\footnote{The fact that, even integrating all the way down to the IR,  we find a {\em local} effective action depends on  the fact
that we work only up to second order in derivatives, and non-local terms are expected to appear at higher orders.}. However, our method applies easily to the computation of the Wilsonian effective action: in particular, we show  that the bare bulk on-shell action can be written as the difference between an UV and an IR contribution, for which we provide explicit expressions in terms of the superpotentials.  This makes straightforward  to move the integration from the far IR to an intermediate cutoff $\Lambda_{IR}$, giving rise to an extra contribution from the IR, which will have the same form as in equation (\ref{introS}).
This is exactly how Wilsonian effective actions were obtained in \cite{holorg}. These works however were either limited to a probe scalar in $AdS$, or in the backreacted case, to no derivative terms. Using the results in the present paper one can easily find the full Wilsonian action up to two derivatives.

\subsection{Outlook}

Although the generating functional of correlation functions has been a central element in the AdS/CFT correspondence, there are many issues that remain obscure concerning its definition and properties especially related to its (local) symmetries. These issues have been recently discussed in \cite{LV,string,anselmi}.
In particular,  the nonlinear realization of local symmetries is subtle as it is not unique, and a stronger principle must be applied in QFT, probably local Weyl invariance.

The issue of local Weyl invariance and the transformation of the QFT source functional under local Weyl transformations was addressed by Osborn in several works \cite{osborn}. The variation was associated, after local redefinitions, to standard $\beta$-functions as well as a local functional of dimension $d$ involving the metric and scalar coupling constants.  In  \cite{nakayama}, a  holographic calculation of vector beta-functions in the probe limit around $AdS$ was also performed (see also \cite{string}).
The above matches holographic approaches to renormalization, and is equivalent to the conformal anomaly in the presence of scalar operators as described in detail in \cite{papa}.

However, what remains again obscure, is to what extend the divergent part of the source action determines all correlation functions, as was suggested in \cite{string}.
Indeed, in holography there  seems to be a 1-1 correspondence between the full bulk action and the boundary counterterms necessary for its renormalization. This has been argued in a different context and language by Lovelace in \cite{lovelace}, and this is indeed the case in 2d QFTs.

In view of this, our calculations in this paper as well as the full set of divergent terms in \cite{papa} go a long way towards fixing the renormalized effective action of holographic scalar flows.

 Another interesting issue that we only touched in this paper is the question: which gravitational bulk actions correspond to renormalizable QFTs. A priori this questions can be systematically answered by comparing the asymptotic solutions to various gravitational actions near scaling boundaries.

As we already remarked, an obvious generalization of this work is the computation of the effective action for less symmetric leading-order solutions, which include black holes and eventually non-trivial bulk
gauge fields. Such backgrounds  display dissipative physics, encoded into retarded finite temperature correlators and non-zero transport coefficients that govern entropy production. It is expected that this will lead to a complex effective action that will trigger the dissipative behavior that arises in this context.

A related direction that is currently pursued is related to including a U(1) symmetry, the associated gauge field in the bulk and finite density in the boundary theory. This is an a priori straightforward application  that is interesting in the context of  condensed matter physics.
The first step in this direction, namely the calculation of the effective potentials for scalar sources has already been achieved in \cite{KN}.

Several non-local holographic observables satisfy similar types of flow that descend from the linear bulk flows of the fields. A generic example is the Wilson loop expectation value, given in holography by the minimization of the 2-surface ending on a given boundary loop, \cite{malda,ry}. In such a case, the RG description of non-local expectation values will be given in terms of a generalized mean-curvature flow of the type described in \cite{bakas}.
Confining and non-confining behavior can then be traced to different behaviors of generalized extrinsic curvatures.
It is an interesting project to derive this flow for different loops and different geometries and try to identify it with other dynamical equations Wilson loops are known to satisfy in gauge theories.

Similar comments apply to another well known non-local observable, namely entanglement entropy as given by the Ryu-Takayanagi conjecture, \cite{tadashi}. From the formulation of the problem we expect also that it will satisfy a mean curvature type flow.

Finally, our work can be very useful for phenomenological model-building using the  holographic approach, for example for holographic realization of cosmological inflation and BSM physics: one can construct models with a strongly coupled sector described holographically, but work directly with a purely $d$-dimensional effective action which however already includes all quantum
corrections coming from the strongly coupled sector. This can subsequently be coupled to an observable sector directly in the lower-dimensional theory, as a function of the energy scale of the interaction.   We would like to stress that for this type of construction to make sense, it is essential to write the effective action up to two derivatives: the effective potential alone contains only information about the ground state, but to understand fluctuations and/or non-static solution (as would be needed e.g. for inflationary model building)
knowledge of the the kinetic term is crucial.

\subsection{Paper structure}
This work is organized as follows.

In Section  \ref{review} we describe our setup, review homogeneous solutions, and provide a simple calculation of the generating functional potential in flat space, as this shows the main idea on which the general covariant calculation is modeled.

In Section \ref{flow} we write Einstein's equations as covariant flow equations up to second order in space-time derivatives, and identify the superpotential equations. Then,  we relate the radial flow to local Weyl transformations and compute the metric and scalar field beta functions.

In Section \ref{effact} we use the flow equations to compute the on-shell action in two different ways.

In Section \ref{renorm} we perform the holographic renormalization procedure and isolate the finite terms, writing the renormalized generating functional as
a sum of covariant terms up to two derivatives. We discuss its renormalization group invariance, and derive the trace identities.  Next, we perform the Legendre transform with respect to the renormalized source and write the quantum effective action  up to two derivatives for the corresponding operator.

In Section \ref{curvature} we use a different  trick to compute the coefficient of the Ricci scalar in the on-shell action: we solve explicitly the bulk Einstein equation in the case of a constant curvature metric on spatial slices, in a perturbation expansion in the spatial curvature.  The result is again consistent with what we obtain in our general discussion, and can in principle be used to compute higher curvature terms.

In Section \ref{examples} we discuss explicit examples: deformations of AdS
 fixed points, exponential potentials and IHQCD, and we give the full  numerical solution of a complete flow between a UV and an IR fixed point.

Several technical details about the ADM formalism, Lie derivatives, flow equations and the identification of the terms in the on-shell action are contained in Appendix A. There, we also derive the constraints on the coefficients of the flow equations coming from the Hamiltonian and momentum constraints, and we show that the dynamical Einstein equations are automatically satisfied by the flow equation ansatz.

In Appendix B we discuss a possible  ambiguity in the identification of the generator of RG transformations, which gives rise to a scheme dependence in the $\beta$-functions. We argue that there is a  unique natural choice of scheme which allows to reproduce the standard form of the field theory trace identities.

In  Appendix C we show that the
on-shell action can be obtained by an alternative method, i.e. by solving the fluctuation for the scalar (gauge-invariant) variable of the system, computing its on-shell action, and covariantizing the result. This gives the same output for the two-derivative action.

\section{The holographic effective potential, revisited}
\label{review}
\subsection{Einstein-Scalar gravity and holography}

Throughout this work, we will consider the holographic calculation of the renormalized effective action for a
scalar operator $\cO$, living in a $d$-dimensional boundary theory which
possesses a $d+1$-dimensional gravitational (holographic) dual. We will assume $d>2$.

To be specific, we consider a minimal holographic setup, where the only degrees of freedom are the $d+1$-dimensional metric $g_{ab}$ with signature $(-+\ldots+)$, and a bulk scalar
field $\phi$ (dual to the operator $\cO$). The bulk theory is Einstein-Scalar gravity with the following action
\begin{equation}
S=S_{bulk}+S_{GH},
\end{equation}
where $S_{bulk}$ is
\begin{equation} \label{action}
S_{bulk}= M^{d-1}\int \,d^d x \,d u\sqrt{-g}\left[R^{(d+1)}-\frac 1 2g^{ab}\partial_a\phi\partial_b\phi+V(\phi)\right],
\end{equation}
and $S_{GH}$ is the boundary Gibbons-Hawking (GH) term required by a
well-defined variational problem with Dirichlet boundary conditions,
\begin{equation}
S_{GH}= 2M^{d-1}  \int \,  \left. d^d x \big(\sqrt {-\gamma} K\big) \right|_{UV}^{IR},
\end{equation}
In these expressions,  $g_{ab}$ is the bulk metric, $V(\phi)$ is the
(bulk) scalar potential, which for now we keep arbitrary. The coordinate  $u$
parametrizes the holographic direction, and $u_{UV}$ and $u_{IR}$ denote the ultraviolet and the
infrared endpoints of this coordinates, which may be the
physical $IR$ and $UV$ of the full theory, or may denote $UV$ and/or
$IR$ cutoffs. In the GH term,  $\gamma_{\mu\nu}$ is the induced
metric on the  slices and $K=\gamma^{\mu\nu}K_{\mu\nu}$ is the trace
of extrinsic curvature.  Here and in the following discussion, the subscripts $UV$ or $IR$ mean  that the quantities are evaluated on the $UV$ or $IR$ slices.

The Planck scale $M_p^{d-1}$ is  considered  very large with respect to the typical curvature scale of the bulk solutions (of order $N^2$ with respect to the large-$N$ parameter of the dual field theory). Since $M_p$  always appears only as an overall factor, we will omit it from now on.

Einstein's equations are obtained by varying  the action $S$ with respect to the metric $g_{ab}$
\begin{equation} \label{AE}
R_{ab}^{(d+1)}-\frac 1 2 R^{(d+1)} g_{ab}=\frac 1 2\partial_a\phi\partial_b\phi-\frac 1 2 \left(\frac 1 2 g^{cd}\partial_c\phi\partial_d\phi-V   \right)g_{ab}.
\end{equation}
The scalar field equation of motion is not independent, and will not
be used here.

\subsection{Homogeneous solutions: the superpotential}

As a first step, we will discuss in detail the calculation of the
holographic renormalized generating functional and quantum effective action in the
case of space-time homogeneous sources $\phi$ (and correspondingly,
constant vacuum expectation values $\<\cO\>$).  In other words, in this
section we will restrict our attention to the quantum effective
potential for $\cO$, and we will postpone the discussion of the full
effective action up to two derivative to the following sections.

Some of the results presented here were already derived in the literature (see
for example \cite{KN}), but we will cast  them in a new form and take one step further in writing the renormalized quantities more explicitly than it has previously been done.

As we are interested in Poincar\'e invariant sources and vevs, we take
the induced metric at constant $u$ to be flat and the scalar field to be
$x$-independent. We take the coordinate $u$ to be the one in which  the  solution of (\ref{AE}) has the
standard domain-wall  metric with homogeneous scalar field:
\begin{equation}
ds^2=du^2+e^{2A(u)}\eta_{\mu\nu} {dx^\mu}{dx^\nu},\,\qquad \phi=\phi(u)
\end{equation}
where $\eta_{\mu\nu}=(-1,\,+1,\,+1,\,...)$ is the Minkowski metric.

The independent Einstein's equations are
\begin{equation} \label{AE1}
2(d-1)\ddot A+\dot\phi^2=0,\qquad
(d-1)\ddot A+d(d-1)\dot A^2=V
\end{equation}
where $.=\frac d {du}$ denotes the radial derivative.

As it is well known, one can obtain any solution of the above equations by introducing  a  superpotential $W(\phi)$, whose scalar derivative
is equal to the radial derivative of the scalar field, $W'=\dot\phi$,
where $'=\frac d {d\phi}$ denotes the derivative with respect to $\phi$.

Einstein's equations then become  equivalent to the first order system:
\begin{equation}\label{AE2}
\dot A=-\frac{1}{2(d-1)}W , \qquad \dot \phi = W'
\end{equation}
plus the condition on $W(\Phi)$:
\begin{equation}\label{SPeq}
V=\frac d{4(d-1)}W^2-\frac 1 2 {W^\prime}^2,
\end{equation}
which will be called the {\em superpotential equation.} Since any
solution can be written in this form,  we can proceed to solve
Einstein's equations in two steps:
first find a solution of the superpotential
equation (\ref{SPeq}), then use it to solve the flow-like equations (\ref{AE2}).

It is sometimes useful to consider $\phi$ as a coordinate instead of
$u$, and one can  use the superpotential to write an equation for the scale
factor as a function of $\phi$,
\begin{equation}\label{AE3}
A'(\phi)=-\frac 1 {2(d-1)}\frac W{W'}.
\end{equation}
 One can use $\phi$ as a
coordinate at least piece-wise, in any region where $W(\phi)$ is
monotonic. In this language the reparametrization invariance of the
solution is completely fixed (unlike in the original system, where we
could still shift $u$ by a constant without affecting the equations).

The regularity of the solutions in this language implies regularity of $V$ {\em and} $W$ at any given point of the flow, \cite{bourdier}.

{\em The superpotential is the most important object in this
  discussion}  and as we will see, completely determines all the
properties of the renormalized effective action. It is worth to pause
for a moment and analyse the system (\ref{AE2}-\ref{SPeq}) in more
detail:
\begin{itemize}
\item All the intricacies of Einstein's equations are confined to the
  superpotential equation (\ref{SPeq}): different solutions of the latter
  correspond to qualitatively different geometries. Since the equation
  is first order, there is a one-parameter family of solutions which
  classify the possible geometries. Notice that this equation is
  insensitive to a choice of bulk coordinate, but depends only on
  scalar quantities.
\item On the other hand, after one choice is made for the superpotential, i.e. once a
  particular solution of (\ref{SPeq}) is chosen, the rest of the
  system is trivial and can be integrated by quadratures. In fact,
  after eliminating the coordinate $u$,   all solutions for a given
  $W(\phi)$  are identical up to a constant shift in the
  function   $A(\phi)$ which solves equation (\ref{AE3}), and can be
  written explicitly as:
\begin{equation}\label{A-phi}
A(\phi) = -\frac {1}{2(d-1)}\int^\phi_{\phi_0} {W(\psi)\over W'(\psi)} d\psi + A(\phi_0)
\end{equation}
All solutions corresponding to the same  $W(\phi)$ coincide up to a
constant shift of the scale factor, i.e. up to a choice of an initial
condition $A(\phi_0)=A_0$ (or equivalently $\phi(A_0) = \phi_0$).
\end{itemize}
The above discussion indicates that all solutions to the system (\ref{AE1})
are classified  by
\begin{enumerate}
\item  one integration constant which picks a solution
of the superpotential equation $W(\phi)$

\item a choice of initial
condition for the flow equations (\ref{AE2}). This  choice however can only
affect the solution by a trivial overall rescaling of the
4-metric.
\end{enumerate}

In a sense, all solutions corresponding to the same
superpotential are equivalent, and all the physics is contained in the
choice of the superpotential.

Let us translate what we have seen above in terms of the dual field
theory. In holography, it is natural to  regard $\phi(u)$ as the running coupling associated to a
dual operator $\cO$, and to consider the warp factor $e^{A(u)}$ as
measuring the  energy scale $\mu$ corresponding to the position in the bulk
parametrized by the coordinate $u$, i.e. (up to an overall normalization):
\be\label{mu1}
\log \mu(u) \equiv A(u)
\ee
 With this identification, the
$\beta$-function of the theory is
\be\label{mu2}
\beta(\phi) \equiv {d\phi \over d \log \mu} = -2(d-1) {W' \over W}
\ee
and equation (\ref{A-phi}) can be read as the solution to the RG-flow
equation, written as an implicit function of the coupling,
\be
\mu = \mu_0 + \int_{\phi_0}^{\phi(\mu)} {d\phi \over \beta(\phi)}
\ee
Thus, in the field theory language, fixing  the superpotential
corresponds to fixing the $\beta$-function, whereas fixing the
solution for a given $W$ corresponds to picking an RG-flow trajectory.
Thus, all the non-trivial physics is encoded in $W$.

The  identification (\ref{mu1}) has been justified  heuristically in the past, but as we will see in Section \ref{beta} the identification $\log \mu \propto A$ can be {\em derived} by writing renormalized trace identities.

\subsection{The renormalized generating functional}

In holography, the generating functional of connected correlators of
the dual operator is the effective action evaluated on-shell. For the
homogeneous solutions we are considering, this can be easily
calculated: taking the trace of  equation (\ref{AE}),  substituting it
in   (\ref{action}),  using equations (\ref{AE1}) and (\ref{AE2}) and
adding the contribution from the GH term, we
obtain \cite{KN}:

\begin{equation}\label{gf1}
S=\int d^d x\int_{u_{UV}}^{u_{IR}}\frac d {du}(-e^{dA}W)
\end{equation}

Therefore, the on-shell action is essentially given by the
superpotential evaluated at the UV and IR endpoints.

At this point we have to be more specific by what we mean by UV and
IR, and how to isolate the divergences. \\

{\em We define the UV (IR) limit of the bulk geometry as the
  regions where $e^{A(u)}$ asymptotes to infinity (zero)}. \\

We will assume that both such regions exist in the solution, although
in general this is not guaranteed.
These limits will be reached as $u\to u_{UV}, u_{IR}$
respectively. From equation (\ref{A-phi}), we see that given a
superpotential, the UV and IR will correspond to specific values
$\phi_{UV}$, $\phi_{IR}$ (which may be finite or infinite) of the scalar field, which moreover will be the same for all solutions with a given $W$.

The second assumption we make is that asking for regularity in the IR picks out  a  superpotential solution such that  the IR endpoint does not contribute to equation (\ref{gf1}). This issue will be discussed in more detail in  later sections, and in the explicit examples.   With this  assumption in mind, we can write the  bare on-shell action  (using $A$ as a coordinate) as
\be
S = \lim_{A\to \infty}\int d^d
x \,  e^{dA}W(\phi(A))
\ee
This quantity is typically divergent and requires renormalization. This can be done as usual, by  replacing the strict $UV$ limit by a cut-off endpoint $u_{\epsilon}$, adding a counterterm, and then taking $\epsilon\to 0$.  The counterterm action, when evaluated on a solution,  must subtract the divergence for any choice of the solution $\phi(A)$, therefore it must have the form:
\be
S^{ct} =  \int d^d
x \,  e^{dA(\epsilon)}W^{ct}(\phi)
\ee
where $W^{ct}$ is a given, fixed solution of the superpotential equation (\ref{SPeq}).  The renormalized generating functional is defined by:
\be\label{Sren}
Z^{(ren)} = e^{iS^{(ren)}}, \qquad S^{(ren)} = \lim_{A\to \infty}\int d^d
x \,  e^{dA}\left[W(\phi(A)) - W^{ct}(\phi(A))\right]
\ee
Let us stress that, in this expression, $W(\phi)$ is the superpotential associated to a bulk solution, whereas $W^{ct}$ is a fixed counterterm superpotential that defines the bulk theory.

In general, given a potential $V(\phi)$, there may be more than one value $\phi_{UV}$ corresponding to inequivalent UV limits. Choosing the $UV$ limit
means choosing a boundary condition for the bulk theory, but this  may still correspond to different possible choices for the superpotential of the bulk solution.  Since the counterterm must subtract the divergence no matter which superpotential we pick for the bulk solution  the renormalization procedure only works if all solutions of equation (\ref{SPeq}) have the same limit as $\phi \to \phi_{UV}$, i.e.\\

{\em The UV point $\phi_{UV}$ must be an attractor for the superpotential equation}.\\

More explicitly, we require that any two solutions $W, \tilde{W}$  of (\ref{SPeq}) satisfy the condition
\be\label{attractor}
{W(\phi) - \tilde{W}(\phi) \over W(\phi)} \to 0 \quad \phi \to \phi_{UV}
\ee

If this condition fails, then one cannot determine the counterterm before specifying the solution, and holographic renormalization cannot be performed.
Here, we will analyze in full generality the superpotential equation, nor specify in details  the requirements for this conditions to be realized. However, this is the case in  most interesting  examples, e.g. :
\begin{enumerate}
\item the case of a UV $AdS$ fixed point at a  finite $\phi_{UV}$, corresponding to a regular  extremum of the potential $V(\phi)$ (see section \ref{adsUV} );
\item the case of an asymptotically free $AdS$ solution, realized as $\phi \to -\infty$ with:
$$V \sim V_0 + V_1 e^{a \phi}+ \ldots $$ ;
This is the case that mimics logarithmic UV running as in \cite{gkn} (see section \ref{ihqcd})
\item runaway exponential potential which have scaling UV solutions,
$V \simeq V_0 e^{b\phi}$ as $\phi \to -\infty$, if the exponent $b$ satisfies the bound $b < \sqrt{2d/(d-1)}$ (section \ref{expo}).
\end{enumerate}

Under the assumption (\ref{attractor}), one can  write explicitly  the renormalized generating functional:  close to $\phi_{UV}$
\be\label{WUV}
\tilde{W}(\phi) \simeq W(\phi) + W_1(\phi)
\ee
with  $W_1$  small compared to $W$. Then, $W_1$ satisfies the {\em linear} equation
\be
{W_1' \over W_1 }= {d \over 2(d-1)} {W\over W'}
\ee
whose solution is
\be\label{w1}
W_1(\phi) = C \exp[-d\A(\phi)], \qquad  \A(\phi) = -{1\over2(d-1)} \int^\phi_{\bar{\phi}}  {W \over W'}
\ee
where $C$ is an integration constant.  We can fix arbitrarily  the reference point $\bar{\phi}$ in the definition of $\A(\phi)$, since any change can be reabsorbed in a redefinition of $C$ (for example we can take $\bar{\phi}$ to be close to $\phi_{UV}$, so we can use the UV asymptotics of the superpotential in the integrand of (\ref{w1})).

 It is important to make clear the distinction between the function $\A(\phi)$, which  does not depend on the specific bulk solution, and the scale factor $A(u)$ of a given solution $(A(u), \phi(u))$, which is specified by an initial condition for equations (\ref{AE2}).   On the other hand,  when evaluated on a given solution $\phi(u)$,   $\A(\phi(u))$ coincides with the  scale factor   {\em up to an additive constant}, which depends on the specific solution chosen (i.e. on the specific radial flow), as one can see by comparing  equations  (\ref{A-phi}) and (\ref{w1}).

At this point we can write explicitly the renormalized on-shell action: since both $W$ and $W^{ct}$ in (\ref{Sren}) have the same UV expansion (\ref{WUV}), as $\phi \to \phi_{UV}$, $W - W^{ct} \simeq (C-C^{ct}) \exp[-d\A(\phi)]$, and  we find:
\be
S^{(ren)} = \lim_{A\to \infty}\int d^d
x \,  C_R \, e^{dA}\, e^{-d\A(\phi(A))}    = \lim_{\phi\to \phi_{UV}}\int d^d
x \,  C_R \, e^{dA(\phi)} e^{-d\A(\phi)}
\ee
The ``renormalized'' constant  $C_R= (C- C^{ct})$  contains the scheme dependence implicit in the choice of the counterterm $W^{ct}$

The UV limit is now manifestly finite as one can see using the expressions (\ref{A-phi}) and (\ref{w1}).


\subsection{The quantum effective potential and the  holographic $\beta$-function} \label{beta}

The renormalized generating functional of the previous subsection is clearly independent of scale, i.e. it is a constant on any given solution $\phi(A)$. In fact, we can observe that it is a constant even before we take the limit: using equations (\ref{A-phi}) with $\bar{\phi} = \phi_0$ and (\ref{w1}) we note that
\be
S^{(ren)}[A(\phi_0)] = \int d^d
x \,  C_R \, e^{dA(\phi_0)} \, .
 \ee
In this form, we can interpret $S^{(ren)}$ as a functional of the bulk solution specified by an initial condition $A(\phi_0)$. Using equation (\ref{A-phi}) we can write the same quantity at an arbitrary point  of the RG-flow, as a function  of the holographic  RG scale $\mu$ and of the  running coupling $\phi$ at the scale $\mu$:
\be\label{SrenRG}
S^{(ren)}[\mu,\phi(\mu)] = C_R \int d^d x \, \exp\left[{dA(\mu)} + {d\over 2(d-1)}\int_{\phi_0}^{\phi(\mu)} d\phi {W\over W'}  \right]
\ee
In writing this expression, we have assumed a generic relation between the scale factor and the RG-scale, but this relation will be fixed shortly.

RG-invariance of (\ref{SrenRG}) is  expressed by the fact that
\be
\mu{d\over d\mu} S^{(ren)}[\mu,\phi(\mu)] =0 ,
\ee
as can be immediately checked by using the bulk equation (\ref{AE3}).

On the other hand, we observe that  $S^{(ren)}$ is an independent  function of the energy scale and the  coupling at that scale, i.e. by keeping $\mu$ fixed and making $\phi(\mu) = \phi$ vary:
\be
S^{(ren)} [\mu,\phi] = C_R \int d^d x \, \exp[{dA(\mu)} - d\A(\phi)]
\ee
where $\A(\phi)$ is given in (\ref{w1}). In this form, $S^{(ren)}$ gives information about all RG-flow trajectories. We can extract the renormalized
expectation value of the  dual operator $\cO$ by the definition:
\be\label{O}
\<\cO\> = {\delta \over \delta \phi} S^{(ren)}[\mu,\phi] = C_R {d\over 2(d-1)}{W\over W'} e^{dA(\mu)}\, e^{-d\A(\phi)}
\ee
Also, taking  a Legendre transform  with respect to $\phi$, we obtain the quantum effective potential as a function  of $\<\cO\>$ and of the renormalization scale (the analog to the Coleman-Weinberg effective potential in ordinary QFT):
\be
\Gamma[\mu,\<\cO\>] = \left[\int d^dx \, \<\cO \> \phi -  S^{(ren)}[\mu,\phi]\right]_{\phi=\phi(\<\cO\>,\mu)}.
\ee

Finally from $S^{(ren)}[\mu,\phi]$  we can derive the renormalized  trace identities and identify the relation between RG scale and radial coordinate. First, as it is standard in d-dimensional field theory, we can obtain the renormalized trace of the stress tensor by the variation with respect to a rescaling $x^\mu \to \lambda x^\mu$:
\be
\< {T^\mu}_\mu \> = \lambda {d\over d\lambda} S^{(ren)} [\mu,\phi]= d\, C_R\, e^{dA(\mu)}\, e^{-d\A(\phi)}
\ee
On the other hand, the standard field theory trace identity reads:
\be
\< {T^\mu}_\mu \> = -\beta \<\cO\>
\ee
 comparing the right hand side with (\ref{O}), we see that the standard trace identity is recovered if
\be
\beta(\phi) = -2(d-1) {W' \over W}
\ee
which is exactly what we obtain if we identify $\exp A = \mu$ (up to a normalization constant), see equations  (\ref{mu1}-\ref{mu2}).

An alternative, but equivalent, derivation will be given in Section  \ref{renorm} , when we will obtain the generating functional for the dual field theory coupled to an arbitrary space-time metric $\gamma_{\mu\nu}$, and we will be able to define the stress tensor in a standard way by taking derivatives with respect to $\gamma_{\mu\nu}$.

\section{The flow equations} \label{flow}

In the previous section we have seen how to obtain the renormalized quantum effective action associated to a bulk solution, in terms of the superpotential of that solution. Here, we will generalize this discussion to the full covariant effective action up to two derivatives, for an arbitrary scalar field  and an arbitrary spatial metric.
Now, the data specifying the solution are $\phi(x,u)$ and the four-dimensional induced metric $\gamma_{\mu\nu}(x,u)$, corresponding to a dual field theory  with a space-time dependent coupling, on a space-time with non-trivial metric. Both the coupling and the metric will be scale-dependent, and their RG-evolution will be governed by RG-flow equation with beta functions given in terms of boundary covariant quantities. The goal of this section  will be to obtain these covariant RG-flow equations starting from the bulk dynamics. This was done previously using the Hamilton-Jacobi formalism,
\cite{deboer,papa} but we will use a different technique which will prove useful to write the finite part of the quantum effective action explicitly, as will be done in the next section.

\subsection{ADM Einstein's equations as Flow equations}

We  start with  the ADM decomposition of the bulk metric:
\begin{eqnarray} \label{ADM}
d s^2&=&(N^2+\gamma_{\mu\nu}N^\mu N^\nu)du^2+2\gamma_{\mu\nu}N^\mu du dx^\nu+\gamma_{\mu\nu}dx^\mu dx^\nu,
\end{eqnarray}
$u$ is the holographic coordinate, and $\gamma_{\mu\nu}(x,u)$ is the induced metric on the slices  $\Sigma_{u}$ orthogonal to the normal vector $n^a=(\frac 1 N, -\frac {N^\mu} N)$.  In this section, we keep the metric (\ref{ADM}) generic. Starting from the next section, and until the end of this paper, we will work in  the gauge  $\partial_\nu N=0$.

The constraints and the dynamical equations can be derived by projecting the Einstein's equations  (\ref{AE})  in different ways.

The $G_{ab}^{(d+1)}n^an^b$ projection of the Einstein's equation is
\begin{eqnarray}
R^{(d)}-K^2+K_a^b K_b^a=\frac 1 2 (\gamma^{ab}-n^a n^b)\partial_a\phi\partial_b\phi -V\label{HC},
\end{eqnarray}
which is the radial Hamiltonian constraint $\frac {\delta S}{\delta N}=0$. Here, $R^{(d)}$ is the Ricci scalar of the induced metric, and  $K_{ab}$ and $n_a$ are respectively  the extrinsic curvature, and unit normal vector to the constant-$u$ slices $\Sigma_{u}$.

The $G_{ab}^{(d+1)}n^a\gamma^{b}_c$ projections of the Einstein's equations are
\begin{eqnarray}
\nabla_a^{(d)}K_c^a-\nabla_c^{(d)}K=\frac 1 2 n^a \partial_a\phi  \gamma^b_c\partial_b\phi ,
\end{eqnarray}
which is the transverse momentum constraint $\frac {\delta S}{\delta N^a}=0$, and in which $\nabla^{(d)}$ is the covariant derivative associated to $\gamma_{\mu\nu}$.

Subtracting the trace part $R^{(d+1)}g_{ab}$, the $G_{ab}^{(d+1)}\gamma^a_c\gamma^{b}_d$ projection of the Einstein's equations are
\begin{eqnarray}
&&R^{(d)}_{ab}-\pounds_nK_{ab}-K K_{ab}+2K_{ac}K^c_b-\frac 1 N\nabla_a^{(d)}\partial_b N\nonumber\\
&=&\frac 1 2\gamma^c_a\gamma^d_b\partial_c\phi\partial_d\phi-\frac 1 {d-1} V\gamma_{ab}, \label{dyn}
\end{eqnarray}
which is the dynamical equation since we have a second order radial derivative term $\pounds_n K_{ab}=\frac 1 2\pounds_n(\pounds_n \gamma_{ab})$.

In the following sections, we will use the Lie derivatives $\pounds_n$ to simplify our notations. Intuitively, the Lie derivative along the normal vector $n^a$ can be considered as the generalization of the radial derivative $\frac {\partial}{\partial u}$ in a general coordinate. The precise definition is in  appendix \ref{app-Lie}

We will solve the Einstein's equations by writing  flow equations order by order in a derivative expansion with respect to the boundary space-time coordinates $x^\mu$.  We will work up to two derivatives.  To this order, we assume an ansatz for  general covariant flow equations of the form:
\begin{eqnarray}
\pounds_n \gamma_{\mu\nu}&=&g_1\gamma_{\mu\nu}+g_2 R^{(d)}_{\mu\nu}+g_3 R^{(d)} \gamma_{\mu\nu}+g_4\partial_\mu\phi\partial_\nu\phi+g_5(\gamma^{\rho\eta}\partial_\rho\phi\partial_\eta\phi)\gamma_{\mu\nu}\nonumber\\
&&+g_6\nabla_\mu^{(d)}\partial_\nu\phi+g_7(\gamma^{\rho\eta}\nabla_\rho^{(d)}\partial_\eta\phi)\gamma_{\mu\nu}+...\,,\\
\pounds_n\phi&=&h_1+h_2 R^{(d)}+h_3 \gamma^{\rho\eta}\partial_\rho\phi\partial_\eta\phi+h_4\gamma^{\rho\eta}\nabla_\rho^{(d)}\partial_\eta\phi+...\,,
\end{eqnarray}
where $g_n(\phi)$ and $h_n(\phi)$ are functions of the scalar field $\phi$ and they will be determined by the equations of motion.

 The leading terms are the first terms without derivative and the subleading terms are those with two derivatives. All possible slice diffeomorphism invariant terms are included. The radial evolutions (in the normal direction of slices) of the induced metric and the scalar field are controlled by these flow equations. As we will see, one can obtain the on-shell action by these flow equations without actually solving them.

\bigskip
The whole setup is making two basic assumptions:

\begin{itemize}
\item The first is that the leading order metric is homogeneous and only radially dependent.

\item The second is that the corrections due to transverse inhomogeneities are small.
    \end{itemize}

         Therefore, to leading order,  the solution is given by solving the leading order equations
\be
\pounds_n \gamma_{\mu\nu}=g_1(\phi)\gamma_{\mu\nu}, \qquad \pounds_n\phi=h_1(\phi)\;.
\ee

The metric flow equation, to lowest order in the derivative expansion,  tells us that only metrics that are conformal to a given transverse metric $\gamma_{\mu\nu}^{(0)}(x)$ are allowed in this formalism, i.e. that all components of the metric obey the same flow. To lowest order, neglecting the space-time dependence, taking $\gamma_{\mu\nu}^{(0)}(x)$ to be the flat metric
means that we are studying vacuum solutions which obey boundary Poincar\'e invariance.

For example, the black hole solutions do not satisfy the leading order equations (even at the homogeneous level $\gamma_{00}$ and $\gamma_{ij}$ have a different radial evolution), so they can not be obtained by this ansatz. The physical reason is that space-time Poincare\'e covariance is broken in the black hole and one should take the non-covariant terms into consideration. To study black hole physics, one should design another set of flow equations whose leading order equations solve the homogeneous black hole solutions and then consider possible two-derivative terms.

\subsection{Solving the flow equations: the superpotential functions}

The Hamiltonian constraint and the momentum constraints are related to the bulk diffeomorphism invariance. After imposing these constraints,  the number of independent scalar functions in the flow equations is greatly reduced.
The details of this calculation are
in  Appendix \ref{a2}. The  idea is to insert the flow equations in the constraints and then require the coefficients of the covariant terms to be such that the bulk equations are obtained. These coefficients are made up of $g_n$ and $h_n$, so the number of independent functions is reduced and one can derive the equations for $g_n$ and $h_n$. We have imposed the gauge fixing $\partial_\nu N=0$, so the lapse function is constant on the hypersurface $\Sigma_u$.

After imposing the constraints,   the flow equations take the following form:
\begin{eqnarray}
\pounds_n \gamma_{\mu\nu}&=&-\frac 1 {d-1}\gamma_{\mu\nu}\left(W+U R^{(d)}+\frac W{2W'} U'(\gamma^{\rho\eta}\partial_\rho\phi\partial_\eta\phi)\right) \nonumber \\
&&+2U R_{\mu\nu}^{(d)}+\left(\frac W{W'} U'-2U''\right)\partial_\mu\phi\partial_\nu\phi-2U'\nabla^{(d)}_\mu\partial_\nu\phi\,,\label{flow-eq-1}\\
\pounds_n\phi&=&W'- U'R^{(d)}+\frac 1 2\left(\frac W{W'} U'\right)'(\gamma^{\rho\eta}\partial_\rho\phi\partial_\eta\phi)+\frac W{W'} U'(\gamma^{\rho\eta}\nabla^{(d)}_\rho\partial_\eta\phi)\label{flow-eq}
\end{eqnarray}
where $W(\phi)$ and $U(\phi)$ are the solutions of
\begin{eqnarray}
V&=&\frac d {4(d-1)}W^2-\frac 1 2{W^\prime}^2, \label{flow-eq-3}\\
1&=&W'U'-\frac {d-2} {2(d-1)}WU. \label{flow-eq-4}
\end{eqnarray}

It remains to check what conditions impose the remaining dynamical Einstein equation (\ref{dyn}).
In fact, one finds that {\em a solution of the the flow equations (\ref{flow-eq-1}-\ref{flow-eq}) solves automatically the dynamical Einstein equation.} This is shown explicitly in Appendix \ref{Appdyn}.
Therefore, the flow equations, together with the conditions (\ref{flow-eq-3}-\ref{flow-eq-4}), accomplish what the superpotential formulation did in the homogeneous case, i.e. to reduce Einstein's equation
to a first order system, in a derivative expansion with respect to the
transverse coordinates.

Equation (\ref{flow-eq-3}) is merely the superpotential equation in the domain wall solution. Equation (\ref{flow-eq-4}) can be interpreted as the second superpotential equation for two-derivatives terms.

The solution of the second superpotential $U(\phi)$ can be written explicitly in terms of the superpotential $W(\phi)$:
\begin{equation} \label{supf}
U(\phi)= e^{-(d-2)\A(\phi)}\left( c_1+\int_{\phi_{0}}^\phi d\tilde\phi\frac 1 {W'(\tilde\phi)} e^{(d-2)\A(\tilde\phi)}\right)
\end{equation}
where $c_1$ is an integration constant. and  $\A(\phi)$  is defined by
\be\label{A}
\A(\phi) = -{1\over 2(d-1)}\int^{\phi}_{\bar{\phi}} d\tilde{\phi} {W(\tilde\phi) \over W'(\tilde{\phi})},
\ee
which is, up to a constant,  the homogeneous solution scale factor for any solution with superpotential $W$,  written as a function of $\phi$. We have fixed the integration constants in $\A(\phi)$ by choosing an arbitrary reference point $\bar{\phi}$, which we can change by redefining the constant $c_1$.

\subsection{Holographic flow  and local Weyl transformations} \label{RGflowsec}

The radial evolution equations (\ref{flow-eq-1}-\ref{flow-eq}) have the form of a geometric flow for the four-dimensional quantities $\gamma_{\mu\nu}(x)$ and $\phi(x)$. We interpret these fields as space-time dependent couplings in the boundary field theory,  as we have seen in Section 2. Then, to lowest order, the flow equations can be put in the form  of geometric RG-flow equations by switching from the variable $u$ to the holographic energy scale $\mu$. This can be identified to lowest order with the scale factor.

If we want to generalize this to higher orders however, things become more complex, since the  dependence on $x^\mu$ of the metric cannot be neglected. There are several reparametrizations mapping the coordinate $u$ to the scale factor.
Instead of thinking about the RG scale, it is more convenient to think in terms of  local Weyl transformations, as in  \cite{osborn}. In what follows we find the relation between Weyl transformations and an infinitesimal motion along the bulk flow.

 We consider two slices at  $u$, and $u+\epsilon$, with $\e\to 0$, where the
induced metrics are $\gamma_{\mu\nu}(x)$ and $\gamma_{\mu\nu}^{(\epsilon)}(x)$ respectively, with
\be\label{weyl1}
\gamma_{\mu\nu}^{(\epsilon)}(x) - \gamma_{\mu\nu}(x) = \epsilon \, \pounds_n \gamma_{\mu\nu}(x)+{\cal O}(\e^2).
\ee
We can separate  the change in $\gamma_{\mu\nu}$ in going  from $u$ to $u+\epsilon$ into two parts: a local Weyl transformation  and a residual volume preserving transformation. This is achieved by writing:
\be\label{weyl2}
\gamma^\epsilon_{\mu\nu}(x) = e^{2\,\epsilon\,\sigma(x)}\,\tilde{\gamma}^{(\epsilon)}_{\mu\nu}(x), \qquad e^{\epsilon\,\sigma(x)} \equiv  \left(\gamma^{(\epsilon)} \over \gamma \right)^{1\over 2d}.
\ee
where $\gamma^{(\epsilon)}$ and $\gamma$ are the absolute values of the metric determinants and $\sigma$ is chosen so that $\tilde{\gamma}^{(\epsilon)}_{\mu\nu}$ and $\gamma_{\mu\nu}$ have the same determinant. To first order in $\epsilon$, we can write equation (\ref{weyl1}) as
\be\label{weyl3}
\epsilon \pounds_n \gamma_{\mu\nu} = \epsilon\left[2\sigma(x) \gamma_{\mu\nu} + \hat{\beta}_{\mu\nu}\right]+{\cal O}(\e^2), \qquad \hat{\beta}_{\mu\nu} \equiv {\tilde{\gamma}^{\epsilon}_{\mu\nu} - \gamma_{\mu\mu}\over \epsilon},
\ee
obtained by substituting (\ref{weyl2}) in (\ref{weyl1}) and expanding in $\epsilon$.
From equation (\ref{flow-eq-1}) we can identify the  $\sigma(x)$ and  $\hat{\beta}_{\mu\nu}$. First, notice that under the holographic flow (\ref{flow-eq-1}), the metric determinant evolves as:
\be\label{weyl31}
\pounds_n \gamma = \gamma \gamma^{\mu\nu} \pounds_n \gamma_{\mu\nu},
\ee
which following  the definitions (\ref{weyl2}-\ref{weyl3}) leads to
\be\label{weyl4}
\sigma = {1\over2d}\gamma^{\mu\nu} \pounds_n \gamma_{\mu\nu}, \qquad \hat{\beta}_{\mu\nu} =  \pounds_n \gamma_{\mu\nu} - {1\over d}\gamma_{\mu\nu}\gamma^{\rho\sigma}\pounds_n \gamma_{\rho\sigma}.
\ee

From equation (\ref{weyl3}), we can relate the generator of the bulk flow  to the generator of local Weyl transformations on the induced metric:
\be\label{weyl5}
\pounds_n \gamma_{\rho\sigma} = \int d^d x\, \sigma(x)\left[ 2 \gamma_{\mu\nu}{\delta \over \delta \gamma_{\mu\nu}}  + {\hat{\beta}_{\mu\nu}\over \sigma} {\delta \over\delta \gamma_{\mu\nu}} \right] \gamma_{\rho\sigma}
\ee\label{weyl6}
This equation shows that the effect of the motion along the holographic flow is equivalent to a Weyl rescaling by a parameter $\sigma$ plus an extra transformation of the metric which is volume preserving. Similarly we have:
\be
\pounds_n \phi = \int d^d x \sigma(x)\left[ {\hat{\beta}_{\phi} \over \sigma} {\delta \over \delta \phi}\right]
\ee
where $\hat{\beta}_\phi$ is the right-hand side of equation (\ref{flow-eq}). Therefore, the Lie derivative  acts on   any  functional of the induced metric and scalar field as:
\bea
&&\pounds_n = \int d^d x \,\sigma (x) \Delta(x) ,\label{weyl7-0} \\
 &&\Delta(x) \equiv 2\gamma_{\mu\nu}(x){\delta \over \delta \gamma_{\mu\nu}(x)} + \beta_{\mu\nu}{\delta \over \delta \gamma_{\mu\nu}(x)} + \beta_\phi {\delta \over \delta \phi(x)},\label{weyl7} \\
&& \beta_{\mu\nu}\equiv {\hat{\beta}_{\mu\nu} \over \sigma}, \qquad \beta_\phi \equiv {\hat{\beta}_\phi \over  \sigma}.   \label{weyl7-1}
\eea

  It is natural to identify the operator
\be\label{weyl8}
\Delta = \int d^d x ~\Delta(x)
\ee
as the generator of RG transformations of the quantum field theory. Then,  quantities $\beta_{\mu\nu}$ and $\beta_{\phi}$ are identified as the $\beta$-functions of the metric and coupling. More precisely, $\beta_{\mu\nu}$ represents the {\em anomalous} change
in the space-time metric beyond a simple Weyl rescaling. Equation (\ref{weyl7-0}) connects the change under the holographic flow to the change under a Weyl rescaling in the field theory plus the additional running of the metric and the coupling constant.

It is convenient to introduce the second order quantities:
\bea
&& X = U R^{(d)} + {W\over2W'}U'\gamma^{\rho\eta}\partial_\rho\phi\partial_\eta\phi, \label{X}\\
&& Y_{\mu\nu}=U R_{\mu\nu}^{(d)}+\left(\frac W{2W'} U'-U''\right)\partial_\mu\phi\partial_\nu\phi-U'\nabla^{(d)}_\mu\partial_\nu\phi, \quad Y = \gamma^{\mu\nu}Y_{\mu\nu}, \label{Y}\\
&& Z = - U'R^{(d)}+\frac 1 2\left(\frac W{W'} U'\right)'(\gamma^{\rho\eta}\partial_\rho\phi\partial_\eta\phi)+\frac W{W'} U'(\gamma^{\rho\eta}\nabla^{(d)}_\rho\partial_\eta\phi). \label{Z}
\eea
In terms of these quantities, the metric and scalar flow equations (\ref{flow-eq-1}-\ref{flow-eq}) read, up to second order in derivatives:
\be\label{flowXYZ}
\pounds_n \gamma_{\mu\nu} = -{1\over d-1}\gamma_{\mu\nu}\left( W + X\right) + 2Y_{\mu\nu}, \qquad  \pounds_n \phi = W' + Z.
\ee
The Weyl parameter $\sigma$,  volume-preserving transformation $\hat{\beta}_{\mu\nu}$, and scalar flow function $\hat{\beta}_\phi$   are then given by:
\bea
\sigma && = -{1\over 2(d-1)}\left[W + X - {2(d-1)\over d} Y\right] \label{weyl9} \\ 
\hat{\beta}_{\mu\nu} &&= 2Y_{\mu\nu} - {2\over d} \gamma_{\mu\nu}Y, \qquad \hat{\beta}_\phi = W' + Z . \label{weyl9-1}
\eea

Explicitly, up to second order in derivatives,   from the definitions (\ref{weyl7}-\ref{weyl7-1}) and the from (\ref{weyl9}-\ref{weyl9-1}) we derive the RG-flow equations for the running coupling and the metric:
\be
\Delta \phi = \beta^{(0)}_\phi + \beta^{(2)}_\phi,  \qquad \Delta \gamma_{\mu\nu}  = 2 \gamma_{\mu\nu} + \beta^{(2)}_{\mu\nu},
\ee
where:
\bea
\beta^{(0)}_\phi &=& -2(d-1) {W' \over W},\label{beta0} \\
\beta^{(2)}_\phi &=& -{2(d-1)\over W}\left(Z - {W'\over W} X + {2(d-1)\over d}{W'\over W} Y\right), \label{beta2-1}\\
\beta^{(2)}_{\mu\nu} &=& -{4(d-1)\over W} \left(Y_{\mu\nu} - {1\over d}\gamma_{\mu\nu}Y \right). \label{beta2-2}
\eea

Note that to zero-th order, under $\Delta$  the metric changes as under a   Weyl rescaling. The anomalous variation starts at second order and it is traceless (therefore it does not affect the metric determinant, see equation (\ref{weyl31}). )

We conclude this section by noting  that there is some freedom in the relation  between the radial bulk  flow and the RG-flow generators. In Appendix \ref{alpha} we define a family of operators $\Delta_{\alpha}$, which differ at second order in derivatives but can all, in principle, be used as defining an RG-flow of the metric and scalar. This corresponds to different choices of scheme for  the definition of the holographic RG scale. A generic choice however produces field theory trace idendtities which are not of the standard form.

\section{Two roads to the two-derivative quantum effective action} \label{effact}
 We are now ready to write down the on-shell action, using Einstein's equations in  first order form provided by the covariant flow equations. We will
do it in two ways, which will lead to the same result, but highlight different
aspects of the procedure:
\begin{enumerate}
\item First we use the fact that the on-shell action
is the generating functional of canonical momenta. This is similar to the
technique used in the Hamilton-Jacobi approach.
\item Alternatively, one can use the flow equations to write the bulk action as a total derivative, as it was the case for  homogeneous solutions.  This will give explicit control over the IR contribution.
\end{enumerate}

From now on, we will suppress the index $(d)$ from the  slice intrinsic curvature and covariant derivative and denote them  simply by $R$ and $\nabla$. We will keep the notation  $R^{(d+1)}$, $\nabla^{(d+1)}$ for bulk covariant quantities.

\subsection{The on-shell  action as a solution of the canonical momentum equation}

We can  derive the (regularized) on-shell action by the variational formula of canonical momentum \cite{deboer}:
\begin{eqnarray} \label{gradient1}
\pi_\phi=\frac {\delta S_{on-shell}}{\delta \phi},\qquad \pi^{\mu\nu}=\frac {\delta S_{on-shell}}{\delta \gamma_{\mu\nu}}
\end{eqnarray}
where $\pi_\phi$ and $\pi^{\mu\nu}$ are the canonical momenta of the scalar field $\phi$ and the induced metric $\gamma_{\mu\nu}$ on the boundary. As the bare on-shell action is generically divergent in the UV, these expressions are defined at a regulated UV boundary. The variations are with respect to the corresponding boundary quantities.

The canonical momenta are related to the flow equations in the following way
\begin{eqnarray} \label{gradient2}
\pi_\phi=\sqrt{-\gamma}\pounds_n\phi,\qquad \pi^{\mu\nu}=\frac 1 2\sqrt{-\gamma}( \gamma^{\mu\rho}\gamma^{\nu\eta}\pounds_n\gamma_{\rho\eta}-\gamma^{\mu\nu}\gamma^{\rho\eta}\pounds_n\gamma_{\rho\eta})
\end{eqnarray}
where the second equation can be written as $\pi^{\mu\nu}=\sqrt{-\gamma}( K^{\mu\nu}-K\gamma^{\mu\nu})$ and $K_{\mu\nu}$ is the extrinsic curvature.

These variational formulae are valid only when one uses Dirichlet boundary conditions ($\phi^{UV},\phi^{IR},\gamma_{\mu\nu}^{UV},\gamma_{\mu\nu}^{IR}$) to solve the equations of motion and then evaluate the on-shell action on these solutions. As the solutions of Einstein's equations are determined by the boundary conditions, the integration constants in the flow equations are non-trivial functions of both the UV and IR boundary data. The variations of the integration constants with respect to the boundary quantities, i.e. $\frac {\delta C}{\delta \phi^{UV}}$, are not zero. Therefore, one can not naively deduce an on-shell action whose variations give the canonical momenta without knowing the dependence of integration constants on the boundary data.

However, if one imposes the IR regularity of the flow equations, the explicit expressions of the integration constants will be fixed, so the variational formulae are useful again. Technically, the IR regularity of the flow equations means that the scalar functions in front of the derivative terms, for example $U(\phi)$ and $U'(\phi)$, stay finite in the IR limit. In fact, the IR regularity of the flow equations is necessary if one wants to do derivative expansions in IR region.  Otherwise, the subleading terms will dominate due to the divergent coefficients, the derivative expansion will break down in the IR and the perturbative solutions will not make sense.

For two derivatives terms, the IR regularity of the flow equations requires
\begin{equation} \label{ir-reg}
 U_{\ast}(\phi)=e^{-(d-2)\A(\phi)}\int_{\phi_{IR}}^\phi d\tilde\phi\frac 1 {W'(\tilde\phi)} e^{(d-2)\A(\tilde\phi)}.
\end{equation}
where the integration constant determined by IR regularity is ${c_1}_{\ast}=\int_{\phi_{IR}}^{\bar{\phi}} d\tilde\phi\frac 1 {W'} e^{(d-2)\A(\tilde\phi)}$ and $\phi_{IR}$ is the value of scalar field in the IR.

The regular solution $U_\ast$ is IR finite because the divergence of $e^{-(d-2)\A(\phi)}$ is cancelled by the vanishing of $\int_{\phi_{IR}}^\phi d\tilde\phi\frac 1 {W'(\tilde\phi)} e^{(d-2)\A(\tilde\phi)}$. Other solutions diverge in the IR. To see this, one can write the general solutions as $U=U_\ast+ e^{-(d-2)\A(\phi)}(c-c_\ast)$. The second term $e^{-(d-2)\A(\phi)}(c-c_\ast)$ diverges in the IR limit $\A\rightarrow -\infty$, so the only regular solution is $U_\ast$.

The IR regularity condition $U|_{IR}<\infty$ seems to be inconsistent with the assertion that the variational formula of canonical momenta is only valid for Dirichlet boundary conditions. In fact, the IR regularity condition is equivalent to fixing the IR boundary value of $U$ to a special finite value, so IR regularity is in accord with the Dirichlet conditions.  For example, for an AdS IR fixed point, $U_{\ast}\rightarrow -\frac {2(d-1)}{d-2}\frac 1 {W^{IR}}$ and for an IR exponential potential, $U_{\ast}\rightarrow 0$. The analytic results of these two examples are verified in the numerical study.

The integration constant ${c_1}_{\ast}$ is now independent of the UV data, so we can use the variational formula of the canonical momenta on the UV boundary without worrying about the non-zero variation of integration constant.

Using the variation formula of canonical momenta, the on-shell action is easily derived:
\begin{equation} \label{onshell-1}
S_{on-shell}=\int d^d x\sqrt\gamma\left(W-  U_{\ast}R-\left(\frac W{W'}  {U_{\ast}}'\right)\frac 1 2\gamma^{\mu\nu}\partial_\mu\phi\partial_\nu\phi+...\right)_{UV}+S^{IR}
\end{equation}
where $S^{IR}$ is a constant IR contribution independent of the UV data and usually it is just set to zero. This expression coincides with the one found in \cite{papa1} using the Hamilton-Jacobi method.

\subsubsection{Radial flow as a gradient flow}

The construction of the regularized on-shell action in the previous section allows to write the radial flow for the induced metric and scalar field as a gradient flow, in a way which is consistent  with the analysis of  \cite{sslee}. Indeed, inverting equation (\ref{gradient2}) for the Lie derivatives of $\phi$ and $\gamma_{\mu\nu}$, and using the definitions (\ref{gradient1})  we can write the radial flow as:
\be
\pounds_n \phi = {1\over \sqrt{\gamma}} \frac {\delta S_{on-shell}}{\delta \phi},\qquad \pounds_n \gamma_{\mu\nu}={1\over \sqrt{\gamma}} {\cal G}_{\mu\nu;\rho\sigma}\frac {\delta S_{on-shell}}{\delta \gamma_{\rho\sigma}},
\ee
where :
\be
 {\cal G}_{\mu\nu;\rho\sigma} = \gamma_{\mu\rho}\gamma_{\nu\sigma} - {1\over d-1} \gamma_{\mu\nu}\gamma_{\rho\sigma}
\ee
is the de Wit metric. This gradient property was shown to be a general feature of holographic RG flows in \cite{sslee}. In particular, leaving aside the scalar field, one can see directly from equation (\ref{flowXYZ})  and the definitions (\ref{X}-\ref{Y}) that the curvature term in the second order contribution to the metric radial flow has exactly the same tensor structure that was found in \cite{sslee}, i.e.
\be
(\pounds_n \gamma_{\mu\nu})^{(2)} \propto R_{\mu\nu} - {1\over 2(d-1)}\gamma_{\mu\nu}R.
\ee

\subsection{The on-shell  effective action as a bulk total derivative}

An alternative way to compute the generating functional  is to write the bulk on-shell Lagrangian as a total derivative, as we did in Section 2 for homogeneous solutions:
\begin{eqnarray}
\mathcal L_{on-shell}=\partial_a (\mathcal L_{eff}\tilde n^a)
\end{eqnarray}
where $\tilde n=Nn=(1,-N^\mu)$ in ADM formalism, so the $(d+1)$-dimensional volume integral is transformed into a $d$-dimensional surface integral over the boundary slices after one uses the Gauss's theorem
\begin{eqnarray}
S_{on-shell}=\int d^dx\int _{UV}^{IR} du\, \mathcal L_{on-shell}=\int d^dx \left. \mathcal L_{eff}\right|_{UV}^{IR}
\end{eqnarray}
where the UV limit corresponds to $u\rightarrow -\infty$, so the integration interval is $[u^{UV},u^{IR}]$. We have assumed that the slices have no space-time boundary. To be precise, the effective Lagrangian is the difference of $\mathcal L_{eff}$ between the IR and the UV.

Using the language of Lie derivative, the total derivative equation becomes
\begin{eqnarray}
N\pounds_n\mathcal L_{eff}=\mathcal L_{on-shell},
\end{eqnarray}
where for a scalar density $B$, we have $\partial_a (B\tilde n^a)=N\pounds_nB$ in the gauge $\partial_\nu N=0$.

In the Einstein-Scalar gravity theory, we can write the above equation explicitly
\begin{eqnarray}
\frac 1 2 \gamma^{-\frac 1 2}\pounds_n\mathcal L_{eff}=V+R -\frac 1 2\gamma^{\mu\nu}\partial_\mu\phi\partial_\nu\phi,
\end{eqnarray}
where we have used the Hamiltonian constraint to simplify the equation.

On the left hand side, the general form of a slice covariant effective Lagrangian density is, up to two derivatives:
\begin{eqnarray}
\mathcal L_{eff}=\sqrt{-\gamma}(F_0+F_1R+F_2\gamma^{\mu\nu}\partial_\mu\phi\partial_\nu\phi+F_3\gamma^{\mu\nu}\nabla_\mu\partial_\nu\phi+...)
\end{eqnarray}
where $F_n(\phi)$ are functions of the scalar field $\phi$.

To derive the effective Lagrangian by the total derivative equations, one should calculate the Lie derivative $\pounds_n$ of the effective Lagrangian density $\mathcal L_{eff}$ by acting with the Lie derivative on all possible terms and substituting in the flow equations $\pounds_n \gamma_{\mu\nu}$ and $\pounds_n \phi$, then match the scalar functions in front of the covariant terms on both sides to obtain the equations of $F_n(\phi)$.

The details of the calculation are given in Appendix \ref{App3}, and the  result is:
\begin{equation} \label{onshellbare}
S_{on-shell}=\int d^d x\sqrt\gamma\left(W-  UR-\left(\frac W{W'}  {U}'\right)\frac 1 2\gamma^{\mu\nu}\partial_\mu\phi\partial_\nu\phi+...\right)_{UV}-(...)_{IR}
\end{equation}
where $U(\phi)=e^{-(d-2)\A}\left( c_1+\int_{\bar{\phi}}^\phi d\tilde\phi\frac 1 {W'} e^{(d-2)\A}\right)$ and $(...)_{IR}$ is the IR contribution. The $\gamma^{\mu\nu}\nabla_\mu\partial_\nu\phi$ term has been transformed into the $\partial_\mu\phi\partial_\nu\phi$ term by integrating by parts. For an IR regular solutions, $W\rightarrow W_\ast$,  $U \rightarrow U_\ast$,  and the on-shell action is the same as the one derived in the previous section.

The on-shell action in this section is more general because  we don't need to take the far IR limit $e^A|_{IR}\rightarrow 0$. Without this restriction, one can calculate the Wilsonian effective action by integrating out part of the geometry, which are the high energy modes in holographic language. In holography, the high energy modes correspond to $u<u^\Lambda$ where $\Lambda$ is the UV cut-off. In this setting, the condition of IR regularity is replaced by an IR boundary condition on the coupling.

\subsection{Conserved quantities and integration constants} \label{conserved2}
In writing equation (\ref{onshellbare}) we have used a particular solution for the functions $F_n(\phi)$,  determined by the matching conditions given  in Appendix \ref{App3}. The general solution however
 contains some  integration constants $C_n$, which contribute to the on-shell Lagrangian extra terms of the form:
\be
\delta \mathcal L\sim \sum_n C_n \, G_n(\phi, \gamma_{\mu\nu}),
\ee where  $G_n$ are covariant terms. However, these terms drop out of the final result   due to an interesting fact: the covariant terms $G_n$ are conserved quantities.  Thus, $\delta {\mathcal L}_{UV} =\delta {\mathcal L}_{IR}$, and these contributions disappear from the final result (\ref{onshellbare}). While the  details are given in Appendix \ref{conserved}, here we summarize the main results.

To two-derivative order, there are three such contributions, $\delta \mathcal L= C_0 G_0 + C_1 G_1 + C_2 G_2 + \mathcal O(\partial^4)$, where $C_0\ldots C_2$ are constants and:
\begin{eqnarray}
G_0&=&\int d^dx\,\sqrt{-\gamma}\left[e^{-d\A}+G_0^{(1)}e^{-(d-2)\A}R+W'^{-2}e^{-(d-2)\A}G_0^{(2)}\frac 1 2\gamma^{\mu\nu}\partial_{\mu}\phi\partial_\nu\phi+\mathcal O(\partial^4)\right],\nonumber\\
G_1&=&\int d^dx\sqrt{-\gamma}\left[e^{-(d-2)\A}R+\frac W{W'}(e^{-(d-2)\A} )'\frac 1 2\gamma^{\mu\nu}\partial_\mu\phi\partial_\nu\phi+\mathcal O(\partial^4)\right],\nonumber\\
G_2&=&\int d^dx\sqrt{-\gamma}\left[W'^{-2}e^{-(d-2)\A} \frac 1 2\gamma^{\mu\nu}\partial_\mu\phi\partial_\nu\phi+\mathcal O(\partial^4)\right],\label{conserved-quantities}
\end{eqnarray}
where
\begin{eqnarray}
G_0^{(1)}( \phi)&=& G_0^{(1)}(\bar \phi)+\frac 1 {2(d-1)}\int_{\bar\phi}^\phi d\tilde \phi \,e^{-2\A}\,W'^{-2}\left(dWU'-(d-2)W'U\right) \label{G1} \\
G_0^{(2)}( \phi)&=& G_0^{(2)}(\bar \phi)+2\int_{\bar\phi}^\phi d\tilde \phi \,e^{(d-2)\A}\,W'\left[\left(e^{-d\A}\right)'' \frac W {W'}U'+\left(e^{-d\A}\right)' \left( \frac W {2W'}U'\right)'\right] \nonumber\\
&&\!\!\!\!\!\!\!-\frac {3d-2}{2(d-1)}\int_{\bar\phi}^\phi d\tilde \phi \,e^{-2\A}\,W U'+2\int_{\bar\phi}^\phi d\tilde \phi \,e^{(d-2)\A}\,W'^2\left(G_0^{(1)}e^{-(d-2)\A}\right)' \label{G2}
\end{eqnarray}
where $ G_0^{(1)}(\bar \phi)$ and $ G_0^{(2)}(\bar \phi)$ are integration constants and are fixed by imposing $G_0^{(1)}(\phi_{UV})=G_0^{(2)}(\phi_{UV})=0$ so that the  three quantities in (\ref{conserved-quantities}) are linearly independent in the UV limit.

As shown in Appendix (\ref{conserved}),  the radial changes of $G_n$ are slice total derivatives $\partial_u G_n=\partial_\mu(...)$ and thus they are  {\em space-time} boundary terms to the effective action.  Since we are assuming the slices have no space-time boundary, the quantities $G_n$ are constant along the radial direction. The on-shell action is the difference between the IR and the UV, thus the terms  $C_nG_n$ are cancelled, so the final result is independent of $C_n$. It would be interesting to understand what these conserved quantities
represent from the bulk GR theory standpoint.


\section{Holographic renormalization} \label{renorm}

\subsection{UV counterterms}

As usual, the on-shell action computed in  the previous section is generically UV-divergent. Therefore, equation (\ref{onshellbare}) only makes sense if it is evaluated at a finite  UV cut-off coordinate $u_{\epsilon}$.  In this section we will identify the appropriate counterterms and analyse  the finite term, in order to write down the  renormalized generating functional explicitly to two-derivative order.

The UV-divergent  on-shell action we have found in the previous section reads, to second order in derivatives:
\be
S_{UV-div}=\int d^dx\sqrt{-\gamma}\left(W\, - \, U\, R - \,\frac {W }{W'}{U }'\frac 1 2\gamma^{\mu\nu}\partial_\mu\phi\partial_\nu\phi\right)_{u_{\epsilon}}\label{div}
\ee
where both $W(\phi)$ is specified by the bulk solution, and  $U(\phi)$ is written in terms $W$  as in equation (\ref{supf}).

 One can count the degree of divergence by the factors of $e^A$ (keeping in mind that  $\A(\phi)$ defined in (\ref{A}) has the same UV behavior as the metric scale factor). From the first term of metric flow equation (\ref{flow-eq-1}), one can deduce that the metric diverges as $\gamma_{\mu\nu}\sim e^{2A}$.
For example,  for a UV $AdS$ fixed point, the potential term in (\ref{div}) diverges as $e^{dA}$ and the second-derivative terms as $e^{(d-2)A}$ when $A\to +\infty$. This can be suspected on dimensional analysis grounds. We will derive this in section \ref{adsUV}.

The covariant divergent terms in equation (\ref{div})  coincide   with those found by Papadimitriou's in  \cite{papa} at orders zero and two in the derivative expansion, and they are found in Table 2 of  that work.  The divergence of the potential and  Ricci terms are manifestly the  same, and  the scalar kinetic term can also be transformed into the same form after some manipulations  (see Appendix \ref{kinetic}):
\be
\frac W{W'} {U}' \simeq -\A'^2e^{-(d-2)\A}\int^\phi d\tilde\phi \frac 1 {W'} e^{(d-2)A}\A'^{-2}\label{div-kin}.
\ee
where $\A(\phi)$ is defined in (\ref{A}) and the approximation holds in the UV limit

In order to subtract the divergences we need to add covariant boundary counterterms to the bulk action. If the superpotential equations (\ref{flow-eq-3}-\ref{flow-eq-4}) allow  for multiple solutions with the same UV behavior as the bulk solutions appearing in (\ref{div}), we can use these new solutions as counterterms:
\be
S^{ct}=-\int d^dx\sqrt{-\gamma}\left(W^{ct}-U_1^{ct}R-\frac {W^{ct}}{{W^{ct}}'}{U_2^{ct}}'\frac 1 2\gamma^{\mu\nu}\partial_\mu\phi\partial_\nu\phi\right)_{u_{\epsilon}}\label{ct}
\ee
where $W^{ct}$ satisfies the superpotential equation (\ref{flow-eq-3}) and $U_1^{ct}$ and $U_2^{ct}$ are different solutions of (\ref{flow-eq-4}). As we have already mentioned in section 2,  for the renormalization procedure to work, we need the UV point to be an {\em attractor}, i.e.  that there exists  continuous classes of solutions $\{W_{c}(\phi)\}$, $\{U_c(\phi)\}$ of equations  (\ref{flow-eq-3}-\ref{flow-eq-4}), labeled by a continuous parameter $c$,   such that:
\be\label{attractor5}
{W_c(\phi) - W_{c'}(\phi) \over W_c(\phi)} \to 0, \quad  {U_c(\phi) - U_{c'}(\phi) \over U_c(\phi)} \to 0, \quad \phi \to \phi_{UV}
\ee
for any pair of values  $c,c'$. If this condition is violated, then it is impossible to find a  set of counterterms which is universal, i.e. independent of the bulk solution.

In fact, it is easy to check  from the  explicit solution (\ref{supf})  of (\ref{flow-eq-4}) that  once the requirement that the UV is an attractor is satisfied for the equation for $W$, it will be automatically true for $U$, since we have:
\be \label{attractors5-1}
U(\phi) \simeq -2[(d-1)/(d-2)] \, W^{-1}(\phi), \qquad  \phi \to \phi_{UV}.
\ee
Therefore, the attractor condition is automatic, once is satisfied by the superpotential equation in (\ref{flow-eq-3}).

Since both $W$ and $U$ solve first order equations, each of the counterterm functions depends on one integration constant, which encodes the  the scheme dependence as in the standard renormalization procedure of quantum field theory.

\subsection{Two-derivative renormalized generating functional }
With the identification of the appropriate counterterns, the renormalized generating functional $S^{(ren)}$ is now defined as the sum of on-shell action  (\ref{onshellbare}) and counter-terms:
\begin{eqnarray}
S^{(ren)}&=&\lim_{\epsilon\to 0}(S_\ast+S^{ct})\nonumber\\
&=&\lim_{\epsilon\to 0}\int d^dx\sqrt{-\gamma}\left(W^R-U_1^R R-\frac W{W'}{U_2^R}'\frac 1 2\gamma^{\mu\nu}\partial_\mu\phi\partial_\nu\phi\right)_{u_{\epsilon}}\label{renormalized-action}
\end{eqnarray}
in the first line, a $*$ indicates that we are evaluating the on-shell action on an IR-regular solution.  In the second line, we have defined the renormalized quantities $W^R=(W-W^{ct})$, $U_1^R=(U -U_1^{ct})$,  $U_2^R=(U -U_2^{ct})$. We stress that $W$ is the one specified by the bulk solution (for example, it is $W_*$ if one uses the IR-regular solution). All these functions are being calculated  on the solution $( \gamma_{\mu\nu}(u), \phi(u))$,  evaluated at $u_{\epsilon}$,  and we are assuming $\lim_{\epsilon\to 0}u_{\epsilon} = u_{UV}$.  In writing the last term, we have used the attractor condition to write the coefficient simply as $W/W'$.

Finally, in equation (\ref{renormalized-action}) we have dropped  the IR contribution coming from the bulk, i.e. we are assuming implicitly  an IR regular solution is being used.

Thanks to the attractor condition (\ref{attractor5}), we can write an explicit expression for  the UV limits of   $W^R$ and $U_i^R$: indeed, these functions  satisfy the linearized versions of the superpotential equations (\ref{flow-eq-3}-\ref{flow-eq-4}) around the solution $W_*(\phi)$ and $U_*(\phi)$:
\begin{eqnarray}
{W^R}'&=&\frac d {2(d-1)}\frac {W} {W'}W^R\label{renormalized-eq-1},\\
{U_i^R}'&=&\frac {d-2} {2(d-1)}\frac {W} {W'}U_i^R\label{renormalized-eq-2}
\end{eqnarray}
In fact, as (\ref{flow-eq-4}) is linear, the equation for $U^R$ is exact without taking the UV limit.

The equations above can be solved straightforwardly and we find that in the  UV limit   $W^R$, $U^R_1$ and $U^R_2$ are simply given by:
\begin{eqnarray}
W^R(\phi)
&\simeq&c_0^R e^{-d\A(\phi)}\label{renormalized-solution-1}\\
&&\nonumber \\
U_1^R(\phi)
&\simeq&c_1^R e^{-(d-2)\A(\phi)}\label{renormalized-solution-2}\\
&&\nonumber\\
U_2^R(\phi)
&\simeq&c_2^R e^{-(d-2)\A(\phi)}\label{renormalized-solution-3}
\end{eqnarray}
where $c^R_i$ are three constants that depend on which counterterms $W^{ct}, U^{ct}$ have been chosen in the subtraction (appearing here as integration constants of the linear equations  (\ref{renormalized-solution-1}-\ref{renormalized-solution-3}) ),  and we recall  the definition:
\be\label{A-phi-2}
\A(\phi) = -{1\over 2(d-1)} \int^\phi_{\bar{\phi}} {W \over W'} d\tilde{\phi}
\ee
where $\bar{\phi}$ is a reference point which can be picked arbitrarily, and {\em independently} of the particular solution for the bulk flow $(\gamma_{\mu\nu}(x,u),\phi(x,u))$. We stress that $\A(\phi)$ is a function of the single variable $\phi$, and does not depend on the specific bulk solution. On the other hand, as explained in Section  \ref{review}, at the homogeneous level, when evaluated on a given solution $\phi(u)$, the function  $\A(\phi(u))$ coincides with the  scale factor  (\ref{A-phi})  {\em up to an additive constant}, which depends on the specific solution chosen (i.e. on the specific radial flow).

Using the UV asymptotics of renormalized superpotentials, the renormalized generating functional becomes
\be
S^{(ren)} =\lim_{u\to u_{UV}}\int d^dx\sqrt{-\gamma}\left[c_0^R e^{-d\A}-e^{-(d-2)\A}\left(c_1^R  R+c_2^R \frac{d-2}{2(d-1)} \left(\frac {W}{W'}\right)^2\frac 1 2\gamma^{\mu\nu}\partial_\mu\phi\partial_\nu\phi\right)\right]\label{renormalized-action-2}
\ee
where $\A(\phi)$ is defined as (\ref{A-phi-2}).

Notice that in equation (\ref{renormalized-action-2}) the functional form is completely determined and the scheme dependence is manifest and only appears in the  coefficients $c_0^R$, $c_1^R$ and $c_2^R$.

It remains to check explicitly  that the result (\ref{renormalized-action-2}) is finite. This was already established in section \ref{review} for the potential term, which is manifestly finite in the UV limit since $\sqrt{\gamma} \sim e^{dA}$ and for any solution of the low $A(u)$ and $\A(\phi(u))$ differ by a finite, $u$-independent constant.

To show the finiteness of (\ref{renormalized-action-2})  we will  express the renormalized generating functional in terms of the the conserved quantities that were introduced in Section \ref{conserved2},  equations (\ref{conserved-quantities}). In doing so, we will also give an explicit expression of $S^{(ren)}[\gamma,\phi]$ in terms of the running coupling and metric at {\em any} point $u$ in the bulk.

In the UV, the renormalized functions in the generating functional have simple asymptotic behavior
\be
W^R\sim e^{-d\A}, \qquad U^R\sim e^{-(d-2)\A},
\ee
These  are the same as the leading terms in the conserved quantities (\ref{conserved-quantities}).

One of the basic properties of the renormalized generating functional is the RG invariance. The quantity $S^{(ren)}$ is a functional of the flow solution $(\gamma_{\mu\nu}(x,u),\phi(x,u))$, but  although the running coupling $\phi$ and  metric $\gamma_{\mu\nu}$ are  $u$-dependent, the generating functional (\ref{renormalized-action-2})  is manifestly independent of $u$, which translates into RG-invariance in the dual field theory language.
Similarly, the conserved quantities (\ref{conserved-quantities}) are also $u$-independent,  by definition. Therefore, it is natural to try to  write the renormalized generating functional as a linear combination of the conserved quantities,
\be
S^{(ren)} = D_0 G_0\, +\, D_1 G_1\, + \,D_2G_2 ,\label{GF-conserved}
\ee
If we succeed in finding finite coefficients $D_i$ that realize the above identity,  this will show the finiteness of $S^{(ren)}$, and moreover we  will obtain a covariant expression for it at any scale $u$.

The coefficients $D_i$ can be easily obtained by assuming (\ref{GF-conserved}) holds and looking at  its  UV limit: from the definition (\ref {renormalized-action}) and (\ref{renormalized-solution-1}-\ref{renormalized-solution-3}),  the constants $D_0,\,D_1,\,D_2$ should satisfy
\begin{eqnarray}
0&=&\lim_{u\to u_{UV}}\sqrt{-\gamma}\left(W^R-D_0e^{-d\A}\right) \label{D1}\\
0&=&\lim_{u\to u_{UV}}\sqrt{-\gamma}\left(-U_1^R-D_0 G_0^{(1)} e^{-(d-2)\A}-D_1e^{-(d-2)\A}\right)R\label{D2}\\
0&=&\lim_{u\to u_{UV}}\sqrt{-\gamma}\left(-\frac W{W'}{U_2^R}'-D_0W'^{-2}e^{-(d-2)\A}G_0^{(2)}\right.\nonumber\\
&&\left.-D_1\frac W{W'}(e^{-(d-2)\A} )' -D_2W'^{-2}e^{-(d-2)\A}\right)\frac 1 2\gamma^{\mu\nu}\partial_\mu\phi\partial_\nu\phi,  \label{D3}
\end{eqnarray}
from which we can read-off:
\begin{eqnarray}
D_0&=&c_0^{R},\label{D-c-1}\\
D_1&=&-c_1^R,\label{D-c-2}\\
D_2&=&-\frac {d-2}{2(d-1)}W_{UV}^2(c_2^R+D_1),\label{D-c-3}
\end{eqnarray}
where the subscript UV denotes that the quantities are evaluated in the UV limit. The expression for $D_2$ assumes that $W(\phi)$ has a finite UV limit. If this is not the case, then one should be more careful in solving equations (\ref{D1}-\ref{D3}). We will see an explicit example of this in the case of asymptotically exponential potentials, in Section \ref{expo}.

Thanks to (\ref{GF-conserved}) and using the explicit expression for $G_0,G_1,G_2$ in equations (\ref{conserved-quantities}), we can finally write the  generating functional as a functional of the running coupling and metric on an arbitrary slice:
\begin{equation}
S^{(ren)}[\gamma_{\mu\nu},\phi]=\int d^dx\,\sqrt{-\gamma}\left(Z_0(\phi)+ Z_1(\phi)R
+ Z_2(\phi) \frac 1 2 \gamma^{\mu\nu}\partial_\mu\phi\partial_\nu\phi\right),\label{running}
\end{equation}
where $\phi$ and $\gamma_{\mu\nu}$ are evaluated at an arbitrary point in the bulk,  the coefficient functions are:
\begin{eqnarray}
Z_0(\phi_0)&=&D_0 e^{-d\A},\\
Z_1(\phi_0)&=&D_0 G_0^{(1)}e^{-(d-2)\A}+D_1e^{-(d-2)\A},\\
Z_2(\phi_0)&=&\left(D_0 G_0^{(2)}+D_2\right)W'^{-2}e^{-(d-2)\A}+D_1\frac W{W'}\left(e^{-(d-2)\A} \right)',
\end{eqnarray}
where $\A(\phi)$ is given in (\ref{A-phi-2}) and the functions   $G_0^{(i)}$ are defined in equations (\ref{G1}-\ref{G2}). The coefficients $D_n$ are related to the integration constants in superpotential solutions by (\ref {D-c-1}-\ref {D-c-3}).

The $e^{-d\A}$ and $e^{-(d-2)\A}$ behavior can be guessed from the UV asymptotics, but they are valid only in the UV region. We need some non-trivial functions to maintain the RG invariance along the whole flow. They are $G_0^{(1)}(\phi_0)$ and $G_0^{(2)}(\phi_0)$, which are related to the two-derivative terms generated by zero-derivative term along the RG flow.

\subsection{Renormalization group invariance and trace identities}

The renormalized  generating functional $S^{(ren)}[\gamma_{\mu\nu},\phi]$ is, by construction, constant on any holographic flow trajectory $\{\gamma_{\mu,\nu}(x,u),\phi(x,u)\}$,  i.e.
\be\label{trace1}
\pounds_n\, S^{(ren)} = \int d^d x \,\sigma(x) \Delta(x) S^{(ren)} = 0,
\ee
where in the first equality  we used equation (\ref{weyl7}).

As we will argue below, we can add a total  derivative terms to $S^{(ren)}$ so that equation (\ref{trace1}) holds locally, without the integral. Let us consider the action of the Lie derivative on the density  ${\cal S}(x)$ defined by:
\be\label{trace1-1}
S^{(ren)} = \int d^d x \,{\cal S}(x).
\ee
By equation (\ref{GF-conserved}), we have:
\be
 {\cal S} = D_0 {\cal G}_0 \,+ \, D_1 {\cal G}_1\,+ \, D_2 {\cal G}_2
\ee
where ${\cal G}_i$ are the local densities of the conserved charges, defined in appendix \ref{conserved}. The local conservation law  for these quantities reads:
\be \label{trace1-2}
\pounds_n {\cal G}_k = \de_\mu (\sqrt{\gamma} J^\mu_k) \qquad k=0,1,2
\ee
where the current densities $J^\mu_k$ are given explicitly in equation (\ref{conserved-1}-\ref{conserved-3}). As shown at the end of Appendix \ref{conserved}, at second order in derivatives it is possible to add total divergence terms to the charge densities ${\cal G}_k$ to cancel the right hand side in equation (\ref{trace1-2}). They are  explicitly given in equations (\ref{conserved-6}) and (\ref{conserved-10}). If we add these total derivative terms to the renormalized action, we can make  its density strictly radially invariant up to higher derivative terms:
\be\label{trace1-3}
\pounds_n {\cal S}(x) = O(\de^4).
\ee
Using the above  equation and equation (\ref{weyl7-0}), it follows easily that the action $S^{(ren)}$  satisfies now  the {\em local} equation:
\be\label{trace2}
\Delta(x) S^{(ren)} =  \left(2\gamma_{\mu\nu}{\delta \over \delta \gamma_{\mu\nu}} + \beta_{\mu\nu}{\delta \over \delta \gamma_{\mu\nu}} + \beta_\phi {\delta \over \delta \phi}\right)  S^{(ren)} =0,
\ee
This equation, translated on the field theory side,  encodes the local renormalization group invariance of $S^{(ren)}$. It of course implies the integrated version $\Delta S^{(ren)}=0$, which is the statement of invariance under a rigid RG-transformation  in the field  theory. Notice that  to obtain it from equation (\ref{trace1}), it was crucial to satisfy the local radial invariance condition (\ref{trace1-3}),   to cancel the factor of $\sigma(x)$. This  could be done by adding suitable boundary terms.

As a side remark, we notice that the operator  in (\ref{trace1}), i.e.
\be
\int d^d x \sigma(x)\Delta(x)  = \int d^d x \left[\left(\pounds_n \gamma_{\mu\nu}\right) {\delta \over \delta \gamma_{\mu\nu}(x)} + \left(\pounds_n \phi\right) {\delta  \over \delta \phi(x)} \right]
\ee
had previously appeared in various works using the Hamilton-Jacobi method,  \cite{ps,ps2}. Close to an $AdS$ boundary  it  reduces to lowest order in derivatives  to $\Delta/\ell$, which  in \cite{ps,ps2} too was  identified  as the  generator of asymptotic dilatations. Furthermore, in  \cite{ps2,papa}  a discussion of the need to add   boundary terms in order to obtain local equations was given, which is similar  to the one presented here. In that case however the analysis   applied to the Hamilton-Jacobi function, here to the renormalized action.

Next, we define  renormalized expectation values $\<T_{\mu\nu}\>$ and $\< \cO\>$ by
\be\label{trace3}
\<T_{\mu\nu}\> =- {2\over  \sqrt{\gamma}} {\delta S^{(ren)} \over \delta \gamma^{\mu\nu}}, \qquad \<\cO\> =  {1\over  \sqrt{\gamma}} {\delta S^{(ren)} \over \delta \phi}.
\ee
Equation (\ref{trace2}) leads to the trace identity:
\be\label{trace4}
\< {T^\mu}_\mu\> = -\beta_{\mu\nu} T^{\mu\nu} - \beta_\phi \< \cO\>.
\ee
The first term on the right hand side of (\ref{trace4}) is somewhat unusual from the field theory point of view. Recalling that $\beta_{\mu\nu}$ was chosen to be traceless (see equations (\ref{weyl4},\ref{weyl7-1}), it represents  a contribution to the trace anomaly from the traceless part of the stress tensor.
To the order at which we are working however, i.e. up to second order in derivatives and curvature invariants, such contribution is absent. Indeed,  $\beta_{\mu\nu}$ starts at second order in derivatives, but the zeroth order part of $\delta S^{(ren)} /\delta \gamma_{\mu\nu}$ is proportional to  $\gamma_{\mu\nu}$. Therefore any contribution from the second term  in (\ref{trace2}) is at least of order four in derivatives (or two in the curvature). We can thus write (\ref{trace2}) as
\be \label{trace5}
 \left(2\gamma_{\mu\nu}{\delta \over \delta \gamma_{\mu\nu}}+ \beta_\phi {\delta \over \delta \phi}\right)  S^{(ren)} = \text{local 4-derivative terms}.
\ee
This matches  the way trace identities are written in field theories with space-time dependent couplings and general metric, for example in \cite{osborn} and it also gives the standard result for the field theory Weyl anomaly in four-dimensional curved space with constant couplings,
\be\label{trace6}
\< {T^\mu}_\mu\> = - \beta_\phi \< \cO\> + \text{($R^2$-terms)}.
\ee

At zeroth order in derivatives, the trace identity (\ref{trace6}) becomes the usual quantum field theory trace anomaly equation for constant coupling, with the $\beta$-function given, according to equation (\ref{beta0}), by:
\be\label{trace7}
\beta^{(0)}(\phi) = -2(d-1) W'/W\;.
\ee
 This can be checked by direct computation of the functional derivatives of $S^{(ren)}$:
\be\label{trace8}
\< {T^\mu}_\mu\> = d Z_0(\phi), \qquad \<\cO\> =  {d\over 2(d-1)}{W\over W'}Z_0(\phi),
\ee
where we have used (\ref{A-phi-2}) in the last equality.
Since, by equations (\ref{AE2}),  the right hand side of (\ref{trace7}) equals $d\phi/dA$, equation (\ref{trace7})  confirms once more the  identification (in the homogeneous case) of the energy scale  with the metric scale factor, as discussed in Section \ref{review}.

To conclude this section, we note  that the renormalized operators defined in equation (\ref{trace3}) are scale dependent: they are defined as functional derivatives of the RG-invariant functional $S^{(ren)}$ with respect to the scale dependent couplings. We can check this explicitly by computing the action of the operator $\Delta$ on these quantities. We find:
\be\label{trace9}
\Delta \< T_{\mu\nu}(x) \> = -d \< T_{\mu\nu}(x) \>, \qquad \Delta \< \cO(x) \> = \left (-d\, - \int d^d y\, {\delta \beta_\phi(x) \over \delta \phi(y)}\right) \< \cO(x) \>.
\ee

On the other hand, we can also construct RG-invariant operators starting from (\ref{trace3}). In order to do this, we first have to define an auxiliary fixed-volume metric $\tilde{\gamma}_{\mu\nu}$, defined for example as the finite part of $\gamma_{\mu\nu}$ in the UV limit:
\be\label{trace10}
\gamma_{\mu\nu} \simeq e^{2A(u,x)}\tilde{\gamma}_{\mu\nu}(x) , \qquad u\to -\infty
\ee
where $A$ diverges in the $UV$ and $\tilde{\gamma}$ stays finite. In  the asymptotically $AdS$ case, this is the usual definition of the metric of the boundary field theory.  Notice that equation (\ref{trace10}) does not fix $\tilde{\gamma}_{\mu\nu}$ uniquely, each different choice corresponding to a different choice for the UV space-time metric.

Since we have taken $\tilde{\gamma}_{\mu\nu}$ to be scale-independent, we can now write an RG-invariant stress tensor:
\be\label{trace11}
\< T^{inv}_{\mu\nu}  \> = \left\<\sqrt{\gamma \over \tilde{\gamma}} T_{\mu\nu}\right\>, \qquad \Delta \< T^{inv}_{\mu\nu} \> =0
\ee

For the scalar operator, to lowest order in the derivative expansion  the RG-invariant combination is:
\be\label{trace12}
\<\cO^{inv} \> = \left\<\sqrt{\gamma\over \tilde{\gamma}} \,\beta_{\phi}\,\cO\right\>, \qquad \Delta \<\cO^{inv} \> = O(\de^2) .
\ee
It would be interesting to find a generalization  that is RG-invariant up to fourth-order terms.

\subsection{The two-derivative quantum effective action}

The renormalized on-shell action (\ref{running}) is dual to the generating functional of the connected correlators of the field theory, as functional of the  renormalized sources at a given holographic  scale $u$.

The classical field (the vev $\cO$) of the scalar operator is given by the variation of the generating functional with respect to the source $\phi(u)$:
\begin{eqnarray}
\<\mathcal O(x)\>_\phi&=&\frac 1{\sqrt{- \gamma}} \frac {\delta S^{(ren)}}{\delta \phi(x)}\nonumber\\
&=&Z_0'+Z_1'R +Z_2'\frac 1 2\gamma^{\mu\nu}\partial_\mu \phi \partial_\nu \phi-\gamma^{\mu\nu}\nabla_\mu (Z_2\partial_\nu \phi)\label{vev}.
\end{eqnarray}

Then, the  1PI effective action is defined by the  Legendre transformation:
\begin{eqnarray}
\Gamma(\<\mathcal O\>_\phi)&=&\int d^d x \sqrt{- \gamma} \phi(x)\<\mathcal O(x)\>_\phi-S^{(ren)}\nonumber\\
&=&\int d^d x \sqrt{- \gamma}\left(\phi Z_0'-Z_0+(\phi Z_1' -Z_1)R+(\phi Z_2' +Z_2)\frac 1 2\gamma^{\mu\nu}\partial_\mu \phi \partial_\nu \phi \right)\label{Gamma}\nonumber\\
\end{eqnarray}
where the source $\phi(x)=\phi(\<\mathcal O(x)\>)$ is a function of the vev. The explicit form can be obtained by inverting (\ref{vev}) order by order. From now on, $\<\mathcal O(x)\>$ will be denoted by ${\cal O}$.

To write the quantum effective action more explicitly, we first assume an ansatz of the form:
\be
\phi(\mathcal  O)=\phi_0(\mathcal  O)+\phi_1(\mathcal O) R + \phi_2(\mathcal O) \frac 1 2\gamma^{\mu\nu}\partial_\mu \mathcal O \partial_\nu \mathcal O + \phi_3(\mathcal O) \gamma^{\mu\nu}\nabla_\mu \partial_\nu \mathcal O
\ee
To leading order,
\be
\mathcal O=Z_0'(\phi_0(\mathcal O)),\label{vev-source}
\ee
so $\phi_0(\mathcal O)$ is the inverse function of $Z_0'(\phi)=\frac d {d\phi}Z_0(\phi)$. To the next  order, we find:
\begin{eqnarray}
0&=&Z_0''(\phi_0(\mathcal O))\left(\phi_1(\mathcal O) R + \phi_2(\mathcal O) \frac 1 2\gamma^{\mu\nu}\partial_\mu \mathcal O \partial_\nu \mathcal O + \phi_3(\mathcal O) \gamma^{\mu\nu}\nabla_\mu \partial_\nu \mathcal O\right)\nonumber\\
&&+Z_1'(\phi_0(\mathcal O)) R+Z_2' (\phi_0(\mathcal O))\frac 1 2\gamma^{\mu\nu}\partial_\mu (\phi_0(\mathcal O)) \partial_\nu (\phi_0(\mathcal O))\nonumber\\
&&-\gamma^{\mu\nu}\nabla_\mu (Z_2(\phi_0(\mathcal O))\partial_\nu (\phi_0(\mathcal O))),\nonumber\\
\end{eqnarray}
which leads to:
\begin{eqnarray}
\phi_1(\mathcal O)&=&-\frac {Z_1'(\phi_0(\mathcal O))}{Z_0''(\phi_0(\mathcal O))},\\
\phi_2(\mathcal O)&=&\frac {\phi_0'(\mathcal O)^2 Z_2'(\phi_0(\mathcal O))+2\phi_0''(\mathcal O) Z_2(\phi_0(\mathcal O))}{Z_0''(\phi_0(\mathcal O))},\\
\phi_3(\mathcal O)&=&\frac {\phi_0'(\mathcal O)Z_2(\phi_0(\mathcal O))}{Z_0''(\phi_0(\mathcal O))}.
\end{eqnarray}

Inserting these expressions back into equation (\ref{Gamma}), the   1PI effective action  takes the simple form:
\begin{equation}
\Gamma[\cO,\gamma] =\int d^d x \sqrt{- \gamma}\left(\mathcal O\phi_0(\mathcal O)-Z_0(\phi_0(\mathcal O))-Z_1(\phi_0(\mathcal O))R-\right.
\label{1PI}
\end{equation}
$$
-
\left.\phi_0'^2(\mathcal O) Z_2(\phi_0(\mathcal O))\frac 1 2\gamma^{\mu\nu}\partial_\mu \mathcal O \partial_\nu  \mathcal O	 \right)
$$
Notice that the two-derivative term coefficient functions are the same as those appearing before  the Legendre transform, evaluated at {\em zeroth-order} expression for $\phi(\cO)$.

\section{The nolinear effective action for constant curvature}\label{curvature}
In the case where one neglects derivatives of the slice curvature $R$, there is a direct way of computing the effective action (for constant $\phi$ source) as a function of $\phi,R$.
This can be achieved by considering the bulk equations in the presence of non-trivial constant curvature along the boundary directions.

We consider again the bulk action (\ref{action}) and the Euclidean bulk metric
\begin{equation}
d^2 s=d u^2+\ell^2e^{2A}d\Omega_d^2,
\end{equation}
where u is the holographic direction, $r$ is the radius of the d-sphere and $d\Omega_d^2=\delta_{ij}\left(\prod_{n=1}^{i-1}\sin^2 \theta_n\right) d \theta^i d\theta^j$ is the Euclidean sphere metric in spherical coordinates with
\be
x^i=r\cos \theta^i\prod_{n=1}^{i-1}\sin \theta^n.
\ee
The scalar is again taken to be a function of $u$, $\phi(u)$.

The Einstein's equations are
\be
d(d-1)\dot A^2-e^{-2A}R_d=\frac 1 2 \dot\phi^2+V(\phi)
\label{eq3},\ee
\be
2(d-1)\ddot A+d(d-1)\dot A^2-\frac {d-2} d e^{-2A}R_d=-\frac 1 2 \dot\phi^2+V(\phi)
\label{eq1},
\ee
where $R_d=d(d-1)\ell^{-2}$ is the curvature of d-sphere $S^d$ of radius $\ell$ and the dot stands for derivative with respect to $u$.

The Klein-Gordon equation is
\begin{equation}
\ddot\phi+d\dot A\dot \phi+V'=0,
\end{equation}
where $'=\frac {d}{d \phi}$ is the derivative with respect to $\phi$.

The on-shell action ( the sum of bulk terms and Gibbons-Hawking term) becomes now
\begin{equation}
S=\ell^{d}\Omega_d \int_{UV}^{IR} du \left[\frac d {du}(2(d-1)e^{dA}\dot A)+\frac 2 d e^{(d-2)A}R_d\right],
\label{eq7}\end{equation}
where $\Omega_d$ is the volume of $S^d$ of radius 1.

We may rewrite the second order equations (\ref{eq3}), (\ref{eq1}) as first order equations  as follows
\be
W(\phi,R_d)=-2(d-1)\dot A\sp
S(\phi,R_d)=\dot \phi\sp
T(\phi,R_d)=e^{-2A}.
\label{eq2}
\ee

In terms of the three ``superpotential" functions, $W,S,T$, the second equations of motion are reduced to a set of first order equations

\be
\frac d {4(d-1)}W^2-W'S-\frac {d-2}d R_dT=-\frac 1 2 S^2+V,\label{cur-2}
\ee
\be
S'S-\frac d {2(d-1)}WS+V'=0\label{cur-3},
\ee
\be
(d-1)ST' =WT\label{cur-4},
\ee
\be
\frac d {4(d-1)}W^2-R_d T=\frac 1 2 S^2+V.
\label{cur-1}
\ee

One of the above equations is redundant.
The equations  can be rewritten as
\be
T={1\over R_d}\left[{dW^2\over 4(d-1)}-{S^2\over 2}-V\right]
\label{eq4},\ee
\be
W'={d-1\over d}S-{2\over d}{V\over S}+{W^2\over 2(d-1)S}\sp
S'={dW\over 2(d-1)}-{V'\over S}.
\label{eq5}\ee
 Therefore, one needs to first solve the system of two first order equations
 (\ref{eq5}), then determine $T$ from (\ref{eq4}) and then solve (\ref{eq2}).
The solution of the system (\ref{eq5}) contains two arbitrary constants. One is fixed by the regularity of the solution. The other is determined by the curvature from
\be
R_d=e^{2A_0-\int_{\phi_0}^{\phi}{W(z)\over (d-1)S(z)}dz}\left({dW(\phi)^2\over 4(d-1)}-{S(\phi)^2\over 2}-V(\phi)\right).
\label{eq6}\ee
Using (\ref{eq5}) it can be shown that the right-hand-side of (\ref{eq6}) is independent of $\phi$ and therefore constant.

We can expand the solutions for small curvature,  derive the analytic form of the solution and the on-shell action, and compare with our previous results that should correspond to the linearized order in the curvature.

To derive the Ricci term in the on-shell action, we first expand $W$, $S$ and $T$  around the flat limit $R_d\rightarrow 0$
\begin{eqnarray}
W&=&W_0(\phi)+R_d W_1(\phi)+...\\
S&=&S_0(\phi)+R_d S_1(\phi)+...\\
T&=&T_0(\phi)+R_d T_1(\phi)+...
\end{eqnarray}

At the zeroth order,
\begin{equation}
S_0=W_0', \qquad T_0=e^{-2\bar A},
\end{equation}
where $W_0$ satisfy the first superpotential equation (\ref{flow-eq-3}) and
\be
\bar A=\bar A(\phi_0)-\frac 1{2(d-1)}\int_{\phi_0}^\phi dz \frac {W_0(z)}{W'_0(z)}.\label{A-bar}
\ee

Expanding (\ref{cur-2}) to first order in small curvature expansion and using the zeroth order solutions of $S_0,T_0$, we obtain a first order equation for $W_1$
\begin{equation}
\frac d {2(d-1)}W_0W_1-W_0'W_1'-\frac {d-2}de^{-2\bar A}=0,
\end{equation}
where $\bar A$ is defined in (\ref{A-bar}). $S_1$ in (\ref{cur-2}) drops out.

The solution of $W_1$ is
\begin{equation}
W_1(\phi)=e^{-d\bar A}\left(c_1-\frac {d-2}d \int_{\phi_{IR}}^\phi \frac {dz}{W_0'(z)} e^{(d-2)\bar A(z)}\right).
\end{equation}
Plugging in the solution of $W_1$, we are able to evaluate the on-shell action to first order in small curvature expansion.

From (\ref{cur-1}) one can derive an algebraic equation for $S_1$,
\be
S_1=\frac 1 {W_0'}\left(\frac d {2(d-1)}W_0W_1-e^{-2\bar A}\right).
\ee

From (\ref{cur-4}) and the zeroth order solutions of $S_0,T_0$, we derive a first order equation for $T_1$
\be
T_1'-\frac {W_0}{(d-1)W_0'}T_1=\frac {T_0}{(d-1)W_0'}\left(W_1-S_1\frac {W_0}{W_0'}\right),
\ee
whose solution is
\be
T_1=e^{-2\bar A}\left(c_2+\frac 1 {d-1}\int_{\phi_{IR}}^\phi \frac {d z}{W_0'(z)}\left(W_1(z)-S_1(z)\frac {W_0(z)}{W_0'(z)}\right)\right).
\ee

We now expand the on-shell action (\ref{eq7}) in the  curvature $R_d$ and substitute the perturbative solution for $W=W_0+R_d W_1+...$.
The on-shell action becomes
\begin{eqnarray}
S&=&\ell^{d}\Omega_d\left[(-e^{dA}W_0)|^{IR}_{UV}+ \int_{\phi_{UV}}^{\phi_{IR}}\, du ~ e^{(d-2)\bar A}R_d\right]\nonumber\\
&=&\int \sqrt{\gamma}\left(W_0- U_\ast(e^{-2 A}R_d)\right)_{UV},
\end{eqnarray}
where we have assumed that $e^{dA}W_0\rightarrow 0$ in the IR for a regular solution of $W_0$. The integration constant $c_1$ does not appear in the final result because the difference between its IR and UV contributions is of higher order in $R_d$ expansion.

The final result is in accordance with our general discussion. $W_0$ is the first superpotential satisfying (\ref{flow-eq-3}) and $U(\phi)=e^{-(d-2)\bar A}\int_{\phi_{IR}}^{\phi}\,\frac {d z}{W_0'(z)} e^{(d-2)\bar A(z)}$ is the IR regular solution of the second superpotential equation (\ref{flow-eq-4}).

\section{Explicit examples} \label{examples}

\subsection{Deformations of AdS fixed points} \label{adsUV}
\subsubsection{UV fixed point}

Consider the case with the following bulk scalar potential which, close to $\phi=0$, takes the form:
\begin{eqnarray}
V(\phi)=\ell^{-2}(d(d-1)-\frac 1 2 m^2\phi^2+O(\phi^3))
\end{eqnarray}
This correspods to a UV $AdS$ fixed point at $\phi=0$.

The UV expansion of the superpotential was discussed in detail in \cite{papa1}, and here we summarize the results.  The power series solution around $\phi=0$ of the superpotential equation (\ref{flow-eq-3}) is:
\begin{equation} \label{Wads}
W=\ell^{-1}\Big(2(d-1)+\frac 1 2 \Delta_\pm \phi^2+O(\phi^3)\Big)
\end{equation}
where
\be
\Delta_\pm=\frac d 2\pm\sqrt{\frac{d^2}4+m^2}
\ee
 are the two solutions of
 \be
 \Delta(\Delta-d)=m^2.
 \ee

For $m^2\ge -\frac {d^2}4+1$, the scaling dimension of the scalar operator in the field theory can only be $\Delta_+$ due to the unitary bound $\Delta> \frac d 2-1$ or the fact that this is the only normalizable solution to be considered as an element of the bulk Hilbert space.

For $-\frac {d^2}4+1> m^2> -\frac {d^2}4$, both $\Delta_+$ and $\Delta_-$ are allowed, so there are two ways to quantize the scalar field.

For $m^2= -\frac {d^2}4$, $\Delta_+=\Delta_-=\frac d 2$. There are $\log |\phi|$ terms due to the degeneracy of the conformal dimensions of the dual operators, \cite{ar}.

For $-\frac {d^2}4>m^2$, the Breitenlohner-Freedman bound is violated and the asymptotic AdS space-time is instable due to the tachyonic field $\phi$.\\

Close to $\phi=0$, we have:
\be
\A = -{1\over 2(d-1)} \int^\phi {W'\over W}  \simeq -{1\over \Delta_{\pm}} \log \phi,
\ee
For a UV fixed point, the right hand side must go to $+\infty$. For $m^2>0$, $\Delta_-<0$ so only $\Delta_+$ is allowed in (\ref{Wads}). This corresponds to a deformation by a vev of an irrelevant operator of dimension $\Delta_+>d$.

We will instead focus on the case of an irrelevant operator, with $m^2<0$ and both $\Delta_\pm >0$ allowed in (\ref{Wads}).  The solution of (\ref{AE2}) in domain-wall coordinates reads at leading order:
\be
e^{A(u)}\simeq e^{-{u\over\ell}} , \qquad \phi(u)\simeq \alpha e^{\Delta_\pm {u\over\ell}}, \qquad u\to -\infty.
\ee
therefore chosing  $\Delta_-$ in (\ref{Wads})  corresponds to deforming by a source, whereas  chosing $\Delta_+$ corresponds to a vev deformation with the source set to zero\footnote{In the range of  $m$  where two quantisations are possible we will always choose the standard quantization, with $\Delta_-$ as the dimension of the source and $\Delta_+$ that of the operator.}.

At subleading order, one finds a possible correction to the superpotential of the form:
\begin{eqnarray} \label{C}
W=\ell^{-1}\Big(2(d-1)+\frac 1 2 \Delta_- \phi^2+O(\phi^3)+C\phi^{\frac d {\Delta_-}}(1+O(\phi))\Big)
\end{eqnarray}
where $C$ is an integration constant.  Notice that  $\frac d {\Delta_-}>2$ from the definition of $\Delta_-=\frac d 2-\sqrt{\frac{d^2}4+m^2}$, so the term proportional to $C$ is always subleading.
The integration constant $C$ is related to the bare vev of the dual scalar operator. We see that the difference between solutions with different constants $C$ becomes negligible as $\phi \to 0$: this means that the $UV$ fixed point is an attractor.
In general, the high order terms will be of order $\phi^{2+(\frac d {\Delta_-}-2)j+2k}$ with non-negative integers $j,k\ge 0$.

The renormalized generating functional (\ref{renormalized-action-2}) takes the form:
\begin{eqnarray}
S^{(ren)}&=&\lim_{u^{UV}\rightarrow -\infty}M^{d-1}\ell^{-1}\int d^d x\sqrt{-\gamma}\left[c_0^R\phi^{\frac d {\Delta_-}}\right.\nonumber\\
&&\qquad\qquad\qquad\left.-\ell^2\phi^{\frac {d-2} {\Delta_-}}\left(c_1^R R +\tilde c_2^R\frac 1 2\gamma^{\mu\nu}\partial_\mu\log\phi\partial_\nu\log\phi\right)\right]
\end{eqnarray}
where
\be
\tilde c_2^R=c_2^R\left(2(d-1)(d-2)\Delta_-^{-2}\right).
\ee

At a finite cut-off $u^{UV}=\ln \epsilon$, the metric sources $\tilde\gamma^\epsilon$ and the scalar field source $\alpha^\epsilon$ are defined as:
\begin{eqnarray}
\gamma_{\mu\nu}&=&e^{-2u^{UV}}\tilde  \gamma_{\mu\nu}^\epsilon,\qquad\phi=\alpha^\epsilon e^{\Delta_- u^{UV}}.
\end{eqnarray}
The UV sources are defined as:
\be
\tilde\gamma(x)=\lim_{u\rightarrow -\infty}\tilde\gamma^\epsilon(x,u^{UV}),\qquad \alpha(x)=\lim_{u\rightarrow -\infty}\alpha^\epsilon(x,u^{UV}).
\ee

Substituting the definition of the sources, the renormalized on-shell action becomes:
\begin{eqnarray}
S^{(ren)}&=&M^{d-1}\ell^{-1}\lim_{u^{UV}\rightarrow -\infty}\int d^d x\sqrt{-\tilde\gamma}e^{-du^{UV}}\left[e^{du^{UV}}c_0^R{\alpha^\epsilon}^{\frac d {\Delta_-}}\right.\nonumber\\
&&\qquad\qquad\qquad\left.-\ell^2  e^{du^{UV}} {\alpha^\epsilon}^{{\frac {d-2} {\Delta_-}}}\left(c_1^R R^{(\tilde\gamma)}+\tilde c_2^R\frac 1 2\tilde\gamma^{\mu\nu}\partial_\mu\log\alpha^\epsilon\,\partial_\nu\log\alpha^\epsilon\right)\right]\nonumber\\
&=&M^{d-1}\ell^{-1}\int d^d x\sqrt{-\tilde\gamma}\left[c_0^R\alpha^{\frac d {\Delta_-}}\right.\nonumber\\
&&\qquad\qquad\qquad\left.- \ell^2\alpha^{{\frac {d-2} {\Delta_-}}}\left(c_1^R R^{(\tilde\gamma)}+\tilde c_2^R\frac 1 2\tilde\gamma^{\mu\nu}(\partial_\mu\log\alpha)\,(\partial_\nu\log\alpha)\right)\right]\nonumber\\
\end{eqnarray}
which is a functional of the UV source $\alpha$, and whose functional form coincides with the result found in  \cite{papa1}.

To the leading order, the vev of the dual scalar operator is:
\begin{eqnarray}
\<\cO\>=\frac 1 {\sqrt{-\tilde\gamma}}\frac {\delta S^{(ren)}} {\delta \alpha} =M^{d-1}\ell^{-1}c_0^R\frac d {\Delta_-}\alpha^{\frac d {\Delta_-}-1}+O(\partial^2)
\end{eqnarray}

The Legendre transform of the renormalized generating functional  (\ref{1PI}) is:
\begin{eqnarray}
\Gamma&=&\int d^d x \sqrt{-\tilde\gamma} \<\cO\> \alpha -S^{(ren)}\nonumber\\
&=&M^{d-1}\ell^{-1}\int d^d x \sqrt{-\tilde\gamma}\left[c_0^R\frac {d-\Delta_-} {\Delta_-} \left(\frac {\Delta_-}{dc_0^R}\right)^{\frac d {d-\Delta_-}} \Big(\frac{\<\cO\>}{M^{d-1}\ell^{-1}}\Big) ^{\frac d {d-\Delta_-}}\right.\nonumber\\
&&\left.+c_1^R\left(\frac {\Delta_-}{dc_0^R}\right)^{\frac {d-2} {d-\Delta_-}} \Big(\frac{\<\cO\>}{M^{d-1}\ell^{-1}}\Big)^ {\frac {d-2} {d-\Delta_-}} R^{(\tilde\gamma)} \right.\nonumber\\
&&\left.+c_2^R\left(\frac {\Delta_-}{dc_0^R}\right)^{\frac {d-2} {d-\Delta_-}} \frac {\Delta_-^2}{(d-\Delta_-)^2}\<\cO\>^ {\frac {d-2} {d-\Delta_-}} \frac 1 2\tilde\gamma^{\mu\nu}\partial_\mu\log\frac{\<\cO\>}{M^{d-1}\ell^{-1}}\partial_\nu \log \frac{\<\cO\>}{M^{d-1}\ell^{-1}}\right]\nonumber\\
\end{eqnarray}

Now we calculate the renormalized generating functional (\ref{running}) as a functional of running coupling. In the UV region, the renormalized generating functional is:
\begin{eqnarray}
S^{(ren)}&=&M^{d-1}\ell^{-1}\int d^dx\,\sqrt{-\gamma}\left(D_0 e^{-d\A}+\ell^2D_1 e^{-(d-2)\A} R \right.\nonumber\\
&&\qquad\qquad\qquad\qquad\qquad\left.
+\ell^2\tilde D_2 e^{-(d-2)\A} \phi^{-2} \frac 1 2 \gamma^{\mu\nu}\partial_\mu\phi\partial_\nu\phi\right)\nonumber\\
&=&M^{d-1}\ell^{-1}\int d^dx\,\sqrt{-\gamma}\left(D_0 \phi^{\frac d{\Delta_-}}+\ell^2 D_1 \phi^{\frac {d-2}{\Delta_-}}R
\right.\nonumber\\
&&\qquad\qquad\qquad\qquad\qquad\left.+\ell^2 \tilde D_2 \phi^{\frac {d-2}{\Delta_-}} \frac 1 2 \gamma^{\mu\nu}\partial_\mu\log\phi\partial_\nu\log\phi\right)\nonumber\\
\end{eqnarray}
where
\be
\tilde D_2=(D_2+ {2(d-1)}(d-2) D_1)\Delta_-^{-2}.
\ee
We have neglected the UV subleading terms. $D_n$ are related to the integration constants in superpotential solutions by (\ref {D-c-1}-\ref {D-c-3}).

To the zero-derivative order, the vev of dual operator of the running coupling is:
\begin{eqnarray}
\<\tilde \cO\>=\frac 1{\sqrt{-\gamma}}\frac {\delta S^{(ren)}}{\delta \phi}=M^{d-1}\ell^{-1}D_0 \frac d {\Delta_-} \phi^{\frac {d-\Delta_-}{\Delta_-}}.
\end{eqnarray}

The Legendre transform of the renormalized generating functional  (\ref{1PI}) as a function of the vev:
\begin{eqnarray}
\Gamma&=&\int d^d x \sqrt{-\gamma} \<\tilde \cO\> \phi -S^{(ren)}\nonumber\\
&=&M^{d-1}\ell^{-1}\int d^d x \sqrt{-\gamma}\left[\tilde E_0\Big(\frac{ \<\tilde \cO\>}{M^{d-1}\ell^{-1}}\Big) ^{\frac d {d-\Delta_-}}-\ell^2\tilde E_1\Big(\frac{ \<\tilde \cO\>}{M^{d-1}\ell^{-1}}\Big) ^ {\frac {d-2} {d-\Delta_-}} R \right.\nonumber\\
&&\left.-\ell^2\tilde E_2 \Big(\frac{ \<\tilde \cO\>}{M^{d-1}\ell^{-1}}\Big) ^ {\frac {d-2} {d-\Delta_-}}\frac 1 2\gamma^{\mu\nu}\partial_\mu\log \Big(\frac{ \<\tilde \cO\>}{M^{d-1}\ell^{-1}}\Big) \partial_\nu \log \Big(\frac{ \<\tilde \cO\>}{M^{d-1}\ell^{-1}}\Big) \right].\nonumber\\
\end{eqnarray}
where
\begin{eqnarray}
\tilde E_0&=&D_0\frac {d-\Delta_-} {\Delta_-} \left(\frac {\Delta_-}{dD_0}\right)^{\frac d {d-\Delta_-}},\\
\tilde E_1&=&D_1 \left(\frac {\Delta_-}{dD_0}\right)^{\frac {d-2} {d-\Delta_-}},\\
\tilde E_2&=&\tilde D_2\left(\frac {\Delta_-}{d D_0}\right)^{\frac {d-2} {d-\Delta_-}} \frac {\Delta_-^2}{(d-\Delta_-)^2} .
\end{eqnarray}

The canonically normalized operator is:
\be
\mathcal O=(M^{d-1}\ell\tilde E_2)^{\frac 1 2}\frac {2(d-\Delta_-)}{d-2}\Big(\frac{ \<\tilde \cO\>}{M^{d-1}\ell^{-1}}\Big)^{\frac {d-2}{2(d-\Delta_-)}}.
\ee

In terms of the canonically normalized operator, the renormalized generating functional is:
\begin{eqnarray}
\Gamma= \int d^d x \sqrt{-\gamma}\left(E_0\mathcal O^{\frac {2d}{d-2}}+E_1\mathcal O^{2}R-\frac 1 2 \tilde\gamma^{\mu\nu}\partial_\mu \mathcal O\partial_\nu\mathcal O\right)
\end{eqnarray}
where
\begin{eqnarray}
E_0&=&(M\ell)^{-\frac {2(d-1)}{d-2}}c_0^R \frac{d-\Delta_-}{\Delta_-}\left(-\frac {8(d-1)}{d-2}c_2^R\right)^{-\frac {d}{d-2}},\\
E_1&=&-\frac {d-2}{8(d-1)}\frac {c_1^R}{c_2^R}.
\end{eqnarray}

\subsubsection{IR fixed point}
Consider the case with the following bulk scalar potential in the IR:
\begin{eqnarray}
V(\phi)=\ell_{IR}^{-2}(d(d-1)-\frac 1 2 m^2(\phi-\phi_{IR})^2+O((\phi-\phi_{IR})^3))
\end{eqnarray}
We will choose $\ell_{IR}=1$ in the following discussion.

The IR regular solutions around $\phi=\phi_{IR}$ are:
\begin{eqnarray}
W_\ast&=&2(d-1)-\frac 1 2 \Delta_-(\phi-\phi_{IR})^2+O((\phi-\phi_{IR})^3)\nonumber\\
U_\ast&=& -\frac 1 {d-2}+O((\phi-\phi_{IR})^1)
\end{eqnarray}
where
\be
\Delta_-=-\frac d 2-\sqrt{\frac{d^2}4+m^2}
\ee
 which are the solutions of
 \be
 \Delta(\Delta+d)=m^2.
 \ee
  The other solutions of the superpotential $W$ will overshoot and can not flow to the AdS IR fixed point. They are showed in the numerical example of AdS-AdS flow.

In the IR limit $u\rightarrow \infty$,
\begin{eqnarray}
\gamma_{\mu\nu}=e^{-2u}\tilde  \gamma_{\mu\nu},\,\phi\simeq\alpha e^{\Delta_+ u} ,
\end{eqnarray}
where $\tilde  \gamma_{\mu\nu}$ stay finite in the IR.

The potential term gives zero in the IR:
\begin{eqnarray}
\sqrt{-\gamma}W_\ast\simeq\sqrt{-\tilde\gamma}e^{-du}d(d-1)\rightarrow 0
\end{eqnarray}

The two derivative terms also vanish in the IR:
\begin{eqnarray}
\sqrt{-\gamma}U_\ast R &\simeq& \sqrt{-\tilde\gamma}e^{-(d-2)u} R^{(\tilde\gamma)}\rightarrow 0\nonumber\\
\sqrt{-\gamma}\left(\frac {W_\ast}{W_\ast'}  {U_{\ast}}'\right)\frac 1 2\gamma^{\mu\nu}\partial_\mu\phi\partial_\nu\phi&\simeq&\sqrt{-\tilde\gamma}e^{-(d-2)u} \, O \left(\phi^{-2}\partial_\mu\phi\partial_\nu\phi\right)\rightarrow 0
\end{eqnarray}
where we have assumed $R^{(\tilde\gamma)}$ and $\phi^{-2}\partial_\mu\phi\partial_\nu\phi$ stay finite.

The IR contribution of the IR regular solutions to the bare on-shell action(\ref{onshellbare}) is just zero.

\subsection{Exponential Potentials} \label{expo}
We consider the scalar potential:
\begin{eqnarray}
V=\ell^{-2}\phi^a e^{b\phi}
\label{vvv}\end{eqnarray}
where $b>0$.

From explicit calculations, the limit $\phi\rightarrow \infty$ corresponds to the IR limit where the scale factor vanishes $e^A\rightarrow 0$ and the limit $\phi\rightarrow -\infty$ corresponds to the UV limit where the scale factor diverges $e^A\rightarrow \infty$.

\subsubsection{UV limit}
To study the UV limit, we consider $\phi\rightarrow -\infty$ and $u\rightarrow -\infty$. We first assume $a=0$ in (\ref{vvv}).

If $b<\sqrt{\frac {2d}{d-1}}$, the superpotential is:
\begin{eqnarray}
W=\ell ^{-1}\Big(W_0 e^{\frac b 2  \phi}\big(1+O(\phi^{-1})\big)+c_0 e^{\frac d {b(d-1)}\phi}\big(1+O(\phi^{-1})\big)\Big)
\end{eqnarray}
where
\be
W_0=\pm\sqrt{\frac {8(d-1)}{(d-1)b^2+2 d}}
\ee
and  $c_0$ is an integration constant and the first term dominates. This is expected because the UV solution is an attractor determined by the first universal term.

For $b>\sqrt{\frac {2d}{d-1}}$,
\begin{eqnarray}
W=\ell^{-1}\Big(W_0 e^{\frac b 2  \phi}\big(1+O(\phi^{-1})\big)+c_0 e^{\sqrt{\frac {d}{2(d-1)}}\phi}\big(1+O(\phi^{-1})\big)\Big)
\end{eqnarray}
where
\be
W_0=\pm \sqrt{\frac {8(d-1)}{(d-1)b^2+2 d}}
\ee
 and $c_0$ in the second term is an integration constant. In the UV, the second term dominates and thus the UV solution is not an attractor. We will not discuss this case any more.

In the critical case $b=\sqrt{\frac {2d}{d-1}}$, one should turn on $a\neq 0$ and study the subleading behavior carefully.

We continue the discussion of the case $b<\sqrt{\frac {2d}{d-1}}$.

In the UV, the function in the renormalized superpotential is:
\be
\A(\phi) = -\frac 1 {b(d-1)} \phi+O(\phi^0)
\ee
where the UV subleading terms are neglected.

The renormalized generating functional  (\ref{renormalized-action-2}) becomes:
\begin{eqnarray}
S^{(ren)}&=&\lim_{u^{UV}\rightarrow -\infty}M^{d-1}\ell^{-1}\int d^d x\sqrt{-\gamma}\left[c_0^R e^{\frac d {b(d-1)}\phi}\right.\nonumber\\
&&\qquad\qquad\qquad\qquad\qquad\left.- \ell^2 e^{\frac {d-2} {b(d-1)}\phi}\left(c_1^RR +\tilde c_2^R\frac 1 2 \gamma^{\mu\nu}\partial_\mu \phi\partial_\nu \phi\right)\right],
\end{eqnarray}
where
\be
\tilde c_2^R=\frac {2(d-2)}{b^2(d-1)}c_2^R.
\ee

At a finite cut-off $u^{UV}=\ln \epsilon$, the metric source $\tilde\gamma^\epsilon$ and the scalar field source $\alpha_\epsilon$ are defined as:
\begin{eqnarray}
\gamma_{\mu\nu}^\epsilon&=&\left(-\frac{\tilde u}\ell\right)^{\frac 4 {b^2(d-1)}}\tilde\gamma^\epsilon_{\mu\nu},\qquad e^{-\frac b 2\phi}=e^{-\frac b 2\alpha_\epsilon}-\frac{\tilde u}\ell,
\end{eqnarray}
where
\be
\tilde u=\frac {b^2} 4 W_0 u^{UV}.
\ee
The UV sources are defined as:
\be
\tilde\gamma(x)=\lim_{u\rightarrow -\infty}\tilde\gamma^\epsilon(x,u^{UV}),\qquad\alpha(x)=\lim_{u\rightarrow -\infty}\alpha^\epsilon(x,u^{UV}).
\ee

Using the definition of the scalar source,
\begin{eqnarray}
e^{-\frac 1 {b(d-1)}\phi}=(e^{-\frac b 2\alpha_{\epsilon}}-\frac{\tilde u}\ell)^{\frac 2 {b^2(d-1)}}.
\end{eqnarray}

Substituting the definition of sources, the renormalized on-shell action becomes:
\begin{eqnarray}
S^{(ren)}&=&\lim_{u^{UV}\rightarrow -\infty}M^{d-1}\ell^{-1}\int d^d x\sqrt{-\tilde\gamma}
\left[\left(\frac {-\tilde u}{e^{-\frac b 2\alpha^\epsilon}-\tilde u}\right)^{\frac {2d} {b^2(d-1)}}c_0^R\right.\nonumber\\
&& \left.\qquad\qquad\qquad\qquad- \ell^2\left(\frac {-\tilde u}{e^{-\frac b 2\alpha^\epsilon}-\tilde u}\right)^{\frac {2(d-2)} {b^2(d-1)}}
\left(c_1^R R^{(\tilde\gamma)}+\tilde c_2^R\frac 1 2 \tilde\gamma^{\mu\nu}\partial_\mu \phi\partial_\nu \phi\right)\right]\nonumber\\
&=&M^{d-1}\ell^{-1}\int d^d x\sqrt{-\tilde\gamma}\left(c_0^R - \ell^2c_1^R R^{(\tilde\gamma)}\right).
\end{eqnarray}
In the UV limit, $\left(\frac {-\tilde u}{e^{-\frac b 2\alpha^\epsilon}-\tilde u}\right)$ asymptotes to 1 and the kinetic term vanishes because:
\be
\partial_\mu \phi=(1-e^{\frac b 2\alpha^\epsilon}\tilde u)^{-1}\partial_\mu \alpha^\epsilon\rightarrow 0.
\ee

The renormalized generating functional  (\ref{running}) as a functional of running couplings can be obtained from  expression (\ref{running}). Since $W\to 0$ in the UV, $D_2=0$, and one must be  careful in solving equations (\ref{D1}-\ref{D3}) for the coefficients: in fact, only the last term in (\ref{D3})  contributes, since it is divergent, while the overall factor $(\de \phi)^2$ can be shown to asymptote to zero. Therefore, the only constraint from that equation is $D_2=0$, and we find:
\begin{equation} \label{Sexp}
S^{(ren)}=M^{d-1}\ell^{-1}\int d^d x\sqrt{-\gamma}\left[D_0 e^{\frac d {b(d-1)}\phi}+\ell^2e^{\frac {d-2} {b(d-1)}\phi}\left(D_1 R+\tilde D_2 \frac 1 2 \gamma^{\mu\nu}\partial_\mu \phi\partial_\nu \phi\right)\right],
\end{equation}
where
\be
\tilde D_2= D_1\frac {2(d-2)}{b^2(d-1)}.
\ee
 Notice that the action is independent of the renormalization coefficient $c_2^R$, but contains only two independent coefficients.

To zero-derivative order, the vev of dual operator of the running coupling is:
\begin{eqnarray}
\<\tilde\cO\>=\frac 1{\sqrt{-\gamma}}\frac {\delta S^{(ren)}}{\delta \phi}=M^{d-1}\ell^{-1}D_0 \frac d {b(d-1)} e^{\frac d {b(d-1)}\phi}.
\end{eqnarray}

The Legendre transform of the renormalized generating functional  (\ref{1PI}) as a function of the vev of the dual operator of the running coupling is:
\begin{eqnarray}
\Gamma&=&\int d^d x \sqrt{-\gamma} \<\tilde \cO\> \phi -S^{(ren)}\nonumber\\
&=&M^{d-1}\ell^{-1}\int d^d x \sqrt{-\gamma}\left[\tilde E_0\frac{\<\tilde\cO\>}{M^{d-1}\ell^{-1}} \ln\frac{\<\tilde\cO\>}{M^{d-1}\ell^{-1}} -\ell^2\tilde E_1\Big(\frac{\<\tilde\cO\>}{M^{d-1}\ell^{-1}} \Big)^{\frac {d-2} d}R\right.\nonumber\\
&&\left. -\ell^2\tilde E_2\Big(\frac{\<\tilde\cO\>}{M^{d-1}\ell^{-1}} \Big)^{\frac {d-2} d} \frac 1 2 \gamma^{\mu\nu}\partial_\mu \ln \frac{\<\tilde\cO\>}{M^{d-1}\ell^{-1}} \partial_\nu \ln \frac{\<\tilde\cO\>}{M^{d-1}\ell^{-1}}  \right]\
\end{eqnarray}
where
\begin{eqnarray}
\tilde E_0&=&D_0\frac {b(d-1)}d ,\\
\tilde E_1&=&D_1\left(\frac {b(d-1)}{D_0 d}\right)^{\frac {d-2}d},\\
\tilde E_2&=&\tilde D_2\left(\frac {b(d-1)}{D_0 d}\right)^{\frac {d-2}d}\left(\frac {b(d-1)}{d}\right)^2.
\end{eqnarray}

The canonically normalised operator is:
\be
\<\cO\>=\frac {2d}{d-2}(M^{d-1}\ell\tilde E_2)^{\frac 1 2}\Big(\frac{ \<\tilde \cO\>}{M^{d-1}\ell^{-1}}\Big)^{\frac {d-2}{2d}}.
\ee

In terms of the canonically normalised operator, the quantum effective action reads:
\be
\Gamma=\int d^d x \sqrt{-\gamma}\left[E_0\<\mathcal O\> ^{\frac {2d}{d-2} }\ln\frac{\<\mathcal O\>}{M^{\frac{d-1}2}\ell^{\frac 1 2}}+E_1\<\mathcal O\> ^{2 }R - \frac 1 2 \gamma^{\mu\nu}\partial_\mu\< \mathcal O\>\partial_\nu \<\mathcal O\> \right]
\ee
where
\begin{eqnarray}
E_0&=&(M\ell)^{-\frac {2(d-1)}{d-2}}{c_0^R}^2\Big(\frac {d-2}{2d}\Big)^{\frac {d+2}{d-2}}\left[-\frac {2(d-1)(d-2)}{d^2}c_1^R\right]^{-\frac d {d-2}},\\
E_1&=&-\frac{d-2}{8(d-1)}.
\end{eqnarray}
After canonical normalisation, the coefficient of the Ricci term is independent of the integration constant.

\subsubsection{IR fixed point and regularity}
To study the IR limit, we consider $\phi\rightarrow \infty$. Assume $a=0$ for the moment.

If $b>\sqrt{\frac {2d}{d-1}}$, the superpotential is
\begin{eqnarray}
W=W_0 e^{\frac b 2  \phi}(1+O(\phi^{-1}))+...+C_0 e^{\frac d {b(d-1)}\phi}(1+O(\phi^{-1}))
\end{eqnarray}
where
\be
W_0=\pm \ell^{-1}\sqrt{\frac {8(d-1)}{(d-1)b^2+2 d}},
\ee
and $C_0$ is an integration constant and the first term dominates.

For $b=\sqrt{\frac {2d}{d-1}}$, one should turn on $a\neq 0$ because this subleading term is important.

 For $b<\sqrt{\frac {2d}{d-1}}$,
\begin{eqnarray}
W=W_0 e^{\frac b 2  \phi}(1+O(\phi^{-1}))+C_0 e^{\sqrt{\frac {d}{2(d-1)}}\phi}(1+O(\phi^{-1}))
\end{eqnarray}
where
\be
W_0=\pm \ell^{-1}\sqrt{\frac {8(d-1)}{(d-1)b^2+2 d}},
\ee
and $C_0$ is an integration constant and the second term dominates if $C_0\neq 0$.
 We will focus on this case in the following discussion.

The IR regular solution of superpotential $W_\ast$ corresponds to zero integration constant $C_0=0$.
The IR regular solution of second superpotential $U_\ast$ is:
\begin{eqnarray}
U_\ast&\simeq&\frac 1 {1+\frac {d-2}{b^2(d-1)}}(u-u_{IR})
\end{eqnarray}
For non-regular solution, the leading behavior of $U$ becomes :
\be
U\simeq  (u_{IR}-u)^{-\frac {d-2} {b^2(d-1)}}.
\ee.

In the IR limit,
\begin{eqnarray}
 \gamma_{\mu\nu}\simeq\tilde \gamma_{\mu\nu}(u_{IR}-u)^{\frac 4 {b^2(d-1)}},\quad
e^{ \phi}\simeq(u_{IR}-u)^{-\frac 2 b},\quad e^A\simeq (u_{IR}-u)^{\frac 2 {b^2(d-1)}}
\end{eqnarray}

The potential term gives zero in the IR:
\begin{eqnarray}
\sqrt{-\gamma}W_\ast\simeq\sqrt{-\tilde\gamma}(u_{IR}-u)^{\frac {2d} {b^2(d-1)}-1}\rightarrow 0
\end{eqnarray}

The two derivative term vanish in the IR also:
\begin{eqnarray}
\sqrt{-\gamma}U_\ast R^{(d)}&\simeq& \sqrt{-\tilde\gamma} (u_{IR}-u)^{\frac {d-2} {b^2(d-1)}+1}R^{(\tilde\gamma)}\rightarrow 0\nonumber\\
\sqrt{-\gamma}\left(\frac {W_\ast}{W_\ast'}  {U_{\ast}}'\right)\frac 1 2\gamma^{\mu\nu}\partial_\mu\phi\partial_\nu\phi&\simeq&\sqrt{-\tilde\gamma}(u_{IR}-u)^{\frac {d-2} {b^2(d-1)}+1}\partial_\mu\phi\partial_\nu\phi\rightarrow 0
\end{eqnarray}
where we have assumed $R^{(\tilde\gamma)}$ and $\partial_\mu\phi\partial_\nu\phi$ stay finite.

The IR contribution of the IR regular solutions to the bare on-shell action(\ref{onshellbare}) is just zero.

For the non-regular solutions, the IR contribution to the bare on-shell action is finite. For positive solution of superpotential $W$, the integration constant is positive $C_0>0$. The IR asymptotics is altered,
\begin{eqnarray}
 \gamma_{\mu\nu}\simeq\tilde \gamma_{\mu\nu}(u_{IR}-u)^{\frac 2 {d}},\quad
e^{ \phi}\simeq(u_{IR}-u)^{-\sqrt{\frac {2(d-1)}d}},\quad e^A\simeq (u_{IR}-u)^{\frac 1 d}
\end{eqnarray}

The potential term is finite in the IR:
\begin{eqnarray}
\sqrt{-\gamma}W_\ast\simeq \sqrt{-\tilde\gamma}C_0(u_{IR}-u)^{1}(u_{IR}-u)^{-1}\simeq\sqrt{-\tilde\gamma}C_0.
\end{eqnarray}

\subsection{IHQCD} \label{ihqcd}
\subsubsection{In the UV}
Assume the bulk potential of IHQCD in the UV region takes the form:
\be
V=\ell^{-2}V_0\left(1+\frac {b_0}{d-1}\lambda+O(\lambda^2)\right)
\ee
where $\lambda=e^\phi$ is the 't Hooft coupling.  In the UV limit, $\phi\rightarrow -\infty$ and $\lambda\rightarrow 0$. We will keep $d$ generic, but throughout this section the relevant case will be $d=4$.

The superpotential solution is:
\be
W=\ell^{-1}W_0\left(1+\frac {b_0}{2(d-1)} \lambda+ O(\lambda^2)\right)
\ee
where
\be
W_0=\pm\sqrt {\frac{4(d-1)}d V_0}.
\ee

$b_0$ is associated with the leading term of the $\beta$-function of $\lambda$
\begin{eqnarray}
\beta_\lambda=\frac{d\lambda}{d\log E}=\frac{d\lambda}{d A}\lambda=-b_0\lambda^2+\mathcal O(\lambda^3). \label{beta-0}
\end{eqnarray}
Here we will keep only the first term, and systematically neglect higher order contributions in $\lambda$, which on the field theory side corresponds to stopping  at one loop order.

The function $\cal A(\phi$) in the renormalized generating functional action in terms of $\lambda$ is
\be
\A(\lambda)=-\frac 1 {2(d-1)}\int^\phi d \tilde \phi \frac W {W'} =\frac 1 {b_0  \lambda}+O(\lambda^0)
\ee

In the UV, the renormalized action (\ref{running}) is:
\begin{eqnarray}
S^{(ren)}&=&M^{d-1}\ell^{-1}\int d^dx\,\sqrt{-\gamma}\left(D_0 e^{-d\A}+D_1 \ell^2e^{-(d-2)\A} R\right.\nonumber\\
&&\left.\qquad\qquad\qquad\qquad+\tilde D_2 \ell^2e^{-(d-2)\A} e^{-2\phi} \frac 1 2 \gamma^{\mu\nu}\partial_\mu\phi\partial_\nu\phi\right)
\end{eqnarray}
where
\be
\tilde D_2=\left(D_2+\frac {d-2}{2(d-1)}W_0^2 D_1\right)\left(W_0\frac {b_0}{2(d-1)} \right)^{-2}.
\ee
 We have neglected the UV subleading terms.

The renormalized action in terms of 't Hooft couping is:
 \begin{eqnarray}
S^{(ren)}&=&M^{d-1}\ell^{-1}\int d^dx\,\sqrt{-\gamma}\left(D_0 e^{ -\frac d{b_0\lambda}}
+ D_1\ell^2 e^{-\frac {d-2}{b_0\lambda}}R \right.\nonumber\\
&&\left.\qquad\qquad\qquad\qquad+\tilde D_2 \ell^2e^{-\frac {d-2}{b_0\lambda}} \lambda^{-4}\frac 1 2 \gamma^{\mu\nu}\partial_\mu\lambda\partial_\nu\lambda\right)\label{ihqcdaction}
\end{eqnarray}

In the dual quantum field theory, the 't Hooft coupling is coupled to the operator as $\frac 1 \lambda Tr F^2$. The vev of $Tr F^2$ can be calculated by varying the renormalized on-shell action with respect to $\lambda ^{-1}$. At the zeroth order,
\begin{eqnarray}
\<\tilde O\>&=&\<Tr F^2\>=\frac 1{\sqrt{-\gamma}}\frac {\delta S^{(ren)}}{\delta \lambda^{-1}}=M^{d-1}\ell^{-1}\left(  -\frac {dD_0}{b_0}\right) e^{ -\frac d{b_0\lambda}} \label{vev}.
\end{eqnarray}

As explained at the end of section 5.3, the operator defined above is scale-dependent, as can be seen from the exponential term containing the running coupling $\lambda(\mu)$. The corresponding RG-invariant operator in this case is, by equation (\ref{trace12}),
\be\label{vev-inv}
\tilde{O}^{inv} = \sqrt{\gamma} {\beta(\lambda) \over \lambda^2} \tilde{O},
\ee
where we have assumed a Minkowski reference metric $\tilde{\gamma}_{\mu\nu}=\eta_{\mu\nu}$. As $\sqrt{\gamma} \sim e^{dA}\sim \mu^d$, from equations (\ref{vev}) and (\ref{vev-inv}) we find:
\be\label{vev2}
\<\tilde{O}^{inv}\> = M_0^d e^{dA} e^{ -\frac d{b_0\lambda(A)}}
\ee
where we have defined the overall  mass scale $M_0$ to collect all constant coefficients in (\ref{vev}).

The right hand side of equation (\ref{vev2})  is in fact independent of $A$ (i.e. scale-independent) at one loop order, as one can see    by solving  (\ref{beta-0}) dropping $O(\lambda^3)$ terms:
\be\label{vev3}
A - {1\over b_0 \lambda(A)}= A_0 -  {1\over b_0 \lambda(A_0)} \equiv \log \Lambda \ell.
\ee
The last equality is our definition  of  $\Lambda$,  the (one-loop) RG-invariant scale of the theory. Up to the choice of an overall energy
normalization, it  is the same as in Yang-Mills theory \cite{gkn}. As in Yang-Mills  theory, where  a choice of $\Lambda$ uniquely defines an RG-trajectory, here a choice of $\Lambda$ picks one particular initial condition $\lambda(A_0)$ for the bulk solution.  Using (\ref{vev3}) in (\ref{vev2}),  we find:
\be
\< \tilde{O}^{inv}\> = (M_0 \ell)^d \Lambda^d.
\ee
As one expects from field theory, the gluon condensate is proportional to the invariant scale $\Lambda$, up to a (scheme-dependent) overall coefficient.

The quantum effective action is defined as the Legendre transform of (\ref{ihqcdaction}):
\be
\Gamma=\int \sqrt {-\gamma}\<\tilde O\>\lambda^{-1}-S^{(ren)}.
\ee
One can see that the Legendre transform is dominated by the first term due to the $\lambda^{-1}$ divergence.

Up to second derivative order, using (\ref{1PI}) we find:
\begin{eqnarray}
\Gamma &=&  {M^{d-1}\ell^{-1}}\int \sqrt {-\gamma} \left[-\frac {b_0} d \frac {\tilde O}{M^{d-1}l^{-1}} \ln \Big(\frac {\tilde O}{M^{d-1}l^{-1}}\Big)\right.\nonumber\\
&&\left.-\ell^2\Big(-\frac {b_0}{d D_0}\frac {\tilde O}{M^{d-1}l^{-1}}\Big)^{\frac {d-2} d} \Big(D_1 R^{(d)}+\tilde D_2\frac {b_0^2}{d^2} \tilde O^{-2}\frac 1 2\gamma^{\mu\nu}\partial_\mu\tilde O\partial_\nu\tilde O\Big)\right].
\end{eqnarray}

We can define a  canonically normalized operator $O$ as:
\be
O=\left[(-D_0)^{-\frac {d-2}{d}}{\tilde D_2}^{\frac 1 2}\Big(\frac {b_0}d\Big)^{\frac {3d-2} {2d}}\frac{2d} {d-2}(M\ell)^{1-\frac 1 d}\right] \tilde O^{\frac {d-2}{2d}}.
\ee
From dimensional analysis, $\tilde O\sim \mu^d$ and $O\sim \mu^{\frac {d-2}2}$.

In four dimension, the canonically normalised operator becomes:
\be
O=\left[2^{-\frac 1 2}  (-D_0)^{-\frac 1 2}\tilde D_2^{\frac 1 2}b_0^{\frac 5 4}(M\ell)^{\frac 3 4}\right]\tilde O^{\frac 1 4},
\ee
and the quantum effective action in terms of $O$ reads:
\be
\Gamma=\int \sqrt {-\gamma} \left[ E_0{O^4} \ln \Big(\frac {O}{M^{\frac 3 2}l^{\frac 1 2}}\Big)+E_1 O^2 R - \frac 1 2\gamma^{\mu\nu}\partial_\mu O\partial_\nu O \right],
\ee
where
\be
E_0=-\left(\frac {c_0^R}{(d-1)(d-2)}\right)^2(M\ell)^{-3},\qquad E_1=-(-c_0^R)^{\frac 1 2}c_1^R\Big(2(d-1)(d-2)\Big)^{-1}.
\ee
where we have written the coefficients in terms of the renormalized integration constants $c^R_i$ governing the subleading behavior of the superpotentials
This expression for the one-loop effective potential  reproduces the field theory expectation based on the conformal anomaly (see e.g. \cite{preskill})

\subsubsection{IR effective potential}
Assume the bulk potential:
\be
V=\ell^{-2}V_\infty(\log\lambda)^{P } \lambda^{2Q}
\ee
 for large $\lambda=e^\phi\rightarrow \infty$.

The solution of superpotential in the IR takes the following form:
\begin{eqnarray}
W(\lambda)\simeq W_0 \ell^{-1}(\log\lambda)^{\frac P 2} \lambda^Q
\end{eqnarray}
where
\be
W_0=V_\infty^{\frac 1 2}\left(\frac d {4(d-1)}-\frac 1 2 Q^2\right)^{-\frac 1 2}
\ee
and $Q=\frac 1 {\sqrt 6}$ is the critical value at $d=4$ dimension.

The function $\A=\A(\lambda)$ in the renormalized generating functional is:
\be
\A(\lambda)=-\frac 1 {6}\int^\phi d \tilde \phi \frac W {W'} =-\frac 1 {6}\left(\frac 1 Q \log \lambda-\frac P{Q^2}\log\log\lambda\right)
\ee

so
\be
e^\A=(\log \lambda)^{P}\lambda^{-Q}
\ee

The renormalized generating functional (\ref{running}) is:
\begin{eqnarray}
S^{(ren)}&=&M^{d-1}\ell^{-1}\int d^dx\,\sqrt{-\gamma}\left(D_0 e^{-4\A}+\ell^{2}\tilde f_1R
\right.\nonumber\\
&&\qquad\qquad\qquad\qquad\qquad\left.+D_3\ell^{2} e^{-4\A} W^{-2} \frac 1 2 \gamma^{\mu\nu}\partial_\mu\phi\partial_\nu\phi\right)\nonumber\\
&=&M^{d-1}\ell^{-1}\int d^dx\,\sqrt{-\gamma}\left(D_0(\log \lambda)^{-4P}\lambda^{4Q} + \ell^{2}\tilde f_1R \right.\nonumber\\
&&\left.\qquad\qquad\qquad\qquad\qquad+D_3\ell^{2} (\log \lambda)^{-5P}\lambda^{2Q}\frac 1 2 \gamma^{\mu\nu}\partial_\mu\phi\partial_\nu\phi	\right)
\end{eqnarray}
where $\tilde f_1(\phi)\sim e^{-2\A}$ for $P>\frac 1 3$, $\tilde f_1(\phi)\sim e^{-4\A}W'^{-2}\phi$ for $P<\frac 1 3$ and $\tilde f_1(\phi)\sim e^{-2\A}\log\phi$ for $P=\frac 1 3$. $D_3$ is a complicated factor and can be computed  from (\ref{running}).

At the zero-derivative order, the vev of dual operator of the running coupling is:
\begin{eqnarray}
\<\tilde{ \mathcal O}\>=\<Tr F^2\>=\frac 1{\sqrt{-\gamma}}\frac {\delta S^{(ren)}}{\delta \lambda^{-1}}=-M^{d-1}\ell^{-1}4QD_0(\log \lambda)^{-4P}\lambda^{4Q+1}
\end{eqnarray}

The Legendre transform of the renormalized generating functional  (\ref{1PI}) as a function of the vev of the dual operator of the running coupling is:
\begin{eqnarray}
\Gamma&=&M^{d-1}\ell^{-1}\int d^dx\,\sqrt{-\gamma}\left(\tilde E_0 \Big(\frac{\tilde {\mathcal O}}{M^{d-1}\ell^{-1}}\Big)^{\frac {4Q}{4Q+1}}\Big(\log\frac{\tilde {\mathcal O}}{M^{d-1}\ell^{-1}}\Big)^{-\frac{4P}{4Q+1}}\right.\nonumber\\
&&\left.-\ell^2\tilde E_2\Big(\log \frac{\tilde {\mathcal O}}{M^{d-1}\ell^{-1}}\Big)^{-\frac{12Q+5}{4Q+1}P}\frac 1 2 \gamma^{\mu\nu}\partial_\mu\Big(\frac{\tilde {\mathcal O}}{M^{d-1}\ell^{-1}}\Big)^{\frac Q{4Q+1}}\partial_\nu\Big(\frac{\tilde {\mathcal O}}{M^{d-1}\ell^{-1}}\Big)^{\frac Q{4Q+1}} \right)\nonumber\\
\end{eqnarray}
where
\begin{eqnarray}
\tilde E_0&=&(-4QD_0)^{\frac 1 {4Q+1}}(4Q+1)^{\frac {4P}{4Q+1}},\\
\tilde E_2&=&\frac{D_3 }{Q^{2}}(-4QD_0)^{-\frac {2Q} {4Q+1}}\Big(4Q+1\Big)^{-\frac{12Q+5}{4Q+1}P} .
\end{eqnarray}

The canonically normalised operator is:
\be
\mathcal O=(M^{d-1}\ell \tilde E_2)^{\frac 1 2}\left(\frac{\tilde {\mathcal O}}{M^{d-1}\ell^{-1}}\right)^{\frac Q{4Q+1}}\left(\log \frac{\tilde {\mathcal O}}{M^{d-1}\ell^{-1}}\right)^{-\frac {12Q+5}{2(4Q+1)}P}.
\ee

In terms of the canonically normalised operator, the Legendre transform of the renormalized generating functional reads:
\be
\Gamma=\int d^dx\,\sqrt{-\gamma}\left(E_0\frac{\mathcal O^4}{M^{d-1}\ell^3}\Big(\log \frac{\mathcal O}{M^{\frac {d-1}2}\ell^{\frac 1 2}} \Big)^{6P}- \frac 1 2 \gamma^{\mu\nu}\partial_\mu\mathcal O\partial_\nu\mathcal O\right),
\ee
where
\be
E_0=-4D_0D_3^{-2}Q^{5-6P}(4Q+1)^{12P}.
\ee

\subsection{Numerical example of complete flow: AdS-AdS flow}
\begin{figure}[h!]
\begin{center}
\includegraphics[width=10cm]{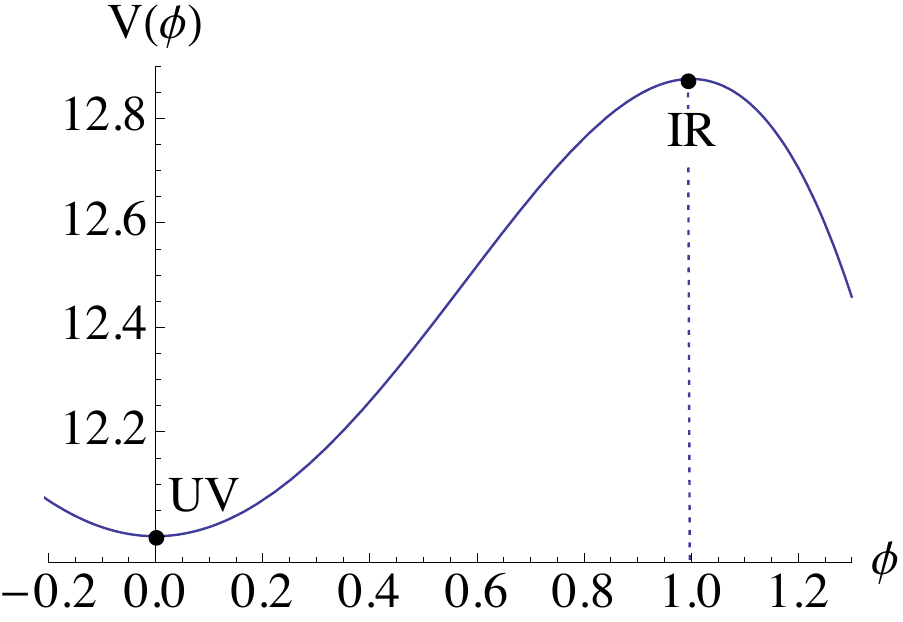}
\caption{The bulk scalar potential $V(\phi)$ for AdS-AdS flow. There are two stationary points at $\phi=0$ and $\phi= 1$ where $\phi=0$ corresponds to the UV AdS fixed point and $\phi=1$ the IR AdS fixed point.} \label{figure1}
\end{center}
\end{figure}

We consider as an example the bulk scalar potential (see Figure \ref{figure1}):
\begin{eqnarray}
V(\phi)=\ell^{-2}\left(d(d-1)-\frac {m^2}2\phi^2+\frac k 4 \phi^4\right)
\end{eqnarray}
where we set $\ell=1$, $d=4$ and $m^2=k=-\frac 7 2$.  There are two kinds of quantizations corresponding to this mass. The scaling dimension is
\be
\Delta_-=\frac d 2-\sqrt{\frac{d^2}4+m^2}=2-2^{-\frac 1 2}\sim 0.40.
\ee

\begin{figure}[h!]
\begin{center}
\includegraphics[width=12cm]{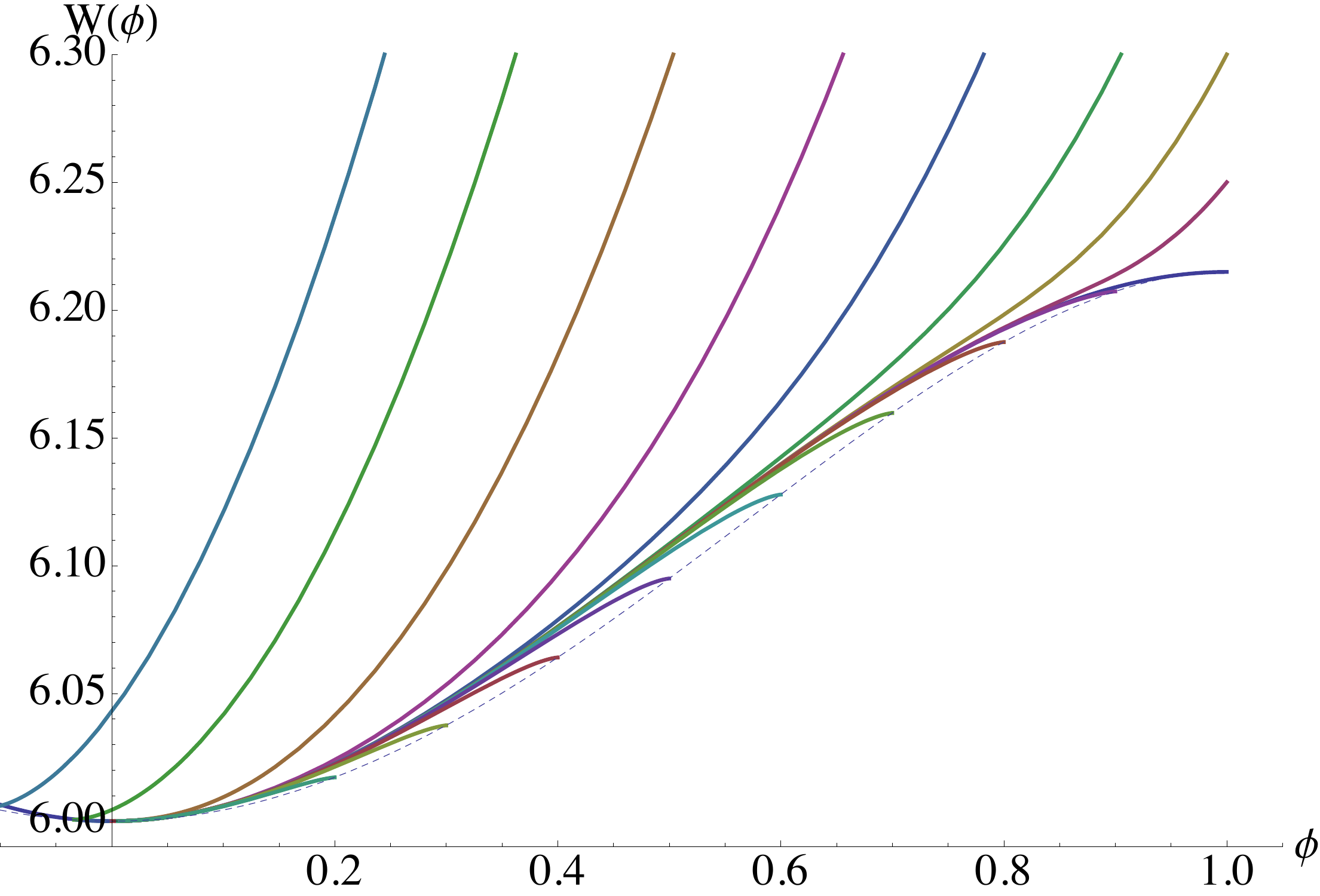}
\caption{AdS-AdS flow: The solutions of the superpotential equation  $V=\frac d {4(d-1)}W^2-\frac 1 2{W^\prime}^2$. There are three kinds of solutions: the regular flow $W^\ast$, the overshoot flows and the truncated flows. The lowest dashed curve is $W=\sqrt{\frac {2(d-1)}d V}$, instead of a solution. 1) The regular one $W^\ast$ flows  from UV AdS  fixed point to IR AdS fixed point . 2) The solutions above the regular one overshoot the IR fixed point. Their IR behavior is determined by the bulk potential with $\phi>1$. As one can see, the highest curve does not flow to the UV fixed point. The UV attractor is only for some range of the solutions. 3) The solutions below the regular one start at the UV fixed point and stop when they intersect with the lowest curve. Below the curve $W=\sqrt{\frac {2(d-1)}d V}$, the solutions become imaginary number, so they stop. To flow to the IR fixed point, one should continue the solution with the other branch of solutions.
} \label{figure2}
\end{center}

\end{figure}
\begin{figure}[h!]
\begin{center}
\includegraphics[width=12cm]{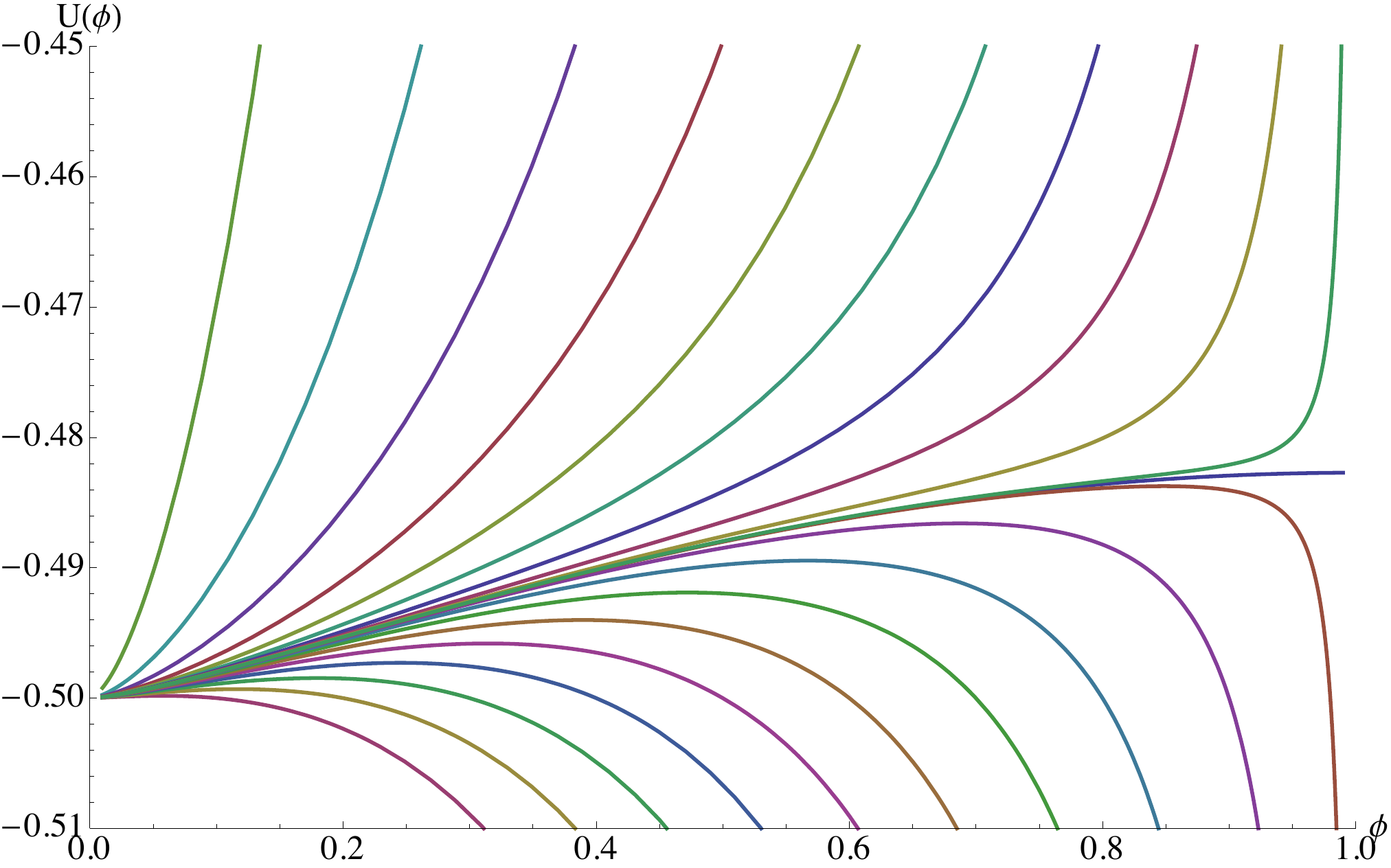}
\caption{AdS-AdS flow: The solutions of the second superpotential $1=U'{W_\ast}'-\frac {d-2} {2(d-1)}U W_\ast$, which are the coefficients of the Ricci term of bare on-shell action. $W_\ast$ is the regular solution of the first superpotential. There is only one IR regular solution. The other solutions diverge in the IR. The UV solution is an attractor and all the solutions flow to UV fixed point.} \label{figure3}
\end{center}
\end{figure}

\begin{figure}[h!]
\begin{center}
\includegraphics[width=12cm]{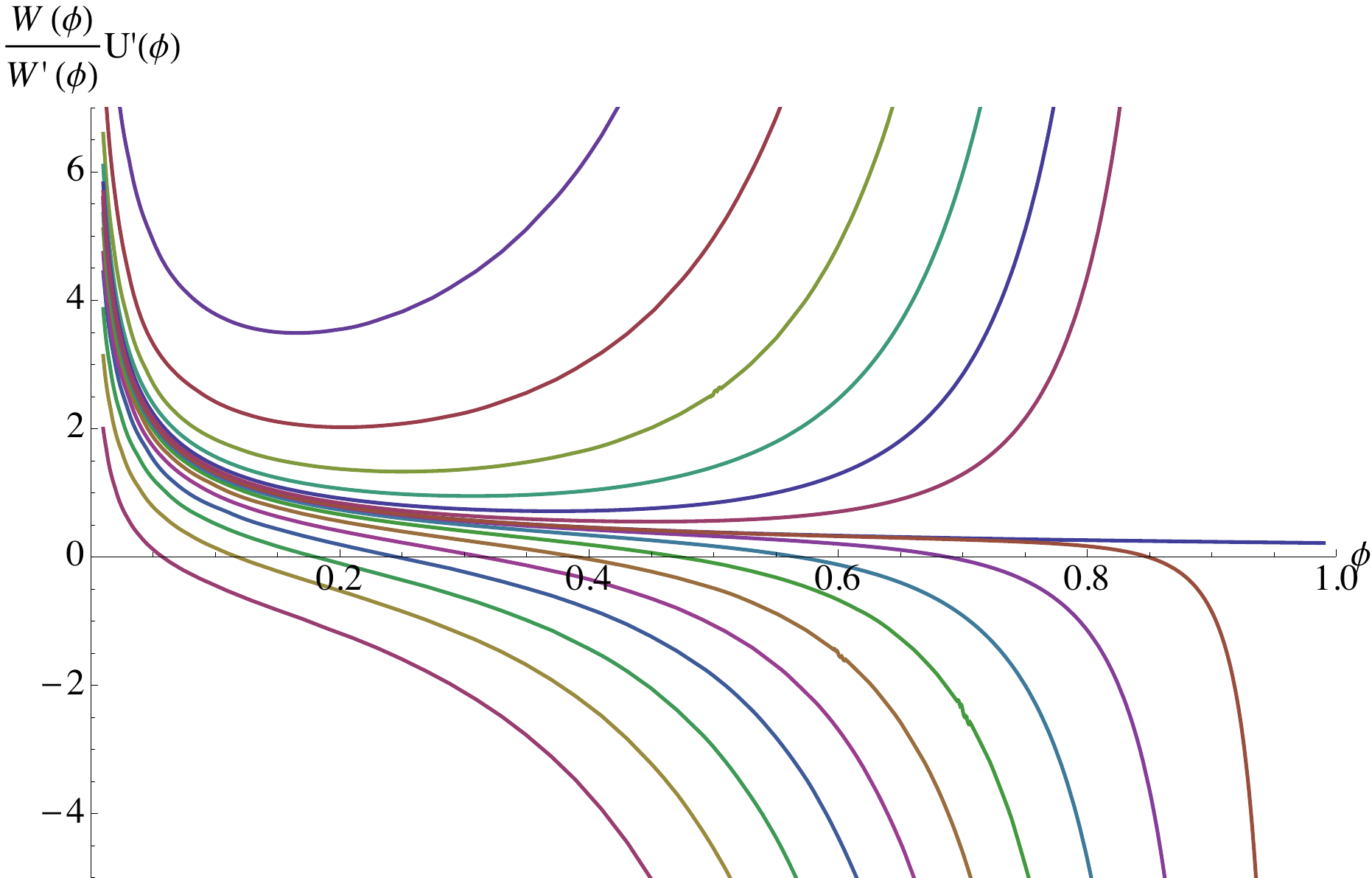}
\caption{AdS-AdS flow: The coefficient of the kinetic term: $\frac W {W'}U'$. There is only one IR regular curve $\frac W {W'}U'_\ast$ that stay finite in the IR. $U_\ast$ is the same regular solution of second superpotential as the previous figure. All the curves diverge in the UV. } \label{figure4}
\end{center}
\end{figure}

\begin{figure}[h!]
\begin{center}
\includegraphics[width=12cm]{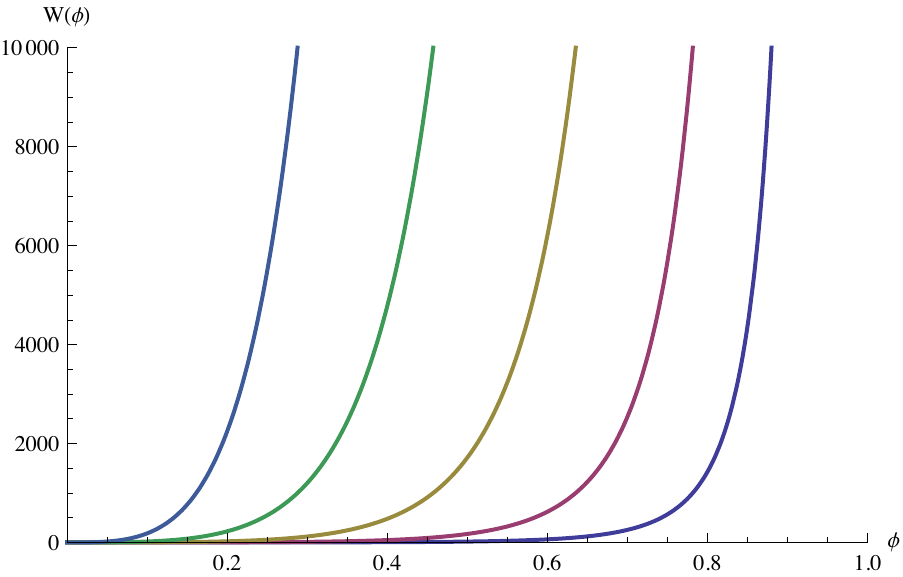}
\caption{AdS-AdS flow: The effective potential of the running coupling at different energy scales. The higher the energy scale, the steeper the potential is.}\label{figure5}
\end{center}
\end{figure}

\begin{figure}[h!]
\begin{center}
\includegraphics[width=12cm]{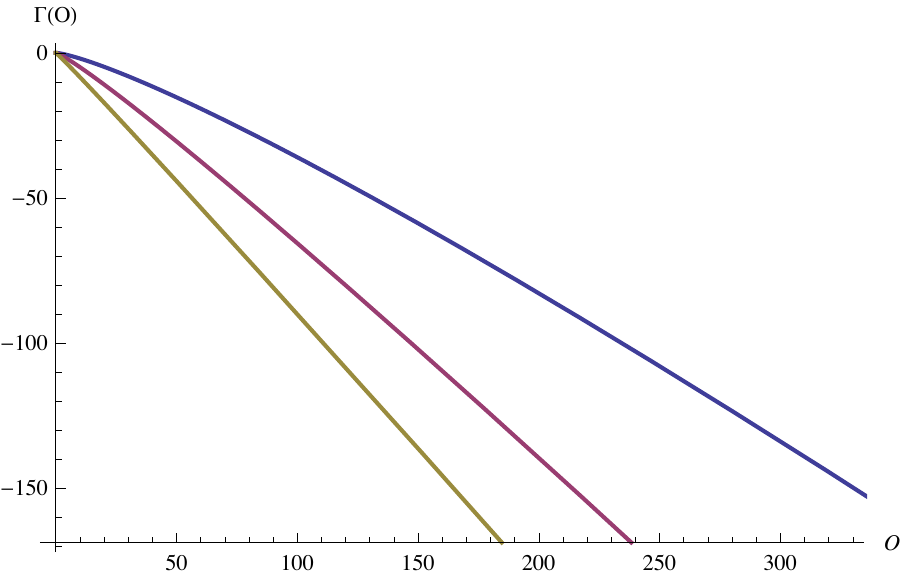}
\caption{AdS-AdS flow: The 1PI effective potential of the dual operator of the running coupling at different energy scales. The potential is lowered as one increases the energy scale. The UV fixed point ($\<\mathcal O\>=0$) is an unstable extrema and will flow to the IR fixed point ($\<\mathcal O\>\rightarrow \infty$) by deformation.}\label{figure6}
\end{center}
\end{figure}

In Figure \ref{figure2}, we draw the numerical solutions of the superpotential equation (\ref{flow-eq-3}). There are two branches of solutions: $W'>0$ and $W'<0$. We focus on the first branch $W'>0$ in Figure \ref {figure2}, where $\phi=0$ is an UV attractor for some solutions. At the AdS fixed points $W'=0$, so the numerical value of the IR regular solution at the fixed points are
\be
W_\ast(0)=\sqrt{\frac {4(d-1)}{d} V(0)}=6 ,\qquad W_\ast(1)=\sqrt{\frac {4(d-1)}{d} V(1)}\simeq 6.21 .
\ee

One can obtain the numerical value of $C$ by subtracting the leading integer power terms in the superpotential
\be
W^{(0)}=2(d-1)+\frac {\Delta_-}2\phi^2+O(\phi^4)
\ee
and drawing the Log-Log plot of
\be
W-W^{(0)}=C\phi^{\frac d{\Delta_-}}(1+O(\phi^2))+O(\phi^4)
\ee
 to read out the constant $C$ and the exponent ${\frac d{\Delta_-}}$. Here the integration constant $C$ is positively correlated to the boundary value $W(\phi=\frac 1 {100})$. For the IR regular superpotential $W_\ast$, $C_\ast$ is about -0.542.

In Figure \ref{figure3}, we have used the regular solution of superpotential to calculate the solutions of second superpotential $U(\phi)$. The blue curve is the regular solution of $U_\ast(\phi)$. At the fixed points,
\be
U_\ast=-\frac  {2(d-1)} {(d-2)W_\ast},
\ee
 so the numerical value of the IR regular solution at the fixed points are
 \be
 U_\ast(0)=-0.5,\qquad U_\ast(1)\sim -0.483.
\ee
In Figure \ref{figure4}, we calculate the coefficient of the kinetic term for different solutions of $U$.\\

In Figure \ref{figure5}, we calculate the renormalized effective potential $W_{ren}(\phi,\mu)$ as a function of the running coupling $\phi$ and energy scales $\mu$. We vary the potential with respect to the running coupling to derive the vev $\<\cO\>$ as a function of the running coupling and energy scales. Then we calculate the Legendre transform of $W_{ren}$ and compute the effective potential for the vev $\<\cO\>$ in Figure \ref{figure6}.

\section{Acknowledgements}\label{ACKNOWL}

We would like to thank S. Kuperstein and Y. Nakayama for interesting discussions. We especially thank I. Papadimitriou for patiently explaining various aspects of his work and for constructive comments on the manuscript.

This work was supported in part by European Union's Seventh Framework Programme under grant agreements (FP7-REGPOT-2012-2013-1) no 316165,
PIF-GA-2011-300984, the EU program ``Thales'' MIS 375734, by the European Commission under the ERC Advanced Grant BSMOXFORD 228169 and was also co-financed by the European Union (European Social Fund, ESF) and Greek national funds through the Operational Program ``Education and Lifelong Learning'' of the National Strategic Reference Framework (NSRF) under ``Funding of proposals that have received a positive evaluation in the 3rd and 4th Call of ERC Grant Schemes''. We also thank the ESF network Holograv for partial support.

\section*{Note added}\label{na}

While this work was being written we became aware of \cite{verlinde} which has some overlap with this work, in particular the discussion of the RG flow of the effective action and the ensuing Ricci flow. We note that the metric $\beta$-function obtained in that work disagrees at second derivative order with the one presented here. This discrepancy  is related to the scheme dependence discussed in Appendix \ref{alpha}.
Reference \cite{verlinde} does describe also the holographic flow of entanglement entropy as mean curvature flow.

The metric $\beta$-function and the gradient flow property where recently discussed in \cite{naka2}.

\newpage
\appendix
\renewcommand{\theequation}{\thesection.\arabic{equation}}
\addcontentsline{toc}{section}{Appendix}
\section*{Appendices}

\section{Technical Details}

In this appendix,  we use the index $(d+1)$ to denote  bulk covariant quantities, related to the bulk metric $g_{ab}$ (e.g. $R^{(d+1)}, \nabla^{(d+1)}$...), while covariant quantities  with no index are those intrinsic to the constant-$u$ slices, and are defined with respect to  the induced metric $\gamma_{\mu\nu}$.

\subsection{Lie derivative} \label{app-Lie}
As the standard definition, the Lie derivative of a tensor for a torsion free connection is
\begin{eqnarray}
\pounds_n A_{\mu_1\mu_2...}^{\nu_1\nu_2...}=n^a\nabla_a^{(d+1)} A_{\mu_1\mu_2...}^{\nu_1\nu_2...}
-(\nabla_a^{(d+1)} n^{\nu_1})A_{\mu_1\mu_2...}^{a\nu_2...}-(\nabla_a^{(d+1)} n^{\nu_2})A_{\mu_1\mu_2...}^{\nu_1 a...}-...&&\nonumber\\
+(\nabla_{\mu_1}^{(d+1)} n^{a})A_{a\mu_2...}^{\nu_1\nu_2...}+(\nabla_{\mu_2}^{(d+1)} n^{a})A_{\mu_1 a...}^{\nu_1\nu_2...}+...,&&\
\end{eqnarray}
where the Lie derivative is taken along $n^a$ and the vector $n^a$ is the normal vector of the slices $\Sigma_u$.

In a coordinate system adapted to $n^a$, the components of the Lie derivative of $A_{\mu_1\mu_2...}^{\nu_1\nu_2...}$ are
\begin{eqnarray}
\pounds_n A_{\mu_1\mu_2...}^{\nu_1\nu_2...}=\frac {\partial}{\partial {X}}A_{\mu_1\mu_2...}^{\nu_1\nu_2...},
\end{eqnarray}
where $X$ is the parameter along the integral curves of $n^a$. Therefore, the Lie derivative $\pounds_n$ is simply the radial derivative $\partial_X$ in the Fefferman-Graham coordinate where $N=1$ and $N^\mu=0$.

\subsection{Derivation of the flow equations\label{a2}}

In the language of Lie derivative, the radial Hamiltonian and transverse momentum constraints become
\begin{eqnarray}
R-\frac 1 2 \gamma^{\mu\nu}\partial_\mu\phi\partial_\nu\phi+V&=&\frac 1 4 (\gamma^{\mu\nu}\gamma^{\rho\eta}-\gamma^{\mu\rho}\gamma^{\nu\eta})(\pounds_n\gamma_{\mu\nu})(\pounds_n\gamma_{\rho\eta})-\frac 1 2 (\pounds_n\phi)^2,\nonumber\\
\\
0 &=&\nabla_\rho(\gamma^{\rho\nu}\pounds_n\gamma_{\mu\nu}-\gamma_\mu^\rho\gamma^{\nu\eta}\pounds_n\gamma_{\nu\eta})-(\partial_\mu\phi)\pounds_n\phi ,\nonumber\\
\end{eqnarray}

and the dynamical equations are
\begin{eqnarray}
R_{\mu\nu}-\frac 1 2\partial_\mu\phi\partial_\nu\phi+\frac 1 {d-1} V\gamma_{\mu\nu}&=&\frac 1 2\pounds_n(\pounds_n\gamma_{\mu\nu})+\frac 1 4(\gamma^{ab}\gamma_\mu^c-2\gamma_\mu^a\gamma^{bc})(\pounds_n\gamma_{ab})(\pounds_n\gamma_{c\nu})\nonumber\\&&+\frac 1 N\nabla_\mu\partial_\nu N,
\end{eqnarray}
We will impose the gauge fixing $\partial_\nu N=0$ so that the last term vanishes and the lapse function is constant on the hypersurface $\Sigma_u$.

\subsubsection{Momentum constraint}
Pluging the flow equations in the momentum constraint,
\begin{eqnarray}
&&(\partial_\mu\phi)\pounds_n\phi\nonumber\\
&=&\partial_\mu\phi\left[-(d-1)g_1'-(g_2'+(d-1)g_3') R-(d-1)g_5'(\gamma^{\rho\eta}\partial_\rho\phi\partial_\eta\phi)\right.\nonumber\\
&&\left.-(-g_4+g_6'+(d-1)g_7')(\gamma^{\rho\eta}\nabla_\rho\partial_\eta\phi)+...\right]\nonumber\\
&&+g_2'{R_\mu}^\nu\partial_\nu\phi+g_2\partial_\nu R_\mu^\nu-(g_2+(d-1)g_3)\partial_\mu R\nonumber\\
&&-(g_4+2(d-1)g_5-g_6')(\partial^\nu\phi)(\nabla_\nu^{(d)}\partial_\mu\phi)\nonumber\\
&&+g_6\nabla_\nu\nabla_\mu\partial^\nu\phi-(g_6+(d-1) g_7)(\nabla_\mu\nabla_\nu\partial^\nu\phi)+O(\partial^5)\, ,
\end{eqnarray}
where the last three lines are unwanted terms because $\partial_\mu\phi$ can not be seperated to match $(\partial_\mu\phi)\pounds_n\phi$. Most of them will disappear after imposing constraint equations among the scalar functions. However, one still have two kinds of unwanted terms $ R^\nu_\mu\partial_\nu\phi$ and $\nabla^3\phi$. In order to cancel them, one should be careful about the order of $\nabla$ and use the definition of Riemann ternsor.

Using the definition of Riemann tensor, the first unwanted term is just some three derivatives terms
\begin{eqnarray}
g_2'{R_\mu}^\nu\partial_\nu\phi=g_2'{R_{\rho\mu}}^{\rho\nu}\partial_\nu\phi=g_2'(\nabla_\nu\nabla_\mu-\nabla_\mu\nabla_\nu)\partial^\nu\phi,
\end{eqnarray}
so they cancel out the last two unwanted terms if
\begin{eqnarray}
g_2'+g_6=0,\,g_2'+g_6+(d-1)g_7=0.
\end{eqnarray}

The second and the third unwanted term vanish due to the Bianchi identity $\nabla^\nu (R_{\nu\mu}-\frac 1 2 R \gamma_{\nu\mu})=0$ if
\begin{eqnarray}
g_3=-\frac 1 {2(d-1)}g_2.
\end{eqnarray}

The fourth term vanishes if
\begin{eqnarray}
g_4+2(d-1)g_5-g_6'=0.
\end{eqnarray}

Assuming all the unwanted terms cancel by these relations, the flow equation of the metric is
\begin{eqnarray}
\pounds_n \gamma_{\mu\nu}&=&g_1\gamma_{\mu\nu}+g_2 R_{\mu\nu}+g_4\partial_\mu\phi\partial_\nu\phi-g_2'\nabla_\mu\partial_\nu\phi\nonumber\\
&&-\frac 1 {2(d-1)}\gamma_{\mu\nu}\left[g_2 R+(g_2''+g_4)(\gamma^{\rho\eta}\partial_\rho\phi\partial_\eta\phi)\right]+O(\partial^4)\,\,\,.
\end{eqnarray}

The momentum constraint becomes
\begin{eqnarray}
&&\nabla_\rho^{(d)}(\gamma^{\rho\nu}\pounds_n\gamma_{\mu\nu}-\gamma_\mu^\rho\gamma^{\nu\eta}\pounds_n\gamma_{\nu\eta})\nonumber\\
&=&\partial_\mu\phi\left(-(d-1)g_1'-\frac 1 2g_2'R+\frac 1 2(g_2'''+g_4')(\gamma^{\rho\eta}\partial_\rho\phi\partial_\eta\phi)\right.\nonumber\\
&&\left.+(g_2''+g_4)(\gamma^{\rho\eta}\nabla_\rho\partial_\eta\phi)+O(\partial^4)\right)\,,\nonumber\\
&=&\partial_\mu\phi\pounds_n\phi
\end{eqnarray}

so
\begin{eqnarray}
h_1=-(d-1)g_1',\,h_2=-\frac 1 2g_2',\,h_3=\frac 1 2(g_2'''+g_4'),\,h_4=g_2''+g_4.
\end{eqnarray}

and the flow equation of the scalar field is
\begin{eqnarray}
\pounds_n\phi=-(d-1)g_1'-\frac 1 2g_2'R+\frac 1 2(g_2''+g_4)'(\gamma^{\rho\eta}\partial_\rho\phi\partial_\eta\phi)\nonumber\\
+(g_2''+g_4)(\gamma^{\rho\eta}\nabla_\rho\partial_\eta\phi)+O(\partial^4)\nonumber\\
\end{eqnarray}

Introducing $g_8=g_2''+g_4$, the flow equations are further simplified as
\begin{eqnarray}
\pounds_n \gamma_{\mu\nu}&=&g_1\gamma_{\mu\nu}+g_2 R_{\mu\nu}+(g_8-g_2'')\partial_\mu\phi\partial_\nu\phi-g_2'\nabla_\mu\partial_\nu\phi\nonumber\\
&&-\frac 1 {2(d-1)}\gamma_{\mu\nu}\left(g_2 R+g_8(\gamma^{\rho\eta}\partial_\rho\phi\partial_\eta\phi)\right)+O(\partial^4)\,,\\
\pounds_n\phi&=&-(d-1)g_1'-\frac 1 2g_2'R+\frac 1 2g_8'(\gamma^{\rho\eta}\partial_\rho\phi\partial_\eta\phi)+g_8(\gamma^{\rho\eta}\nabla_\rho\partial_\eta\phi)+O(\partial^4)\nonumber\\
\end{eqnarray}

There are only three independent unknown scalar functions $g_1,\,g_2,\,g_8$ in the flow equations. They will be determined by the Hamiltonian constraint.

\subsubsection{Hamiltonian constraint}

Pluging the flow equations in the Hamiltonian constraint, the right hand side of the Hamiltonian constraint is
\begin{eqnarray}
&&\frac 1 4 (\gamma^{\mu\nu}\gamma^{\rho\eta}-\gamma^{\mu\rho}\gamma^{\nu\eta})(\pounds_n\gamma_{\mu\nu})(\pounds_n\gamma_{\rho\eta})-\frac 1 2 (\pounds_n\phi)^2\nonumber\\
&=& \frac 1 4d(d-1)g_1^2+\frac 1 4(d-2)g_1 g_2R+\frac 1 4((d-2)g_1g_8-2(d-1)g_1g_2'')(\partial_\mu\phi\partial^\mu\phi)\nonumber\\
&&-\frac 1 2(d-1)g_1g_2'(\nabla^\mu\partial_\mu\phi)\nonumber\\
&&-\frac 1 2(d-1)^2g_1'^2-\frac 1 2(d-1)g_1'g_2'R+\frac 1 2(d-1)g_1'g_8'(\partial_\mu\phi\partial^\mu\phi)\nonumber\\
&&+(d-1)g_1'g_8(\nabla^\mu\partial_\mu\phi)+O(\partial^4),
\end{eqnarray}
and should be equal to the left hand side $R-\frac 1 2 \partial_\mu\phi\partial^\mu\phi+V$ for any solutions, so we have the following equations.

For zero derivative terms,
\begin{eqnarray}
V= \frac 1 4d(d-1)g_1^2-\frac 1 2(d-1)^2g_1'^2\label{HC_1}
\end{eqnarray}

For $R$ terms,
\begin{eqnarray}
R=\left(\frac 1 4(d-2)g_1 g_2-\frac 1 2(d-1)g_1'g_2'\right)R\label{HC_2}
\end{eqnarray}

For $\partial_\mu\phi\partial^\mu\phi$ terms,
\begin{eqnarray}
-\frac 1 2\partial_\mu\phi\partial^\mu\phi=\left(\frac 1 4(d-2)g_1g_8-\frac 1 2(d-1)g_1g_2''+\frac 1 2(d-1)g_1'g_8'\right)\partial_\mu\phi\partial^\mu\phi\label{HC_3}
\end{eqnarray}

For $\nabla^\mu\partial_\mu\phi$ terms,
\begin{eqnarray}
0=\left(-\frac 1 2(d-1)g_1g_2'+(d-1)g_1'g_8\right)\nabla^\mu\partial_\mu\phi\label{HC_4}
\end{eqnarray}

Introduce superpotential $W(\phi)$ as the solution of the following equation,
\begin{eqnarray}
V=\frac d {4(d-1)}W^2-\frac 1 2{W^\prime}^2,
\end{eqnarray}
where $W$ is assumed to be positive.

From (\ref{HC_1}) one obtains the solution of $g_1$
\begin{eqnarray}
g_1=-\frac 1 {d-1}W,
\end{eqnarray}
where we have chosen the sign to be minus so that the metric is decreasing in the IR direction.

Using the solution of $g_1$, the equation for the $R$ terms (\ref{HC_2}) becomes
\begin{eqnarray}
\frac 2 {W'}=g_2'-\frac {d-2}{2(d-1)}\frac W{W'}g_2,
\end{eqnarray}
so $g_2$ is
\begin{eqnarray}
g_2(\phi)=e^{-(d-2)\A(\phi)}\left(2 c+\int_{\bar\phi}^\phi d\tilde\phi\frac 2 {W'(\tilde\phi)} e^{(d-2)\A(\tilde\phi)}\right),
\end{eqnarray}
where $\tilde c$ is the integration constant and $\A$ is defined as (\ref{A}).

Introduce $U(\phi)=\frac 1 2 g_2(\phi)$, so
\begin{eqnarray}
U(\phi)=e^{-(d-2)\A(\phi)}\left( c+\int_{\bar\phi}^\phi d\tilde\phi\frac 1 {W'(\tilde\phi)} e^{(d-2)\A(\tilde\phi)}\right),
\end{eqnarray}
We will express the solutions of flow equations by $W$ and $U$.

From the equation for the $\nabla^\mu\partial_\mu\phi$ terms (\ref{HC_4}) and the solution of $g_1$, one obtains
\begin{eqnarray}
g_8= \frac W{W'} U'.
\end{eqnarray}

It is not obvious that the solutions of $g_1,g_2,g_8$ solve the equation for the $\partial_\mu\phi\partial^\mu\phi$ terms (\ref{HC_3}), but one can verify that this equation is indeed satisfied, so it is merely a redundant equation.

Therefore, the final results of the flow equations are
\begin{eqnarray}
\pounds_n \gamma_{\mu\nu}&=&2U R_{\mu\nu}+\left(\frac W{W'} U'-2U''\right)\partial_\mu\phi\partial_\nu\phi-2U'\nabla_\mu\partial_\nu\phi\nonumber\\
&&-\frac 1 {d-1}\gamma_{\mu\nu}\left(W+U R+\frac W{2W'} U'(\gamma^{\rho\eta}\partial_\rho\phi\partial_\eta\phi)\right)+...\,,\\
\pounds_n\phi&=&W'- U'R+\frac 1 2\left(\frac W{W'} U'\right)'(\gamma^{\rho\eta}\partial_\rho\phi\partial_\eta\phi)+\frac W{W'} U'(\gamma^{\rho\eta}\nabla_\rho\partial_\eta\phi)+...\nonumber\\
\end{eqnarray}
where $W(\phi)$ and $f(\phi)$ are the solutions of
\begin{eqnarray}
V&=&\frac d {4(d-1)}W^2-\frac 1 2{W^\prime}^2,\\
1&=&W'U'-\frac {d-2} {2(d-1)}WU.
\end{eqnarray}

\subsubsection{Dynamical equation} \label{Appdyn}
In this section, we want to verify that the flow equations determined by the constraints solve the dynamical equation automatically.

The dynamical equation is
\begin{eqnarray}
&&R^{(d)}_{\mu\nu}-\frac 1 2\partial_\mu\phi\partial_\nu\phi+\frac 1 {d-1} V\gamma_{\mu\nu}\nonumber\\
&=&\frac 1 2\pounds_n(\pounds_n\gamma_{\mu\nu})+\frac 1 4(\gamma^{ab}\gamma_\mu^c-2\gamma_\mu^a\gamma^{bc})(\pounds_n\gamma_{ab})(\pounds_n\gamma_{c\nu})\end{eqnarray}
where we have fixed the gauge $\partial_\mu N=0$, so the lapse function $N$ is a constant on the hypersurface.

The second derivative term in the dynamical equation is complicated
\begin{eqnarray}
&&\pounds_n(\pounds_n\gamma_{\mu\nu})\nonumber\\
&=&2W'U' R_{\mu\nu}+2W'\left(\frac W{2W'} U'-U''\right)'\partial_\mu\phi\partial_\nu\phi-2W'U''\nabla_\mu\partial_\nu\phi\nonumber\\
&&-\frac 1 {d-1}\gamma_{\mu\nu}W' \left(W'+\left(\frac W{W'} U'\right)'(\gamma^{\rho\eta}\partial_\rho\phi\partial_\eta\phi)+\frac W{W'} U'(\gamma^{\rho\eta}\nabla_\rho\partial_\eta\phi)\right)\nonumber\\
&&-\frac 1 {d-1}W\left(2U R_{\mu\nu}+\left(\frac W{W'} U'-2U''\right)\partial_\mu\phi\partial_\nu\phi-2U'\nabla_\mu\partial_\nu\phi\right.\nonumber\\
&&\,\,\,\,\,\,\,\,\,\,\,\,\,\,\,\,\,\,\,\,\,\,\,\,\,\,\,\,\,\,\,\,\,\,\,\,\left.-\frac 1 {d-1}\gamma_{\mu\nu}\left(W+2U R+\frac W{W'} U'(\gamma^{\rho\eta}\partial_\rho\phi\partial_\eta\phi)\right)\right)\nonumber\\
&&+2U \pounds_n R_{\mu\nu}+\left(\frac W{W'} U'-2U''\right)\pounds_n(\partial_\mu\phi\partial_\nu\phi)-2U'\pounds_n(\nabla_\mu\partial_\nu\phi)\nonumber\\
&&-\frac 1 {2(d-1)}\gamma_{\mu\nu}\left(2U \pounds_n R+\frac W{W'} U'\pounds_n(\gamma^{\rho\eta}\partial_\rho\phi\partial_\eta\phi)\right)+O(\partial^4)
\end{eqnarray}

where the last two lines require additional calculation.

\begin{eqnarray}
\pounds_n R_{\mu\nu}&=&-\frac 1 2 \gamma^{\rho\eta}\nabla_\mu\nabla_\nu(\pounds_n\gamma_{\rho\eta})-\frac 1 2 \gamma^{\rho\eta}\nabla_\rho\nabla_\eta(\pounds_n\gamma_{\mu\nu})+\gamma^{\rho\eta}\nabla_\rho\nabla_{(\mu}(\pounds_n\gamma_{\nu)\eta})\nonumber\\
&=&\frac {d-2}{2(d-1)}\nabla_\mu\partial_\nu W+\frac 1 {2(d-1)}\gamma_{\mu\nu}\nabla_\rho\partial^\rho W+O(\partial^4)\nonumber\\
\pounds_n(\partial_\mu\phi\partial_\nu\phi)&=&\partial_\mu(\pounds_n\phi)\partial_\nu\phi+\partial_\mu\phi\partial_\nu(\pounds_n\phi)=2W''\partial_\mu\phi\partial_\nu\phi+O(\partial^4)\nonumber\\
\pounds_n(\nabla_\mu\partial_\nu\phi)&=&\nabla_\mu\partial_\nu(\pounds_n\phi)-(\pounds_n\Gamma_{\mu\nu}^\rho)\partial_\rho\phi+O(\partial^4)\nonumber\\
&=&\nabla_\mu\partial_\nu W'-\frac 1 2 \gamma^{\rho\eta}(\nabla_\mu(\pounds_n\gamma_{\nu\eta})+\nabla_\nu(\pounds_n\gamma_{\mu\eta})-\nabla_\eta(\pounds_n\gamma_{\mu\nu}))\partial_\rho\phi+O(\partial^4)\nonumber\\
&=&\nabla_\mu\partial_\nu W'+\frac 1 {2(d-1)}W'(2\partial_\mu\phi\partial_\nu\phi-\gamma_{\mu\nu}\partial^\rho\phi\partial_\rho\phi)+O(\partial^4)\nonumber\\
\pounds_n\gamma^{\mu\nu}&=&-\gamma^{\mu\rho}\gamma^{\nu\eta}\pounds_n\gamma_{\rho\eta}
\end{eqnarray}

The second term on the right hand side of the dynamical equation is
\begin{eqnarray}
&&\frac 1 4(\gamma^{ab}\gamma_\mu^c-2\gamma_\mu^a\gamma^{bc})(\pounds_n\gamma_{ab})(\pounds_n\gamma_{c\nu})\nonumber\\
&=&\frac 1 4 (-\frac d {d-1}W)(2U R_{\mu\nu}+\left(\frac W{W'} U'-2U''\right)\partial_\mu\phi\partial_\nu\phi-2U'\nabla_\mu\partial_\nu\phi\nonumber\\
&&\,\,\,\,\,\,\,\,\,\,\,\,\,\,\,\,\,\,\,\,\,\,\,\,\,\,\,\,\,-\frac 1 {2(d-1)}\gamma_{\mu\nu}\left(2W+2U R+\frac W{W'} U'(\gamma^{\rho\eta}\partial_\rho\phi\partial_\eta\phi)\right))\nonumber\\
&&+\frac 1 4\Big(\frac {d-2}{d-1}UR+\left(\frac {d-2}{2(d-1)}\frac W{W'}U'-2U''\right)(\partial_\rho\phi\partial^\rho\phi) \nonumber\\
&&\qquad\qquad\qquad-2U'(\nabla_\rho\partial^\rho\phi) \Big)(-\frac 1 {d-1})W\gamma_{\mu\nu}\nonumber\\
&&+\frac 1 {d-1}W\Big(2U R_{\mu\nu}+\left(\frac W{W'} U'-2U''\right)\partial_\mu\phi\partial_\nu\phi-2U'\nabla_\mu\partial_\nu\phi\nonumber\\
&&\,\,\,\,\,\,\,\,\,\,\,\,\,\,\,\,\,\,\,\,\,\,-\frac 1 {2(d-1)}\gamma_{\mu\nu}\left(W+2U R+\frac W{W'} U'(\gamma^{\rho\eta}\partial_\rho\phi\partial_\eta\phi)\right)\Big)+O(\partial^4)
\end{eqnarray}

Now one can collect different terms on the right hand side from the previous calculations:

The zero derivative terms on the right hand side are
\begin{eqnarray}
&&\gamma_{\mu\nu} \left(-\frac 1 {2(d-1)}W'^2+\frac 1 {2(d-1)^2} W^2+\frac {d-2}{4(d-1)^2}W^2\right)\nonumber\\
&=&\gamma_{\mu\nu}\frac 1 {d-1}\left(\frac {d}{4(d-1)^2}W^2-\frac 1 2W'^2\right)\nonumber\\
&=&\frac 1 {d-1}V\gamma_{\mu\nu}
\end{eqnarray}

The $R_{\mu\nu}$ terms on the right hand side are
\begin{eqnarray}
&&R_{\mu\nu}\left( W'U'-\frac 1 {(d-1)}WU-\frac {d-4} {2(d-1)}WU\right)\nonumber\\
&=&R_{\mu\nu}\left( W'U'-\frac {d-2} {2(d-1)}WU\right)\nonumber\\
&=&R_{\mu\nu}
\end{eqnarray}

The $\partial_\mu\phi\partial_\nu\phi$ terms on the right hand side are
\begin{eqnarray}
&&(\partial_\mu\phi\partial_\nu\phi)\left(-\left(\frac {W'} {2W} U'-\frac {d-2}{4(d-1)}U\right)'\left(-\frac {W^2}{W'}\right)-\left( W'U'-\frac {d-2}{2(d-1)}WU\right)''\right)\nonumber\\
&=&(\partial_\mu\phi\partial_\nu\phi)\left(-\frac 1 2\left(\frac 1 W\right)'\left(-\frac {W^2}{W'}\right)-0\right)\nonumber\\
&=&-\frac 1 2\partial_\mu\phi\partial_\nu\phi
\end{eqnarray}

The $R\gamma_{\mu\nu}$ terms on the right hand side are
\begin{eqnarray}
R\gamma_{\mu\nu}\frac 2 {(d-1)^2}WU\left( \frac 1 2+\frac d 8-\frac 1 4-\frac {d-2}8-\frac 1 2   \right)=0
\end{eqnarray}

The $\gamma_{\mu\nu}\nabla^\rho\partial_\rho\phi$ terms on the right hand side are
\begin{eqnarray}
&&\gamma_{\mu\nu}(\nabla^\rho\partial_\rho\phi)2\Big(-\frac 1 {4(d-1)}WU' +\frac 1 {2(d-1)}W'U\nonumber\\
&&\qquad\qquad\qquad-\frac 1 {2(d-1)}W'U+\frac 1 {4(d-1)}WU'\Big)=0
\end{eqnarray}

The $\gamma_{\mu\nu}\partial^\rho\phi\partial_\rho\phi$ terms on the right hand side are
\begin{eqnarray}
\gamma_{\mu\nu}\partial^\rho\phi\partial_\rho\phi\frac 1 {2(d-1)}(-WU''+\frac 1 {2(d-1)}\frac {W^2}{W'}U'+WU''-\frac 1 {2(d-1)}\frac {W^2}{W'}U')=0\nonumber\\
\end{eqnarray}

The $\partial_\mu\phi\partial_\nu\phi$ terms on the right hand side are
\begin{eqnarray}
\partial_\mu\phi\partial_\nu\phi(-)\left(W'U'-\frac {d-2}{2(d-1)}WU\right)'=0
\end{eqnarray}

The right hand side of the dynamical equation matches the left hand side exactly by these first order flow equations! Therefore, the dynamical equation is automatically solved by the first order equations determined by the constraints.

\subsection{Lie derivative matching condition} \label{App3}
To derive the effective Lagrangian, we will solve the Lie derivative matching equation \begin{eqnarray}
V+R^{(d)}-\frac 1 2\gamma^{\mu\nu}\partial_\mu\phi\partial_\nu\phi=\frac 1 2 \gamma^{-\frac 1 2}\pounds_n\mathcal L_{eff}
\end{eqnarray}

Using the flow equations
\begin{eqnarray}
&&\gamma^{-\frac 1 2}\pounds_n\mathcal L_{eff}\nonumber\\
&=&-\frac d {2(d-1)}WF_0+W'F_0'\nonumber\\
&&+\left(\frac {d-2}{4(d-1)}fF_0+\left(-\frac d {2(d-1)}W\right)F_1-\frac 1 2 f'F_0'+W'F_1'+\frac 1 {d-1}WF_1\right)R\nonumber\\
&&+\left(\frac {d-2}{4(d-1)}\frac W{2W'}f'F_0-\frac 1 2 f''F_0-\frac d {2(d-1)}WF_2+W'F_2'+F_1W''+F_3W'''\right)(\partial\phi)^2\nonumber\\
&&+\left(\frac 1 2\left(\frac W{2W'} f'\right)'F_0'+\frac 1 {d-1}WF_2+2W''F_2-\frac {d-2}{2(d-1)}W'F_3\right)(\partial\phi)^2\nonumber\\
&&+\Big(\frac W{2W'} f'F_0'-\frac 1 2 f'F_0+\left(-\frac d {2(d-1)}W\right)F_3+W'F_3'+\frac 1 {d-1}WF_3\nonumber\\
&&\qquad\qquad+F_1W'+F_3W''\Big)\Box\phi+O(\partial^4)
\end{eqnarray}

To the leading order, the matching of the Lie derivative matching equation on both sides requires
\begin{eqnarray}
V&=&-\frac d {4(d-1)}WF_0+\frac 1 2 W'F_0'
\end{eqnarray}
whose solution is
\begin{eqnarray}
F_0=e^{-d\A}(C_0-We^{d\A})
\end{eqnarray}
$\A$ is defined in (\ref{A}).

Since $C_0$ is irrelevant to the on-shell action, we will choose $C_0=0$ so that $F_0=W$. This choice will simplify the calculation at subleading order.

For the two-derivative terms, the matching on both sides requires
\begin{eqnarray}
1&=&W'F_1'-\frac {d-2} {2(d-1)}WF_1\\
-\frac 1 2&=&\left(2W''-\frac {d-2} {2(d-1)}W\right)F_2+W'F_2'+F_1W''\nonumber\\
&&\qquad+\left(W''-\frac {d-2}{2(d-1)}W\right)'F_3\nonumber\\\\
-F_1W'&=&W'F_3'+\left(W''-\frac {d-2} {2(d-1)}W\right)F_3
\end{eqnarray}
which is greatly simplified by the choice $C_0=0$.

The solutions are
\begin{eqnarray}
F_1&=&e^{-(d-2)\A}\left(C_1+\int_{\phi_0}^\phi W'^{-1} {e^{(d-2)\A}}\right)\nonumber\\
F_2&=&F_3'+\frac 1 2\left(C_2 W'^{-2}e^{-(d-2)\A} +\frac W{W'}F_1'\right)\nonumber\\
F_3&=&W'^{-1}e^{-(d-2)\A}\left(C_3-\int_{\phi_0}^\phi F_1W'e^{(d-2)\A}\right)\nonumber
\end{eqnarray}

\subsection{Integration constants and conserved currents} \label{conserved}
As we have mentioned in the result of effective Lagrangian, there are some irrelevant terms with new integration constants and they are related to some conserved quantities.

As a warm up, we consider the first term in
\be
\mathcal L_{eff}=\sqrt{-\gamma}(F_0+F_1R+F_2\gamma^{\mu\nu}\partial_\mu\phi\partial_\nu\phi+F_3\gamma^{\mu\nu}\nabla_\mu\partial_\nu\phi+O(\partial^4)).
\ee

The equation of $F_0$ is
\begin{eqnarray}
V=-\frac d {4(d-1)}WF_0+\frac 1 2 W'F_0'
\end{eqnarray}
whose solution is
\begin{eqnarray}
F_0=C_0e^{-d\A}-W
\end{eqnarray}
where $\A$ is defined in (\ref{A}). There is a new integration constant $C_0$ in $F_0$.

In the domain wall solution $\gamma_{\mu\nu}=\eta_{\mu\nu}e^{2(A(\bar\phi)+\A)}$ with $A(\bar\phi)$ being an integration constant of the scale factor, the effective Lagrangian is
\begin{eqnarray}
\mathcal L_{eff}=\sqrt{-\gamma}(F_0+O(\partial^2))=e^{dA(\bar\phi)}(C_0-We^{d\A}+O(\partial^2))\,,
\end{eqnarray}
so $C_0$ is an integration constant in the indefinite integral. $C_0$ will not appear in the result because the final result is the difference between the UV and IR contributions.

However, in the general situation $\gamma_{\mu\nu}\neq \eta_{\mu\nu}e^{2(A(\bar\phi)+\A)}$. $C_0$ is multiplied by $\sqrt{-\gamma} e^{-d\A}$, not by one in a trivial way any more. It is less obvious that the independence of $C_0$ still holds. As we will see, to two derivative order $C_0$ will be multiplied by some Ricci and $\partial \phi\partial \phi$ terms in a way that $C_0$ drops out in the final result.

In the subleading order, there are more integration constants $C_1$,$C_2$ in the higher order coefficients $F_1$ and $F_2$. The final result should not depend on these integrations constants also because they are just some irrelevant constants in the indefinite integral in a generalized sense. These integration constants are different from those in the flow equations $c_0,\,c_1$ which have physical meaning as source and vev and are relevant to the final result.

The dropping out of $C_n$ is due to an interesting fact: they are multiplied by some non-linearly conserved quantities.

To two derivative order, the first three new integrations constants appear in the solution in the following way
\begin{eqnarray}
F_1&=&C_0G_0^{(1)}+C_1e^{-(d-2)\A}+U\nonumber\\
F_2&=&F_3'+\frac 1 2\left((C_0G_0^{(2)}+C_2) W'^{-2}e^{-(d-2)\A} +\frac W{W'}(C_1e^{-(d-2)\A}+U)'\right)\nonumber\\
F_3&=&W'^{-1}e^{-(d-2)\A}\left(C_3-\int_{\phi_0}^\phi (C_1e^{-(d-2)\A}+U)W'e^{(d-2)\A}\right)\nonumber
\end{eqnarray}

From the solution of $F_2$, $F_3$ will disappear from the effective Lagrangian after an integration by parts. The solution of $F_2$ depends on $C_1$ which is the integration constant of $F_1$.

The corresponding conserved quantities are
\begin{eqnarray}
C_0&:&\,G_0=\int d^dx\,\sqrt{-\gamma}\left(e^{-dA}+G_0^{(1)}e^{-(d-2)\A}R+W'^{-2}e^{-(d-2)\A}G_0^{(2)}\frac 1 2\gamma^{\mu\nu}\partial_{\mu}\phi\partial_\nu\phi+O(\partial^4)\right)\nonumber\\
C_1&:&\,G_1=\int d^dx\sqrt{-\gamma}\left(e^{-(d-2)\A}R+\frac W{W'}(e^{-(d-2)\A} )'\frac 1 2\gamma^{\mu\nu}\partial_\mu\phi\partial_\nu\phi+O(\partial^4)\right)\nonumber\\
C_2&:&\,G_2=\int d^dx\sqrt{-\gamma}\left(W'^{-2}e^{-(d-2)\A} \frac 1 2\gamma^{\mu\nu}\partial_\mu\phi\partial_\nu\phi+O(\partial^4)\right)
\end{eqnarray}
where
\begin{eqnarray}
G_0^{(1)}( \phi)&=& G_0^{(1)}(\bar \phi)+\frac 1 {2(d-1)}\int_{\bar\phi}^\phi d\tilde \phi \,e^{-2\A}W'^{-2}(dWU'-(d-2)W'U) \\
G_0^{(2)}( \phi)&=& G_0^{(2)}(\bar \phi)+2\int_{\bar\phi}^\phi d\tilde \phi \,e^{(d-2)\A}W'\left((e^{-d\A})'' \frac W {W'}U'+(e^{-d\A})' ( \frac W {2W'}U')'\right) \nonumber\\
&&-\frac {3d-2}{2(d-1)}\int_{\bar\phi}^\phi d\tilde \phi \,e^{-2\A}W U'+2\int_{\bar\phi}^\phi d\tilde \phi \,e^{(d-2)\A}W'^2(G_0^{(1)}e^{-(d-2)\A})'
\end{eqnarray}
$ G_0^{(1)}(\bar \phi)$ and $ G_0^{(2)}(\bar \phi)$ are integration constants and are fixed by imposing $G_0^{(1)}(\phi_{UV})=G_0^{(2)}(\phi_{UV})=0$ so that the  three quantities in (\ref{conserved-quantities}) are linearly independent in the UV limit.

Their radial changes are described by their Lie derivative $\pounds_n$,
\begin{eqnarray}
&&\pounds_n\mathcal G_0=\sqrt{-\gamma}\nabla_\mu\left(\Big(-U'e^{-d\A} +2\left(\frac W{2W'}U'\right)(e^{-d\A})'+W' G_0^{(1)}e^{-(d-2)\A}\Big)\gamma^{\mu\nu}\partial_\nu\phi\right)\nonumber\\
&&\qquad\qquad+ O(\partial^4)\label{conserved-1}\\
&&\nonumber\\
&&\pounds_n\mathcal G_1=\sqrt{-\gamma}\nabla_\mu( W'e^{-(d-2)\A}\gamma^{\mu\nu}\partial_\nu \phi)+ O(\partial^4)\label{conserved-2}\\
&&\nonumber \\
&&\pounds_n\mathcal G_2= O(\partial^4)\label{conserved-3}
\end{eqnarray}
where $\mathcal G_0,\mathcal G_1,\mathcal G_2$ are charge densities: $G_0=\int d^dx\,\mathcal G_0,\,G_1=\int d^dx\,\mathcal G_1,\,G_2=\int d^dx\,\mathcal G_2$. From (\ref{conserved-1}-\ref{conserved-3}), one can construct three bulk conserved currents. The radial changes of the charge density is a total derivative
\be \label{conserved-4}
\pounds_n\mathcal G_k=\partial_\mu(\sqrt{-\gamma}J_k^\mu).
\ee
To two-derivative order,  the current densities $J_k^\mu(\phi)$ on the right hand sides of  (\ref{conserved-1}-\ref{conserved-3}) have the form:
\be\label{conserved-5}
J_k^\mu(\phi)=J_k(\phi)\partial^\mu\phi+O(\partial^3).
\ee

By adding an appropriate  total derivative to the ${\cal G}_k$'s , we can construct  {\em radially-invariant} charge densities:
\be\label{conserved-6}
\tilde{\mathcal G_k}=\mathcal G_k+\partial_\mu(\sqrt{-\gamma}H_k\partial^\mu\phi+O(\partial^3)),\quad \pounds_n\tilde{\mathcal G_k}=O(\partial^4),
\ee
where the functions $H_k(\phi)$ are required to satisfy:
\be\label{conserved-8}
\pounds_n \partial_\mu(\sqrt{-\gamma}H_k\partial^\mu\phi)=-\partial_\mu(\sqrt{-\gamma}J_k\partial^\mu\phi).
\ee
Using the flow equations (\ref{flow-eq-1}-\ref{flow-eq}), the above equation is simplified to
\be\label{conserved-9}
-\frac {d-2}{2(d-1)}WH_k+W' H_k'+ W''H_k=J_k,
\ee
whose solution is
\be\label{conserved-10}
H_k=W'^{-1}e^{-(d-2)\A}\left(C_3+\int_{\bar\phi}^\phi d\tilde \phi \,J_k(\tilde \phi ) e^{(d-2)\A(\tilde \phi )}\right).
\ee

\subsection{Kinetic term counterterm}\label{kinetic}
Here we show how to transform the divergence of the kinetic term into the same form as Papadimitriou's result, \cite{papa}
\begin{eqnarray}
\frac W{W'} {U^{div}}'&=&\frac W{W'}\left(\frac 1 {W'}+\frac {d-2}{2(d-1)}\frac W{W'}{U^{div}}\right)\nonumber\\
&=&\frac {W^2}{W'^2}e^{-(d-2)\A}\left(\frac 1 We^{(d-2)\A}+\frac {d-2}{2(d-1)}\int^\phi d\tilde\phi\frac 1 {W'} e^{(d-2)\A}\right)\nonumber\\
&=&\frac {W^2}{W'^2}e^{-(d-2)\A}\left(\int^\phi d\tilde\phi\left(\frac 1 W e^{(d-2)\A}\right)'+\frac {d-2}{2(d-1)}\int^\phi d\tilde\phi\frac 1 {W'} e^{(d-2)\A}\right)\nonumber\\
&=&-\frac {W^2}{W'^2}e^{-(d-2)\A}\int^\phi d\tilde\phi\frac  {W'}{W^2} e^{(d-2)\A}\nonumber\\
&=&-\A'^2e^{-(d-2)\A}\int^\phi d\tilde\phi \frac 1 {W'} e^{(d-2)\A}\A'^{-2}\label{divergence}.
\end{eqnarray}

\section{Non-universality and scheme dependence at second order}\label{alpha}

In this appendix we show that,  in a derivative expansion, the second order contributions to the $\beta$-functions are not unique. In fact, at second order in derivatives  we can split in many different ways the  weyl rescaling and the radial evolution of the metric in equation (\ref{weyl5}), by writing:
\be\label{alpha1}
\pounds_n \gamma_{\mu\nu} = \sigma_{\alpha}\left(2\gamma_{\mu\nu} + \beta^{(2)}_{\mu\nu,\alpha} \right), \quad \pounds_n \phi = \sigma_\alpha\left(\beta^{(0)}_\phi + \beta^{(2)}_{\phi,\alpha}\right)
\ee
where $\sigma_\alpha$ and $\beta^{(2)}_{\mu\nu,\alpha}$ are defined by:
\be
\sigma_{\alpha} = -{1\over 2(d-1)}\Big[W + \alpha\left(d X - 2(d-1) Y\right)\Big] , \quad \sigma_{\alpha}\beta^{(2)}_{\mu\nu,\alpha} = \left[{(d\alpha-1)\over (d-1)} X - \alpha Y\right]\gamma_{\mu\nu} + 2Y_{\mu\nu} .
\ee
and $\alpha$ is any real number, and the quantities on the right hand sides are defined in equations (\ref{X}-\ref{Y}).  The choice $\alpha=1/d$ corresponds to the splitting used in Section \ref{RGflowsec}.

Therefore, we can write the Lie derivative  in terms of any one out of a family $\Delta_{\alpha}$ of dilatation operators, all of which consist of a Weyl rescaling to lowest order:
\be\label{weyl10}
\pounds = \int d^d x \sigma_{\alpha}(x)\Delta_{\alpha}(x), \quad \Delta_\alpha(x) \equiv 2\gamma_{\mu\nu}{\delta \over \delta \gamma_{\mu\nu}} + \beta_{\mu\nu,\alpha}^{(2)}{\delta \over \delta \gamma_{\mu\nu}} + (\beta^{(0)}_{\phi}+ \beta^{(2)}_{\phi,\alpha}) {\delta \over \delta \phi}
\ee
with:
\bea
&& \beta^{(2)}_{\mu\nu,\alpha}=-{4(d-1)\over W} \left[Y_{\mu\nu} - \gamma_{\mu\nu}\left(\alpha Y - {(d\alpha -1)\over 2 (d-1)}X \right)\right], \label{weyl11-2}\\
&& \beta^{(2)}_{\phi,\alpha} = -{2(d-1)\over W}\left[Z - \alpha{W'\over W}\Big(d X - 2(d-1)Y\Big)\right],\label{weyl11-3}
\eea
where we have only kept terms of second order in derivatives, and $X,Y,Z$ are defined in equations (\ref{X}-\ref{Z}).

It can be check that, to second order, all the operators $\Delta_\alpha$ lead to the same expression for the Lie derivative when acting on the metric and scalar field. The differences are of fourth order in derivatives.   Moreover, the zero-th order $\beta$-functions are unchanged.  Therefore, we could in principle  identify any of the $\Delta_{\alpha}$'s with  the generator of local scale transformations on the fields.

The dependence on $\alpha$ of the second order  $\beta$-functions can be interpreted as a ``scheme dependence'' in the choice of the relation between scale transformation and radial flow. The difference between different schemes arises at higher orders in the derivative expansion, like it happens in ordinary perturbation theory in field theory.

 For $\alpha=1/d$ we recover equations (\ref{beta2-1}, \ref{beta2-2}).
However, it is only in this case that $\beta^{(2)}_{\mu\nu}$ is traceless. Therefore, for any other value of $\alpha$ the term proportional to $\beta^{(2)}_{\mu\nu}$ contains an additional Weyl rescaling. For general $\alpha$, the determinant of the metric changes in an anomalous way under $\Delta_\alpha$:
\be
\Delta_{\alpha} \gamma = 2d \gamma - (1-d\alpha) {4(d-1)\over W} \left[ Y - {1\over (d-1)}X \right]
\ee

We will see shortly  that the choice $\alpha=1/d$ is special because in this case the trace anomaly in four dimensions receives no geometric  contributions of order of the curvature tensor.  Such contributions start instead at order curvature squared, in agreement with field theory results, \cite{osborn}.

The operators $\Delta_{\alpha}$ given in equation (\ref{weyl10}),  could all in principle be considered as generators of  RG transformations, and by the same line of arguments  used in Section 5.3, we can write  for any value of $\alpha$:
\be\label{trace9-a}
\Delta_{\alpha}(x) S^{(ren)} = 0.
\ee
Therefore, we can write trace identities with all values of $\alpha$. However, for any value other than $\alpha = 1/d$ (in which case  $\Delta_\alpha = \Delta$ discussed above), $\beta^{(2)}_{\mu\nu,\alpha}$ is {\em not} traceless, thus introducing  curvature terms already at second order in the trace identity (\ref{trace5}). Equivalently, in writing (\ref{trace9-a}) explicitly, one can incorporate the trace part of the $\beta_{\mu\nu,\alpha}$ term in the $\<{T^\mu}_\mu\>$ factor, in which case one obtains:
\be\label{trace10-a}
\left( 1 + {\beta_{\alpha} \over 2d}\right) \< {T^\mu}_\mu\> + \left(\beta_{\mu\nu,\alpha} - {1\over d} \gamma_{\mu\nu} \beta_\alpha\right) {1\over\sqrt{\gamma}}{\delta S^{(ren)} \over \delta \gamma_{\mu\nu}} + \beta_{\phi,\alpha} \< \cO\> =0
\ee
where $\beta_{\alpha} \equiv \gamma^{\mu\nu}\beta_{\mu\nu,\alpha}$. As noted in section 5.3, the traceless part of $\beta_{\mu\nu,\alpha}$ contributes only four derivative terms and higher, and to two-derivative order  we can neglect the seond term in (\ref{trace9-a}).  We may now derive the trace identity for $\alpha \neq 1/d$, up to second order in derivatives, as:
\be \label{trace11-a}
\< {T^\mu}_\mu\> = - \beta_{\phi,\alpha}\left(1- {1\over 2d} \beta^{(2)}_\alpha\right) \<\cO\> +  \text{4-derivative terms}
\ee
 This equation is rather awkward, since it is not the scalar $\beta$-function $\beta_{\phi,\alpha}$ which appears in front of $\<\cO\>$. Moreover,  we can compute explicitly the coefficient of $\<\cO\>$ using the definitions (\ref{weyl11-2}-\ref{weyl11-3}). The result is, up to second order:
\be\label{trace12-a}
\beta_{\phi,\alpha}\left(1- {1\over 2d} \beta^{(2)}_\alpha\right) = \beta_\phi
\ee
where the right hand side coincides with the scalar $\beta$-function with $\alpha = 1/d$, i.e. the one appearing in  $\Delta$ and written explicitly  in equation (\ref{beta2-1})\footnote{This is not surprising: both $\<\cO\>$ and $\<{T^\mu}_\mu\>$ are $\alpha$-independent, so the right hand side of (\ref{trace12-a}) must be too, and we have already seen that it is equal to $\beta_\phi$ for $\alpha =1/d$.}.
In other words,  the expression for the trace of the stress tensor in terms of the metric and scalar field invariants is scheme independent. However, for general $\alpha$, the right hand side of the trace identity does not match the standard quantum field theory result $\beta_\alpha \<O\>$ i.e. the scalar $\beta$-function {\em in that scheme} times the vev of the operator, but this happens only for $\alpha = 1/d$.

 Equations (\ref{trace11-a}) and (\ref{trace12-a}) show that in order to match the standard field theory trace identities, the most natural  choice for the generator of RG transformations is $\Delta$ , i.e. to set $\alpha=1/d$ and to have a traceless $\beta_{\mu\nu}$.

\section{Linearized Analysis}

\subsection{Linear gauge invariant variables}
We consider only the scalar modes in linear metric fluctuations around the background domain wall metric,
\begin{equation}
ds^2=e^{2\phi}du^2+e^{2A}(e^{2\psi}\eta_{\mu\nu}+\partial_\mu\partial_\nu E)dx^\mu dx^\nu+2\partial_\nu Bdudx^\nu
\end{equation}

and the scalar field fluctuations $\delta \Phi=\chi$.

Under diffeomorphism $u\rightarrow u+\xi^u$ and $x^\nu\rightarrow x^\nu+\partial^\nu \xi$, the metric fluctuations transform as $\delta g_{ab}\rightarrow\delta g_{ab}-\nabla_a\xi_b-\nabla_b\xi_a$. To the linear order,
\begin{equation}
\psi\rightarrow\psi-\dot A\xi^u,\,\phi\rightarrow\phi-\dot \xi^u
\end{equation}
\begin{equation}
B\rightarrow B-{\frac 1 2}(\xi^u+\dot \xi)+{\dot A}\xi,\,
E\rightarrow E-\xi e^{-2A}
\end{equation}

The scalar field transform as $\delta \Phi\rightarrow \delta \Phi-\dot \Phi \xi^u$, so
\begin{equation}
\chi\rightarrow \chi-\dot \Phi \xi^u
\end{equation}

One can construct two linear gauge invariant variables
\begin{equation}
\zeta_1=\frac \chi z-\psi,\,\zeta_2=\frac 1{2(d-1)}\dot\Phi\chi+\dot\psi-\dot A\phi
\end{equation}

where $z=\frac {\dot\Phi}{\dot A}$. The first one is similar to the Mukhanov-Sasaki variable and the second one is invariant under the use of background equations of motion.

\subsection{The equations of motion}

In the following, we will fix the gauge by requiring $B=E=0$. The metric becomes diagonal
\begin{equation}
ds^2=e^{2\phi}du^2+e^{2(A+\psi)}\eta_{\mu\nu}dx^\mu dx^\nu
\end{equation}
 and thus greatly simplifies the calculation.

The Einstein's equations are
\begin{eqnarray}
(d-1)[(d\dot\psi-\dot \phi)\dot A+\ddot\psi]&=&-\frac 1 2\dot\Phi\dot\chi+V\phi+\frac 1 2 V^\prime\chi\\
(d-1)(d\dot\psi\dot A+\partial^2\psi e^{-2A})&=&\frac 1 2 \dot\Phi\dot\chi+V\phi+\frac 1 2 V^\prime\chi\\
 \partial_\mu\partial_\nu[(d-2)\psi+\phi]&=&0\\
-(d-1)(\partial_\nu\dot \psi-\dot A\partial_\nu\phi)&=&\frac 1 2 \dot\Phi\partial_\mu\chi
\end{eqnarray}

where we have assumed the solution to the third equation is
\begin{equation}
(d-2)\psi+\phi=f_1(u)
\end{equation}

and insert this relation into the first equation to kill an unwanted term $\partial_\nu^2[(d-2)\psi+\phi]$ that will lead to inconsistency.

The fourth equation means that the second gauge invariant variable $\zeta_2$ is homogeneous
\begin{equation}
\zeta_2=f_2(u).
\end{equation}

For the case of homogeneous fluctuations, the linear fluctuations become linear perturbations of the background solutions. For example, if we fix $\phi=0$, their equations of motion are
\begin{eqnarray}
-2(d-1)\ddot \psi-d(d-1)2\dot A\dot \psi&=& \dot\Phi\dot\chi-V'\chi,\\
d(d-1)2\dot A\dot\psi&=& \dot \Phi\dot\chi+V'\chi.
\end{eqnarray}
which are linear perturbations around the background equations of motion. The corresponding linear perturbed superpotential is $\tilde W=W+\delta W$ with $\delta W=-\frac 1 {2(d-1)}\dot\psi$ and $\delta W'=\dot \chi$.

Since we are discussing inhomogeneous fluctuations around the background, it is natural to set $f_1(u)=0$ and $\zeta_2=f_2(u)=0$.

The Klein-Gordon equation is
\begin{equation}
V^{\prime\prime}\chi+2(d-1)\dot\psi \dot\Phi-2(d-1)\psi V^\prime+d\dot A\dot \chi+e^{-2A}\partial^2\chi +\ddot\chi=0.
\end{equation}

Assuming $f_1=0$, after some complicated manipulation, one can get the equation of motion for the linear gauge invariant variables
\begin{equation}
\ddot \zeta_1+(d\dot A+2\frac {\dot z}z)\dot \zeta_1+e^{-2A}\eta^{\mu\nu}\partial_\mu\partial_\nu\zeta_1+\frac {\dot z}z\zeta_2=0
\end{equation}
where the last term vanishes because we have assumed $\zeta_2=0$.\\

Another way to derive the equation of motion for gauge invariant variable is expanding the action to second order perturbations. Setting $f_1=f_2=0$, one can find
\begin{equation}
S=\left(-\frac 1 2\right)\int \,d^d x \int_{uv}^{IR}\,d u z^2 e^{dA}[(\partial_u\zeta)^2+e^{-2A}(\partial_\mu\zeta)(\partial^\mu\zeta)]+S_{IP}+S_{GH}+(...)
\end{equation}

where $S_{PI}$ are complicated boundary terms generated by integration by parts, $S_{GH}$ is the Gibbons-Hawking term and $(...)$ vanishes by constraints.

\subsection{The solution of gauge invariant variable}
The equation of motion for gauge invariant $\zeta=\frac \chi z-\psi$ is

\begin{equation}
\ddot\zeta+(d\dot A + 2 {\frac{\dot z}z})\dot\zeta + e^{-2A}\eta^{\mu\nu}{\partial_\mu \partial_\nu}\zeta=0.
\end{equation}

To solve this equation, one can rewrite the equation of motion in momentum space as
\begin{equation}
\partial_u(z^2e^{dA}\partial_u\zeta)=z^{2}e^{(d-2)A}p^2\zeta
\end{equation}
which is exactly solvable in derivatives expansions of $p^2$ order by order.

Written in an iterated way, the solution is
\begin{eqnarray}
\zeta=C\left[1+D\int_{u^{UV}}^u\, d\tilde u z^{-2}e^{-dA}\right]+p^2\int_{u^{UV}}^u\, d\tilde u z^{-2}e^{-dA}\int_{u^{UV}}^{\tilde u}\, d\tilde{\tilde u}z^2 e^{(d-2)A}\zeta
\end{eqnarray}

The first integration constant $C=\zeta^{UV}$ is the source for the gauge invariant variable and the second integration constant $D$ can be determined by IR regularity.

The integration constants $C$ and $D$ are constant on the slices $\Sigma_u$. However, they are also general functions of momentum $p$. For example, if slice Lorentz invariance is assumed, $D$ is a function of square momentum $D(p^2)=\sum_{i=0}^\infty D_{2i}{(p^2)}^i$ and there is one integration constant $D_{2i}$ at each order in derivative expansion.

The above solution can be written as the standard product of a boundary source $\zeta_{UV}$ and a boundary-to-bulk propagator $K(u)$
\begin{equation}
\zeta=\zeta_{UV}K(u,p^2)=\zeta_{UV}\left[K_1(u,p^2)+D(p^2)K_2(u,p^2)\right]
\end{equation}
 where $K_1=\sum_{i=0}^\infty U_i p^{2i}$, $K_2=\sum_{i=0}^\infty U_{i+1} p^{2i}$. $f_i(u)$ is defined iteratively as
\begin{eqnarray}
U_0&=&1,\,\nonumber\\
U_1&=&\int_{u^{UV}}^u\, d\tilde u z^{-2}e^{-dA},\,\nonumber\\
U_{i+2}&=&\int_{u^{UV}}^u\,d\tilde u z^{-2}e^{-dA}\int_{u^{UV}}^{\tilde u}\,d\tilde{\tilde u}z^2 e^{(d-2)A}U_i.
\end{eqnarray}

To the subleading order, the boundary-to-bulk propagator is
\begin{equation}
K=1+D_0U_1+p^2(D_2U_1+U_2+D_0U_3)+\mathcal O (p^4)
\end{equation}

An IR regular solution can be obtained by fixing the constants $D_{i}$ order by order.

\subsection{On-shell action of the gauge-invariant variable}
The bulk action for the gauge invariant variable is
\begin{equation}
S=\left(-\frac 1 2\right)\int \,d^d x \int_{u^{UV}}^{u^{IR}}\,d u z^2 e^{dA}[(\partial_u\zeta)^2+e^{-2A}(\partial_\mu\zeta)(\partial^\mu\zeta)].
\end{equation}

By integrating by parts and using the equation of motion,
\begin{eqnarray}
S&=&\left(-\frac 1 2\right)\int \,d^d x \int_{u^{UV}}^{u^{IR}}\,d u \partial_u(z^2 e^{dA}\zeta\partial_u \zeta)\nonumber\\
&+&\left(-\frac 1 2\right)\int\,d^d x\int_{u^{UV}}^{u^{IR}}\,d u z^2e^{(d-2)A} [(\partial_\mu\zeta)(\partial^\mu\zeta)+
\zeta(\partial_\mu\partial^\mu\zeta)].
\end{eqnarray}

The second line is a total divergence of transverse derivative and we assume the transverse surface term vanishes on the transverse boundary.

To evaluate the action, we go to the momentum space
\begin{eqnarray}
S_{on-shell}=\left(-\frac 1 2\right)\int \,\frac {d^d p}{(2\pi)^d} \left. z^2 e^{dA}\zeta\partial_u \zeta\right |_{u^{UV}}^{u^{IR}}
\end{eqnarray}

From the solution of boundary to bulk propagator $K(u,p^2)$
\begin{eqnarray}
z^2 e^{dA}\zeta\partial_u \zeta&=&\zeta_{uv}^2 K z^2 e^{dA}\partial_u K\nonumber\\
&=&\zeta_{uv}^2K\Big[D_0+p^2\Big( D_2+\int_{u^{UV}}^{u}\,d u z^{2}e^{(d-2)A}\nonumber\\
&&\qquad\qquad\qquad+D_0\int_{u^{UV}}^{u}\,d \tilde u z^{2}e^{(d-2)A}\int_{u^{UV}}^{\tilde u}\,d \tilde{\tilde u} z^{-2}e^{-dA}\Big)+\mathcal O (p^4)\Big]\nonumber\\
\end{eqnarray}

If we impose $\dot \zeta|_{u^{IR}}=0$, the gauge invariant variable will stay finite in the IR limit $e^A\rightarrow 0$. Under this assumption, the integration constant is determined
\begin{eqnarray}
D(p)=0
\end{eqnarray}
so
\begin{eqnarray}
z^2 e^{dA}\zeta\partial_u \zeta&=&\zeta_{uv}^2\left[p^2 \int_{u^{UV}}^{u}\,d u z^{2}e^{(d-2)A}+\mathcal O (p^4)\right]
\end{eqnarray}

The on-shell action of the gauge invariant variable is
\begin{eqnarray}
S_{on-shell}^{\zeta}&=&\left(-\frac 1 2\right)\int \,\frac {d^d p}{(2\pi)^d}\zeta_{UV}^2\left[p^2 \int_{u^{UV}}^{u^{IR}}\,d u z^{2}e^{(d-2)A}+\mathcal O (p^4)\right]\nonumber\\
&=&\left(-\frac 1 2\right)\int \,d^d x \left(\int_{u^{UV}}^{u^{IR}}\,d u z^{2}e^{(d-2)A}\right)\eta^{\mu\nu}\partial_\mu\zeta_{UV}\partial_\nu\zeta_{UV}+\mathcal O (p^4)\nonumber\\
&=&\int \,d^d x \left(-\frac 1 2{A'}^2\int_{\phi^{UV}}^{\phi^{IR}}\,\frac {d\phi}{W'}  {A'}^{-2}e^{(d-2)A}\right)\eta^{\mu\nu}\partial_\mu\chi_{UV}\partial_\nu\chi_{UV}\nonumber\\
&&+\int \,d^d x \left({A'}\int_{\phi^{UV}}^{\phi^{IR}}\,\frac {d\phi}{W'}  {A'}^{-2}e^{(d-2)A}\right)\eta^{\mu\nu}\partial_\mu\chi_{UV}\partial_\nu\psi_{UV}\nonumber\\
&&+\int \,d^d x \left(\frac 1 2\int_{\phi^{UV}}^{\phi^{IR}}\,\frac {d\phi}{W'}  {A'}^{-2}e^{(d-2)A}\right)\eta^{\mu\nu}\partial_\mu\psi_{UV}\partial_\nu\psi_{UV}+...\nonumber\\
&=&-\int \,d^d x e^{dA}\Big[ \left(\frac W{W'} {U_\ast}'\right)\frac 1 2e^{-2A}\eta^{\mu\nu}\partial_\mu\chi\partial_\nu\chi\nonumber\\
&&\qquad\qquad+2(d-1)  {U_\ast}'e^{-2A}\eta^{\mu\nu}\partial_\mu\chi\partial_\nu\psi+...\Big]_{UV}\nonumber\\
\end{eqnarray}
where $z=\frac {\dot\Phi}{\dot A}=(A')^{-1}$ and $\zeta=\frac \chi z-\psi=A' \chi -\psi$. We have expanded the gauge invariant variable $\zeta$ into scalar field fluctuations $\chi$ and metric fluctuations $\psi$. The covariantization of the first term $(\partial\chi)^2$ is the same as the kinetic term in the previous non-linear discussion.

We can also "derive" the Ricci term by covariantizing the second term $\partial\chi\partial\psi$. After an integrating by parts, the second term becomes
\begin{eqnarray}
&&-\int \,d^d xe^{dA}\left( {U_\ast}'\chi\right)\left(-2(d-1)e^{-2A}\eta^{\mu\nu}\partial_\mu\partial_\nu\psi\right)\nonumber\\
&=&-\int \,d^d x\sqrt{-\gamma} U_\ast(\Phi+\chi)R^{(d)}+...
\end{eqnarray}
which agrees with the Ricci term in the non-linear calculation. To the linear order, \be
R^{(d)}=-2(d-1)e^{-2A}\eta^{\mu\nu}\partial_\mu\partial_\nu\psi+...
\ee
 for $\gamma_{\mu\nu}=e^{2(A+\psi)}\eta_{\mu\nu}$.

Therefore, the covariantization of the linear result gives us the same bare on-shell action as the non-linear calculation
\begin{eqnarray}
S_{on-shell}^{\zeta}
&=&-\int \,d^d x e^{dA}\Big[ ( U_\ast'\chi)\left(-2(d-1)e^{-2A}\eta^{\mu\nu}\partial_\mu\partial_\nu\psi\right)\nonumber\\
&&\qquad\qquad\qquad+\left(\frac W{W'} {U\ast}'\right)\frac 1 2e^{-2A}\eta^{\mu\nu}\partial_\mu\chi\partial_\nu\chi+...\Big]\nonumber\\
&\rightarrow&-\int d^d x\sqrt\gamma\left(U_\ast R+\left(\frac W{W'}  U_\ast' \right)\frac 1 2\gamma^{\mu\nu}\partial_\mu\phi\partial_\nu\phi+...\right).
\end{eqnarray}


\end{document}